\documentclass[twocolappendix,numberedappendix]{emulateapj} 
\usepackage{amsmath}
\usepackage{amssymb}
\usepackage{color}
\usepackage{enumerate}

\newcounter{address}
\def\vector#1{\mbox{\boldmath $#1$}}
\newcommand{\kpc}{\ensuremath{\,\mathrm{kpc}}}
\newcommand{\kms}{\ensuremath{\,\mathrm{km\ s}^{-1}}}

\newcommand{\vlos}{\ensuremath{v_{\rm los}}}

\newcommand{\flag}[1]{\texttt{\lowercase{#1}}}
\newcommand{\emcee}{\flag{emcee}}

\newcommand{\eq}[1]{\begin{align}#1\end{align}}

\newcommand\nar{{New A Rev.}}%

\begin{document}
\shorttitle{Velocity anisotropy of Galactic halo stars} \shortauthors{Hattori, Valluri, Loebman \& Bell }
\title{Reliability of the measured velocity anisotropy of the Milky Way stellar halo}
\author{Kohei~Hattori\altaffilmark{1}, Monica~Valluri\altaffilmark{1}, Sarah~R.~Loebman\altaffilmark{1,2}, Eric~F.~Bell\altaffilmark{1}}
\altaffiltext{\theaddress}{\label{1}\stepcounter{address} Department of Astronomy, University of Michigan, Ann Arbor, MI, 48104, USA;  khattori@umich.edu}
\altaffiltext{\theaddress}{\label{2}\stepcounter{address} Michigan Society of Fellows}

\begin{abstract}
Determining the velocity distribution of halo stars is 
essential for estimating the mass of the Milky Way and for inferring its formation history. 
Since the stellar halo is a dynamically hot system, 
the velocity distribution of halo stars is well described by the 3-dimensional velocity dispersions $(\sigma_r, \sigma_\theta, \sigma_\phi)$, 
or by the velocity anisotropy parameter $\beta=1-(\sigma_\theta^2+\sigma_\phi^2)/(2\sigma_r^2)$. 
Direct measurements of $(\sigma_r, \sigma_\theta, \sigma_\phi)$ consistently suggest $\beta =0.5$-$0.7$ for nearby halo stars.  
In contrast, the value of $\beta$ at large Galactocentric radius $r$ is still controversial, since reliable proper motion data are available for only a handful of stars. 
In the last decade, several authors have tried to estimate $\beta$ for distant halo stars by fitting the observed line-of-sight velocities 
at each radius with simple velocity distribution models (local fitting methods). 
Some results of local fitting methods imply  $\beta<0$ at $r \gtrsim 20 \kpc$, 
which is inconsistent with recent predictions from cosmological simulations. 
Here we perform mock-catalogue analyses to show that the estimates of $\beta$ based on local fitting methods 
are reliable only at $r \leq 15 \kpc$ with the current sample size ($\sim10^3$ stars at a given radius).
As $r$ increases, the line-of-sight velocity (corrected for the Solar reflex motion) becomes increasingly closer to the Galactocentric radial velocity, so that
it becomes increasingly more difficult to estimate tangential velocity dispersion $(\sigma_\theta, \sigma_\phi)$ from line-of-sight velocity distribution.
Our results suggest that the forthcoming Gaia data will be crucial for understanding the velocity distribution of halo stars at $r \geq 20\kpc$.
\end{abstract}


\section{Introduction}\label{section:introduction}

The velocity distribution of halo stars provides a lot of useful information about the Milky Way. 
For example, by regarding the halo stars as the dynamical tracers, 
we can estimate the mass of the Milky Way through Jeans equation, 
if we can determine or assume their density $\rho$ and their 3-dimensional velocity dispersions $(\sigma_r, \sigma_\theta, \sigma_\phi)$ \citep{Dehnen2006,Gnedin2010}. 
Also, since the stellar halo is a collisionless system, the orbital shapes of halo stars are relatively immune to adiabatic change of the gravitational potential. Thus, the current distribution of orbits can provide some insight into how the stellar halo was formed.

A useful quantity to describe the orbital distribution of halo stars is 
the velocity anisotropy parameter \citep{Binney1980},
\eq{
\beta (r) = 1- \frac{\sigma_\theta^2(r) + \sigma_\phi^2(r)}{2 \sigma_r^2(r)} . 
}
By definition, $\beta=0$ corresponds to an isotropic velocity distribution. 
If radial or circular orbits dominate, $\beta$ is positive ($0<\beta<1$) or negative ($-\infty<\beta<0$), respectively. 
Since $\beta$ only depends on the velocity distribution at a given position 
and is independent of the potential,  $\beta$ is a potentially powerful tool with which 
to compare the dynamical state of the Milky Way and that of simulated galaxies.

Interestingly, essentially all the recently published simulations of Milky Way-like galaxies based on $\Lambda$CDM cosmology
show a qualitatively similar radial profile of $\beta$. Specifically,  the $\beta$ profile of most simulated stellar haloes 
almost monotonically increases from $\beta \simeq 0$-$0.5$ near the Galactic center, 
through $\beta \simeq 0.3$-$0.7$ near the Solar circle, 
to $\beta\simeq 0.5$-$1$ at virial radius \citep{Diemand2005,Sales2007}. 
Importantly, most simulated haloes do not show negative value of $\beta$ outside the Solar circle.\footnote{
In a companion paper \cite{Loebman2016} discuss some exceptional situations where this general $\beta$ profile is not attained.}
This characteristic $\beta$ profile was also found in classical simulations of an initially cold stellar system \citep{vanAlbada1982} that collapses and experiences violent relaxation \citep{Lynden-Bell1967}. 
The fact that most cosmological simulations predict a qualitatively similar profile of $\beta$ is intriguing, 
since the radial profile of $\beta$ may be used to test $\Lambda$CDM cosmology.

The observed value of $\beta=0.5$-$0.7$ of the halo stars in the Solar neighborhood is consistent with these simulations \citep{CY1998,CB2000,Smith2009,Bond2010}. 
The direct determination of $\beta$ at larger Galactocentric radii has been hampered by difficulty in obtaining accurate proper motion data, 
and it is only recently that \cite{Deason2013} obtained $\beta=0.0_{-0.4}^{+0.2}$ at $18 \kpc < r < 30 \kpc$ 
by using 13 distant halo stars with proper motion data derived with the Hubble Space Telescope (see \citealt{Cunningham2016} for their updated results). 
Their measurement of $\beta \simeq 0$ at large $r$ is intriguing, since it differs significantly from the characteristic $\beta(r)$ profile found in simulated galaxies.

Prior to these direct measurements of $\beta$, there were several attempts to estimate $\beta$ in the distant halo by using 3-dimensional positions and line-of-sight velocities. 
Broadly speaking, these methods can be classified into two methodologies: global fitting methods and local fitting methods.

The global fitting is based on global models of the stellar halo. In these methods, a functional form of the global model of the stellar halo is assumed. Then the likelihood of the model parameters given the line-of-sight velocity data is evaluated to determine the best-fit parameters. 
For example, 
\cite{SL1994} and \cite{SL1997} chose the functional forms of the radial and tangential velocity dispersion profiles that satisfy the spherical Jeans equation and fitted the 4-dimensional information of halo stars with these models to claim a declining profile of $\beta(r)$.
The results from \cite{SL1997} suggest 
that $\beta \simeq 0.5$ at $r = 8\kpc$, $\beta \simeq 0$ at $r = 20 \kpc$, 
and $\beta \simeq -1.3$ at $r = 50 \kpc$. 
Also, \cite{Deason2012} used a distribution function model to fit the 4-dimensional information for halo stars located at $16 \kpc < r < 48 \kpc$ and claimed that $\beta \simeq 0.5$. 
More recently, \cite{Williams2015} analyzed a similar dataset with a more sophisticated action-based distribution function model and 
derived a radially varying $\beta$ profile with $\beta \simeq 0.4$ at $r=15 \kpc$ and $\beta \simeq 0.65$ at $r=50 \kpc$ -- a result that is qualitatively similar to cosmological simulations, but rises more slowly with increasing radius.

The local fitting methods, on the other hand, interpret the line-of-sight velocity distribution of halo stars with a simple statistical model (e.g., Gaussian distribution) and estimate the velocity moments of halo stars.
The idea behind these studies is simple: if the velocity distribution of halo stars is isotropic ($\beta=0$), we expect the line-of-sight velocity dispersion $\sigma_{\rm los}$ to be independent of the heliocentric line-of-sight direction. On the other hand, if $\beta \neq 0$, we expect the line-of-sight projection of the velocity ellipsoid gradually changes across the sky, due to the off-center location of the Sun in the Milky Way. 
In order to derive the velocity dispersion profile as a function of Galactocentric radius $r$, authors often divide their sample into several bins according to $r$ and perform the analyses for each radial bin. 
Therefore, these methods can be considered as a series of local fits of the velocity distribution. 
For example, 
\cite{Sirko2004} assumed that their sample of halo stars obeys a Gaussian velocity distribution 
and claimed $\beta \simeq 0$. 
\cite{Kafle2012} and \cite{King2015} used a similar formulation as in \cite{Sirko2004} 
for a larger sample of stars and claimed that $\beta<0$ at $r\gtrsim20 \kpc$.
Also, \cite{Hattori2013} and \cite{Kafle2013} even claimed a metallicity-dependence of $\beta$ such that $\beta < 0$ for relatively metal-poor halo stars and $\beta>0$ for relatively metal-rich halo stars.

It is currently unclear whether these discrepant estimates of $\beta$ are a result of the fitting method or difference in the sample of stars used. However, it is important to note that both global fitting and local fitting methods have their own advantages and disadvantages, and these methods are complementary to each other. In order to illustrate this point, we now compare the advantages and disadvantages associated with the (global) distribution function fitting and the local fitting methods.

When a distribution function model is used to fit a given sample of halo stars, is is generally assumed that the stellar halo is in dynamical equilibrium and the functional form of the distribution function is known. These basic assumptions have some advantages for the distribution function fitting. For example, these models are designed to be physical (e.g., non-zero phase-space density), so the best-fit solution for any given data set is guaranteed to be physical. Also, distribution function (combined with the potential model of the Milky Way) contains all the information needed to calculate any velocity moments. Thus, if we have some external knowledge on the stellar halo (such as the density profile estimated from other surveys) in addition to the kinematical data we are trying to fit, we can naturally incorporate additional information to improve the fit. However, the basic assumptions also have disadvantages: it is unclear if the stellar halo is really in dynamical equilibrium and if the assumed functional form is really adequate to model the stellar halo.

Local fitting methods, on the other hand, do not require the stellar halo to be in dynamical equilibrium, since we just need to fit the current velocity distribution at a given Galactocentric radius. In these methods, we need to assume the functional form of the velocity distribution (or at least some  properties of the functions, such as the symmetry of the velocity distribution), but we can assume simple and flexible functions (e.g., Gaussian velocity distribution) to mitigate the arbitrariness of the chosen functions. 
By design, the best-fit model is not guaranteed to be in dynamical equilibrium (especially when the sample size is small), even if there is some external evidence to believe that the stellar halo is in dynamical equilibrium. However, the local fitting methods are useful if the stellar halo is not in dynamical equilibrium. For example, let us consider an idealized situation where the stellar halo is a sum of smooth component and a substructure, and suppose that the contribution from the substructure is only prominent at a certain Galactocentric radius. In this case, local fitting methods may detect the presence of substructure as an anomaly in the velocity dispersion profile, while global fitting methods 
which require smooth distribution functions may be adversely affected 
by this local substructure. Indeed, \cite{Kafle2012} used a local fitting method and claimed that the radial profile of $\beta(r)$ shows a dip-like structure at $r=17 \kpc$. Although we will show in this paper that this dip might not be a real signature of the $\beta$ profile (due to the small sample size), at least it is safe to say that currently proposed stellar halo distribution functions do not accommodate this kind of dip; and hence these two classes of fitting methods are complementary to each other.

In this paper, we focus on the local fitting methods and evaluate how far out in the stellar halo are these methods can reliably estimate $\beta(r)$  from the currently available 4 dimensional data. To this end, we apply the local fitting methods to a set of mock catalogues and compare the estimated profile of $\beta$ with the input profile of $\beta$. Here we report that the widely-used local fitting methods can reliably recover $\beta$  within $r<15 \kpc$, but can only weakly constrain $\beta$ at $r>15 \kpc$, if 1000 sample stars are used for a given radius.
This work is an extension of \cite{Hattori2013}, who applied a matrix-based local fitting method to Sloan Digital Sky Survey (SDSS) data and pointed out that the estimation of $\beta$ from the line-of-sight velocity distribution can be biased at $r\gtrsim$ (16-18) $\kpc$. It is important to note that most of the above-mentioned works of local- and global-fitting methods (except for \citealt{Sirko2004} and \citealt{Hattori2013}) do not present comparison with mock data to validate their methodology. Thus, their results at large $r$ might have been affected by systematic errors that were not accounted for.\footnote{
\cite{Williams2015} confirmed that the observed line-of-sight velocity distribution 
is well reproduced by that of mock data generated from their best-fit model. 
However, this procedure is not enough to guarantee that the $\beta(r)$ profile of their best-fit model is similar to 
the $\beta(r)$ profile of the observed halo population.
}

The outline of this paper is as follows.
In Section 2 we describe the maximum-likelihood and Bayesian formulations for inferring $\beta$ from 4-dimensional data.
In Section 3 we describe our mock catalogues. 
In Section 4, we show the results of our maximum-likelihood analyses of our mock catalogues. 
In Section 5, we show the results of our Bayesian mock-analyses. 
Section 6 presents a discussion, and Section 7 sums up.

\section{Method} \label{section:method}

Here we outline our maximum-likelihood and Bayesian formulations of the local fitting method 
to estimate the 3-dimensional velocity dispersion of halo stars from 4-dimensional phase-space coordinates (3D positions and line-of-sight velocities).

We note that 
we do not take into account any observational errors in our formulation 
(nor in our mock catalogues; see Section \ref{section:mock_catalogues}), 
since the main aim in this paper is to demonstrate 
how the performance of these widely-used local fitting methods deteriorates 
as the Galactocentric radius of the sample stars increases, 
even if we use idealized stellar data.

\subsection{Maximum-likelihood method} \label{section:formulationMaxL}

Following \cite{Sirko2004} and \cite{Kafle2013}, we assume that the distribution of velocity $\vector{v}$ of halo stars 
at a given location $\vector{x}$ 
takes the form\footnote{
This functional form includes the classical distribution function, the Osipkov-Merritt model in a singular isothermal potential (see Appendix \ref{appendix:OM}) as a special case.
} 
\eq{
&f_{\rm Gauss}(\vector{v} | \vector{x}) = 
\frac{1}{(\sqrt{2\pi})^{3/2} \sigma_r \sigma_\theta \sigma_\phi} \nonumber \\
& \times \exp 
\left[ - \left( 
   \frac{v_r^2}{2\sigma_r^2(r) }  
+ \frac{v_\theta^2}{2\sigma_\theta^2(r)}
+ \frac{(v_\phi - V_{\rm rot}(r) )^2}{2\sigma_\phi^2(r) } 
\right) \right]. \label{eq:DF}
}
We adopt a spherical coordinate system $(r, \theta, \phi)$ 
such that $r=| \vector{x} |$ is the Galactocentric radius, 
$\theta$ is the polar angle ($\theta=0$ corresponds to the Galactic disc plane), 
and $\phi$ is the azimuthal angle.
The principal axes of the velocity ellipsoid are assumed to be aligned with this spherical coordinate system, 
which can be justified by the recent work of \cite{Evans2016}.
Also, 3-dimensional velocity dispersions $(\sigma_r, \sigma_\theta, \sigma_\phi)$ 
and 
the mean azimuthal velocity $V_{\rm rot}$ 
are assumed to be functions of $r$ only. 
The last assumption implies
a spherical density distribution of halo stars and 
a spherical potential of the Milky Way. 
This contradicts the claimed flattening of the stellar halo \citep{Deason2011} and the potential \citep{Koposov2010}, 
although the density distribution may be less flattened in the outer part of the stellar halo \citep{Carollo2007}.

Under this velocity distribution model, 
the probability density that a star located at $\vector{x}$ has the line-of-sight velocity $v_{\rm los}$ in the Galactic rest frame 
for a given set of parameters $(\sigma_r, \sigma_\theta, \sigma_\phi, V_{\rm rot})$ is expressed as 
(see Appendix A of \citealt{Sirko2004} for derivation)
\eq{
&P(v_{\rm los} | \vector{x}, \sigma_r, \sigma_\theta, \sigma_\phi, V_{\rm rot} ) \nonumber \\
&= \frac{1}{\sqrt{2\pi} \sigma_{\rm los} (\vector{x})} 
\exp \left[ 
- \frac{(v_{{\rm los},i} - V_{\rm rot}Q_{\phi}(\vector{x}))^2}{2 \sigma_{\rm los}^2 (\vector{x})}
\right] .
\label {eq:GaussLOS}
}
Here, 
\eq{
\sigma_{\rm los}(\vector{x}) = \sqrt{\sigma_r^2(r) Q_r^2(\vector{x}) + \sigma_\theta^2(r) Q_\theta^2(\vector{x}) + \sigma_\phi^2(r) Q_\phi^2(\vector{x})}
\label {eq:LOSVD}
}
is the line-of-sight velocity dispersion of halo stars located at $\vector{x}$. 
Also, 
$(Q_r, Q_\theta, Q_\phi) = (
\vector{e}_{\rm los} \cdot \vector{e}_r , 
\vector{e}_{\rm los} \cdot \vector{e}_\theta, 
\vector{e}_{\rm los} \cdot \vector{e}_\phi
)$ 
are dot products of 
the unit vector along the line-of-sight $\vector{e}_{\rm los}$ 
and each of the unit vectors $(\vector{e}_r, \vector{e}_\theta, \vector{e}_\phi)$ 
of the spherical coordinate system at $\vector{x}$.

Suppose we have a sample of $N$ halo stars 
such that the location and line-of-sight velocity of $i$th star, $(\vector{x}_i, v_{{\rm los},i})$, are known ($i=1,\cdots, N$) 
and that $N$ stars have an identical Galactocentric radius $r=|\vector{x}_i|$. 
Then, the total log-likelihood of the observational data given the parameters is expressed as  
\eq{
&\ln L = \ln \left[ \prod_{i=1}^{N} P(v_{{\rm los},i} | \vector{x}_i, \sigma_r, \sigma_\theta, \sigma_\phi, V_{\rm rot} ) \right] \nonumber \\
&= - \frac{N}{2} \ln (2\pi)
- \sum_{i=1}^{N} \left[ 
\ln \sigma_{\rm los} (\vector{x}_i) 
+\frac{(v_{{\rm los},i} - V_{\rm rot}Q_{\phi}(\vector{x}_i))^2}{2 \sigma_{\rm los}^2 (\vector{x}_i)} 
\right] .
\label{eq:lnL}
}
The maximum-likelihood method (in the limit of no observational errors)
finds the set of parameters that maximizes $\ln L$ 
at each radius $r$.

\subsection{Bayesian method} \label{section:method_Bayes}

Another useful method to estimate $\beta$ from 4-dimensional information is the Bayesian method \citep{Kafle2012}. 
The Bayesian method explicitly incorporates information about prior knowledge or constraints on parameters. 
The Bayesian method has the advantage that it makes it clear whether the data are constraining the parameters or the answers (posterior distribution) are driven by the priors.
Here we briefly outline the formulation of the Bayesian method.

As in the maximum-likelihood method, 
let us assume that the distribution function of the stellar halo is given by equation (\ref{eq:DF}). 
In Bayesian formulation, our aim is to obtain the posterior distribution of the model parameters given the data. 
In our case, the posterior distribution can be expressed as (via Bayes' Theorem)
\eq{
&P(\sigma_r, \sigma_\theta, \sigma_\phi, V_{\rm rot} | \{ (\vector{x}_i, v_{{\rm los},i}) \}_{i=1}^{N}) \nonumber \\
&= \frac{P( \{ v_{{\rm los},i} \}_{i=1}^{N}  | \{ \vector{x}_i \}_{i=1}^{N}, \sigma_r, \sigma_\theta, \sigma_\phi, V_{\rm rot}) P(\sigma_r, \sigma_\theta, \sigma_\phi, V_{\rm rot})}{P( \{ \vector{x}_i, v_{{\rm los},i} \}_{i=1}^{N} )} .
} 
Here the likelihood $P( \{ v_{{\rm los},i} \}_{i=1}^{N}  | \{ \vector{x}_i \}_{i=1}^{N}, \sigma_r, \sigma_\theta, \sigma_\phi, V_{\rm rot})$ is identical to $L$ in equation (\ref{eq:lnL}) 
and the evidence $P( \{ (\vector{x}_i, v_{{\rm los},i}) \}_{i=1}^{N} )$ can be regarded as a constant. 
Thus the only additional task for us is to set a certain prior distribution $P(\sigma_r, \sigma_\theta, \sigma_\phi, V_{\rm rot})$ of the model parameters.

In order to be as objective as possible in judging the performance of the Bayesian method, 
we use three types of relatively uninformative 
priors A, B and C as described below. 
We note that in the limit of an infinite number of sample stars with no error, 
the posterior distribution is expected to be independent of the choice of prior. 
However, in reality we only have a finite number of stars, 
so we need to use appropriate prior information in order to make the best use of the available data.

Prior A is a uniform prior for all the parameters  $(\sigma_r, \sigma_\theta, \sigma_\phi, V_{\rm rot})$  given by
\eq{
&P_A(\sigma_r, \sigma_\theta, \sigma_\phi, V_{\rm rot}) \nonumber \\
&\propto
\begin{cases}
1, \;\; ( \sigma_{\rm tot}<v_{\rm esc}, \; |V_{\rm rot}|<v_0), \\
0, \;\;{\rm (otherwise)}.
\end{cases}
}
Here, we define $\sigma_{\rm tot} \equiv \sqrt{\sigma_r^2+\sigma_\theta^2+\sigma_\phi^2}$. 
Also, $v_0(r) = 220 \kms$ and $v_{\rm esc}(r)$ are the circular velocity and escape velocity 
at the Galactocentric radius $r$ of a truncated singular isothermal potential, respectively 
(although the details of the potential model do not affect the results).

From a mathematical point of view, prior A is not purely uninformative,
since $(\sigma_r, \sigma_\theta, \sigma_\phi)$ in our model are so-called {\it scale parameters} 
(while $V_{\rm rot}$ in our model is a so-called {\it location parameter}). 
The Jeffreys' rule (\citealt{Jeffreys1961}, Section 3.10; \citealt{Ivezic2013}, Section 5.2.1), 
suggests that for a scale parameter, a more appropriate choice of a prior is one
that is inversely proportional to the scale parameter. 
However, since the use of Jeffreys' rule for more than one parameters is controversial 
(\citealt{Robert2009}, Section 4.7), we adopt two additional priors: 
\eq{
&P_B(\sigma_r, \sigma_\theta, \sigma_\phi, V_{\rm rot}) \nonumber \\
&\propto
\begin{cases}
\sigma_r^{-1}, \;\; ( 1\kms<\sigma_{\rm tot}<v_{\rm esc}, \; |V_{\rm rot}|<v_0), \\
0, \;\;{\rm (otherwise)}, 
\end{cases}
}
and 
\eq{
&P_C(\sigma_r, \sigma_\theta, \sigma_\phi, V_{\rm rot}) \nonumber \\
&\propto
\begin{cases}
(\sigma_r \sigma_\theta \sigma_\phi)^{-1}, \;\; ( 1\kms<\sigma_{\rm tot}<v_{\rm esc}, \; |V_{\rm rot}|<v_0), \\
0, \;\;{\rm (otherwise)}.  
\end{cases}
}
We note that the lower limit on $\sigma_{\rm tot}$ is set to be a small but non-zero value ($1\kms$) 
so that the prior distribution can be normalized. 
The upper limit  on $\sigma_{\rm tot}$ is chosen to be $v_{\rm esc}$ 
so that most of the stars are bound to the Milky Way, 
but we have confirmed that our results do not change when a larger value is adopted.

In practice, it is fair to state that these three priors are equally uninformative, 
so the use of any one of them is equally justified. 
It is important to note that priors A and B are independent of $(\sigma_\theta, \sigma_\phi)$, 
which implies that using these prior distributions is equivalent to setting a flat prior on velocity anisotropy $\beta$. 
In contrast, prior C is weighted heavily towards small values of $(\sigma_\theta, \sigma_\phi)$, 
and therefore towards large value of $\beta (\simeq 1)$.

\section{Mock catalogues} \label{section:mock_catalogues}

Here we describe how we generate the mock catalogues 
with which we test the maximum-likelihood and Bayesian methods. 

\subsection{Assumptions on our mock catalogues}

In generating  the mock catalogues we first assume that the sample stars obey the distribution function model in equation (\ref{eq:DF}) 
with no net rotation $(V_{\rm rot} = 0)$. 
Since we want to quantify the error associated with $\beta$, 
we simply assume that $(\sigma_r, \sigma_\theta, \sigma_\phi) = (1, \sqrt{1-\beta_{\rm true}}, \sqrt{1-\beta_{\rm true}}) \times (100\; {\rm {km \; s^{-1}}})$, 
independent of $r$.

Also, we assume that each mock catalogue contains 1000 stars, 
and all of them have an identical Galactocentric radius. 
In reality, 
most of the previous studies used a few thousand halo stars in total, with stars binned 
according to their Galactocentric radii.
Since such a bin typically contains a few hundred stars, 
our mock catalogues are better populated than reality. 
Also, our mock catalogues are much simpler to analyze 
since we can ignore the radial dependence of the halo density.

Furthermore, we assume a simple model for the spatial selection function that mimics the  Sloan Digital Sky Survey (SDSS). 
Specifically, 
all the stars in a given mock catalogue 
are distributed at high Galactic latitude with $|b| > 30^\circ$ and 
are distributed more than $5 \kpc$ away from the Galactic disc plane, 
but otherwise they are distributed uniformly in $(\sin \theta, \phi)$-space. 
We have confirmed that our results are essentially unchanged 
if we do not apply the cuts on Galactic latitude or distance from the disc plane.

Lastly, once we generate mock catalogues, 
we transform the 3-dimensional velocities in the Galactocentric frame to a line-of-sight velocity in the frame of an 
observer (Sun) moving on a circular orbit with radius $8 \kpc$ at a velocity of $220 \kms$.

These assumptions (especially the assumption of no observational errors) are rather simplistic and idealistic.
However, by using mock catalogs with these assumptions, we can be sure that any systematic errors associated with our mock analyses 
are {\it no less serious than} the systematic errors affecting previous local fitting analyses 
\citep{Sirko2004, Kafle2012, Kafle2013, Hattori2013,King2015}. 

\subsection{Parameters of our mock catalogues}

In this paper, we generate 1000 mock catalogues 
for a given set of parameters $(r, \beta_{\rm true})$. 
The Galactocentric radius of the sample stars are assumed to be either $r=6,7, \cdots, 30 \kpc$ in steps of $1 \kpc$. 
Also, we adopt eight values of $\beta_{\rm true} = 0.75, 0.5, \cdots, -1$ in steps of $0.25$. 
Thus we generate in total $25 \times 8 \times 1000=2\times10^5$ mock catalogues, 
each contains 1000 mock stars.

\section{Result 1: maximum-likelihood method} \label{section:resultMaxL}

Here we investigate the reliability of the maximum-likelihood method. 
In this Section we analyze our mock catalogues with the maximum-likelihood method to derive 
$(\sigma_r,\sigma_\theta,\sigma_\phi,V_{\rm rot})$ from 4-dimensional information. 
Then we calculate the corresponding values of velocity anisotropy $\beta_{\rm MaxL}$ 
(the maximum-likelihood solution for the velocity anisotropy).
These calculations were performed by using GNU Scientific Library \citep{Galassi2009GSL}.

\subsection{Illustrative results} \label{section:showCase}

Since the total number of  mock catalogues employed in this paper is huge, 
we begin by presenting results that illustrate the performance of the maximum-likelihood method, focusing first on analyses of mock catalogues with $\beta_{\rm true}=0.5$ and $-1$. 

\begin{figure*}
\begin{center}
	\includegraphics[angle=0,width=1.95\columnwidth]{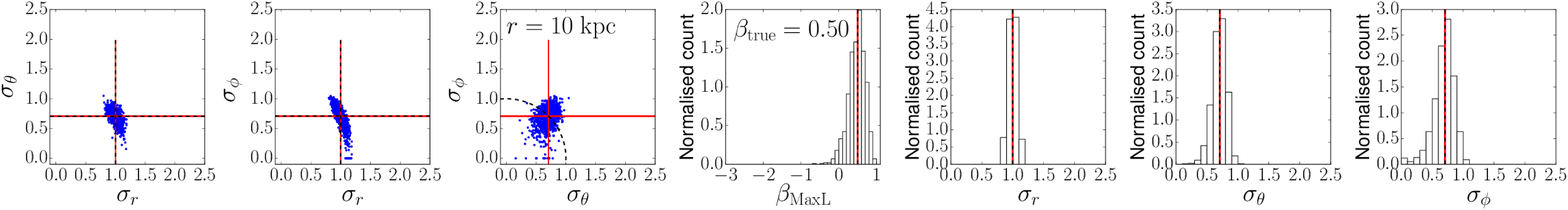} \\
	\includegraphics[angle=0,width=1.95\columnwidth]{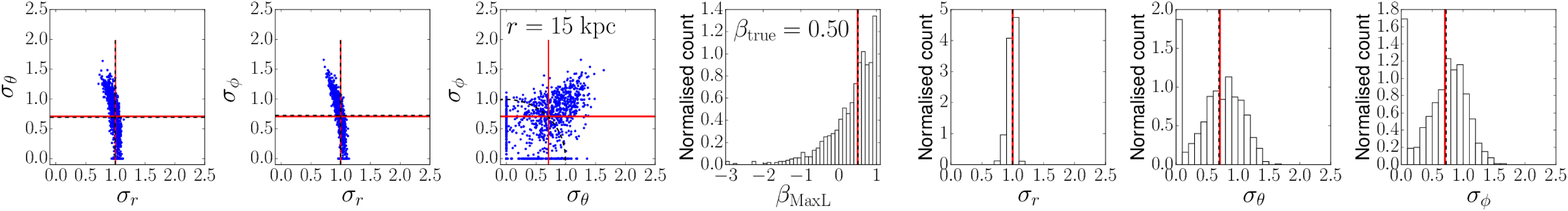} \\
	\includegraphics[angle=0,width=1.95\columnwidth]{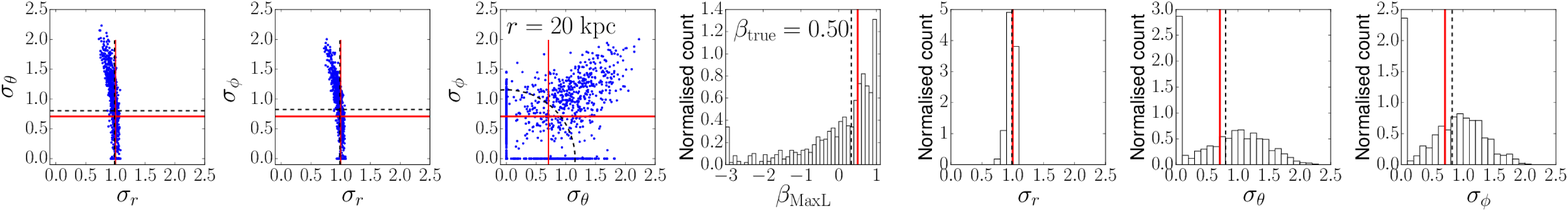} \\
	\includegraphics[angle=0,width=1.95\columnwidth]{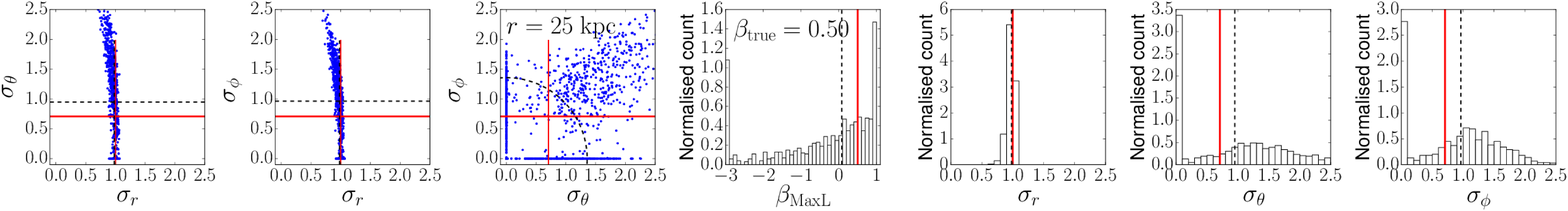}
\end{center}
\caption{
The distribution of the maximum-likelihood solutions for our mock catalogues with $\beta_{\rm true}=0.5$.  Each row shows the solutions  for $(\sigma_r, \sigma_\theta, \sigma_\phi, \beta_{\rm MaxL})$ 
 for our 1000 mock catalogues at a single value of $r$ 
($r/ \kpc =10,15,20$, and $25$). 
We note that $(\sigma_r, \sigma_\theta, \sigma_\phi)$ are normalized by a constant value of $100 \kms$. 
The three scatter plots in each row 
show the distributions of $(\sigma_r, \sigma_\theta)$, $(\sigma_r, \sigma_\phi)$, and $(\sigma_\theta, \sigma_\phi)$. 
The other panels in each row  show the histograms of $\beta_{\rm MaxL}, \sigma_r, \sigma_\theta$, and $\sigma_\phi$. The vertical and horizontal red solid lines show the true values of the mock catalogues. The vertical and horizontal black dashed lines  indicate the median value of the 1000 solutions. On the third panel from the left,  the black dashed arc indicates the values of $(\sigma_\theta, \sigma_\phi)$ that correspond to the median value of $\beta_{\rm MaxL}$ [see equation (\ref{eq:arc})]. 
The performance of the maximum-likelihood method deteriorates as $r$ increases, 
as can be seen in the more broadened distribution of $(\sigma_\theta, \sigma_\phi)$ 
and the more broadened histogram of $\beta_{\rm MaxL}$ at larger $r$.
}
\label{fig:showCaseBetap0.5_MaxL}
\end{figure*}

Figure \ref{fig:showCaseBetap0.5_MaxL}  shows the distributions of the maximum-likelihood solutions for 1000 mock catalogues with $\beta_{\rm true}=0.5$. From top to bottom the panels show the results for $r/\kpc =10, 15, 20$, and $25$, demonstrating how the results of the maximum-likelihood estimation deteriorate with increasing $r$. 

The histograms of $\sigma_r$ (5th column from the left), show that
$\sigma_r$ is well estimated at all radii, 
and the median value of $\sigma_r$ (black dashed line) almost coincides with the true value (red solid line). 
On the other hand, the maximum-likelihood estimates of both tangential velocity dispersion components become broader as $r$ increases and the median and true values increasingly diverge. 
For example, the histograms of $\sigma_\phi$ (right-most column) at $r=10 \kpc$ and $r=15 \kpc$ show that the median values of $\sigma_\phi$ coincide with the true values. Note that for  $r=15 \kpc$ a fraction of solutions are clustered at  $\sigma_\phi=0$, and are unrealistic (the origin of these unrealistic solutions is discussed in the Appendix \ref{appendix:unphysical}.) The fraction of unrealistic solutions increases as $r$ increases. 
Also, we note that the median value of $\sigma_\phi$ begins to deviate from the exact value at $r > 15 \kpc$. 
These properties are also true for $\sigma_\theta$. 
Since $\sigma_r$ is well estimated, 
the error in $\beta_{\rm MaxL}$ is dominated by the errors in $(\sigma_\theta, \sigma_\phi)$. 
As a result, the median value of $\beta_{\rm MaxL}$ begins to deviate from the true value of $\beta_{\rm true}=0.5$ at $r> 15 \kpc$ (see 4th column). 

The performance of the maximum-likelihood method can be well summarized in the 
$(\sigma_\theta, \sigma_\phi)$-space (3rd column). 
In this space, a curve of constant $\beta$ (when $\sigma_r$ is fixed to the true value) 
is described by an arc defined by 
\eq{
\sigma_\theta^2 + \sigma_\phi^2 = (1-\beta) \sigma_{r, {\rm true}}^2  . \label{eq:arc}
}
In the 3rd column of Figure \ref{fig:showCaseBetap0.5_MaxL},  the  arc (black dashed curve) corresponds to $\beta = {\rm median}(\beta_{\rm MaxL})$, while the blue dots show the distribution of the solutions. 
At $r=10 \kpc$, the maximum-likelihood solutions are distributed compactly around the true values (marked by red lines),  and the arc of 
$\beta (\sigma_\theta, \sigma_\phi) = {\rm median}(\beta_{\rm MaxL})$ 
goes through the true location of $(\sigma_{\theta, {\rm true}}, \sigma_{\phi, {\rm true}})$. 
At $r=15\kpc$, the distribution of $(\sigma_\theta, \sigma_\phi)$ is broadened 
and a fraction of solutions are found to be unrealistic ($\sigma_\theta=0$ or $\sigma_\phi=0$). 
At $r > 15 \kpc$, the distribution of the solutions is broadened further and the fraction of unrealistic solution is increased. 
Also, some fraction of solutions attain a large value of $\sigma_\theta^2 + \sigma_\phi^2$,
which corresponds to a highly negative value of $\beta_{\rm MaxL}$. 
As a result, the arc of median $\beta_{\rm MaxL}$ begins to deviate from the 
true location of $(\sigma_{\theta, {\rm true}}, \sigma_{\phi, {\rm true}})$ at $r > 15 \kpc$.

\begin{figure*}
\begin{center}
	\includegraphics[angle=0,width=1.95\columnwidth]{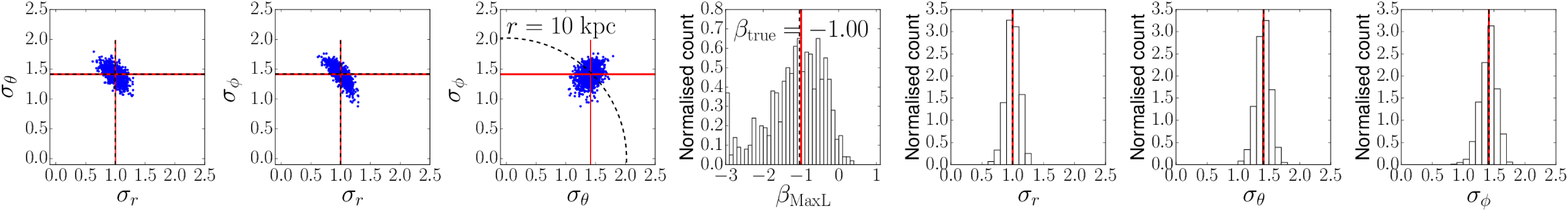} \\
	\includegraphics[angle=0,width=1.95\columnwidth]{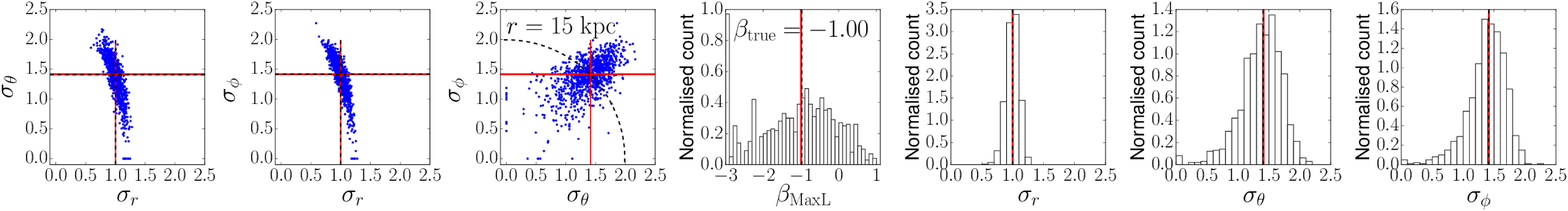} \\
	\includegraphics[angle=0,width=1.95\columnwidth]{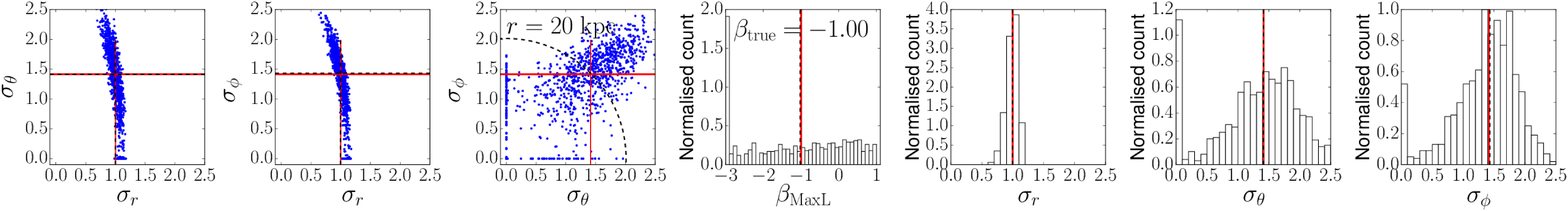} \\
	\includegraphics[angle=0,width=1.95\columnwidth]{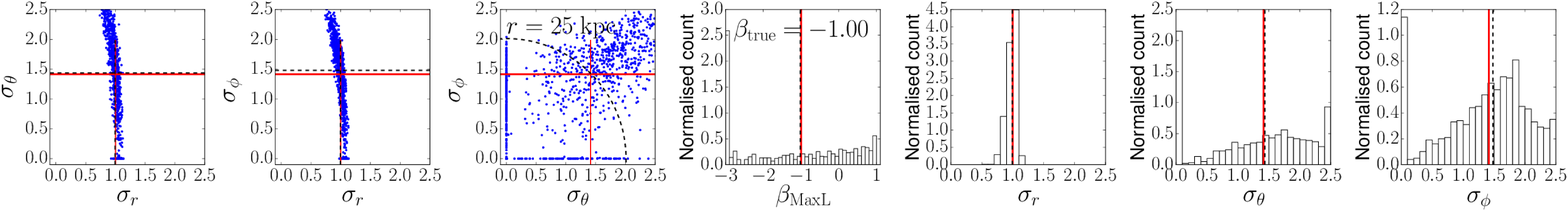}
\end{center}
\caption{
Figure showing the same as in Figure \ref{fig:showCaseBetap0.5_MaxL}, but with $\beta_{\rm true}=-1$. 
}
\label{fig:showCaseBetam1.0_MaxL}
\end{figure*}

Figure \ref{fig:showCaseBetam1.0_MaxL}  shows the same results but with mock catalogues with $\beta_{\rm true}=-1$. 
Again, the maximum-likelihood solutions deteriorate at $r > 15 \kpc$. 
In this case, the median value of $\beta_{\rm MaxL}$ 
happens to stay very close to the exact value of $\beta_{\rm true}=-1$ even at $r=25 \kpc$. 
However, this result only suggests that, 
in the case that $\beta_{\rm true} =-1$, 
the maximum-likelihood method {\it on average} returns the unbiased value of $\beta_{\rm MaxL}$ 
{\it for a large number of independent datasets}. 

As we can see from the highly broadened histogram of $\beta_{\rm MaxL}$ (4th column), 
the maximum-likelihood method hardly ever constrains the true anisotropy at $r > 15 \kpc$ if we only use $\sim 10^3$ stars for a given radius. 
In principle, the quality of the estimate of $\beta$ can be improved by increasing the sample to $\sim 10^4$ stars at a given radius (see Appendix \ref{appendix:N1e4} and Section \ref{section:N1e4}). However, the prospects for obtaining line-of-sight velocities for such a large sample is observationally infeasible in the near future.

\subsection{Detailed analyses of maximum-likelihood method}

In Section \ref{section:showCase}, we demonstrated that the maximum-likelihood method becomes unreliable at large $r$, 
especially at $r > 15 \kpc$. 
Here we have a closer look at this problem.

For each pair of $(r, \beta_{\rm true})$, we have 1000 mock catalogues, 
so we have an {\it ensemble} of 1000 solutions. 
In order to evaluate the statistical properties of the solutions, 
we derive the 2.5, 16, 50, 84 and 97.5 percentiles 
of $\beta_{\rm MaxL}$ 
and investigate how the percentile ranges depend on $r$ and $\beta_{\rm true}$.

\subsubsection{Results for fixed $\beta_{\rm true}$} \label{section:MaxL_fixedBeta}

Here we investigate how the performance of the maximum-likelihood method depends on $r$.

\begin{figure*}
\begin{center}
	\includegraphics[angle=0,width=0.475\columnwidth]{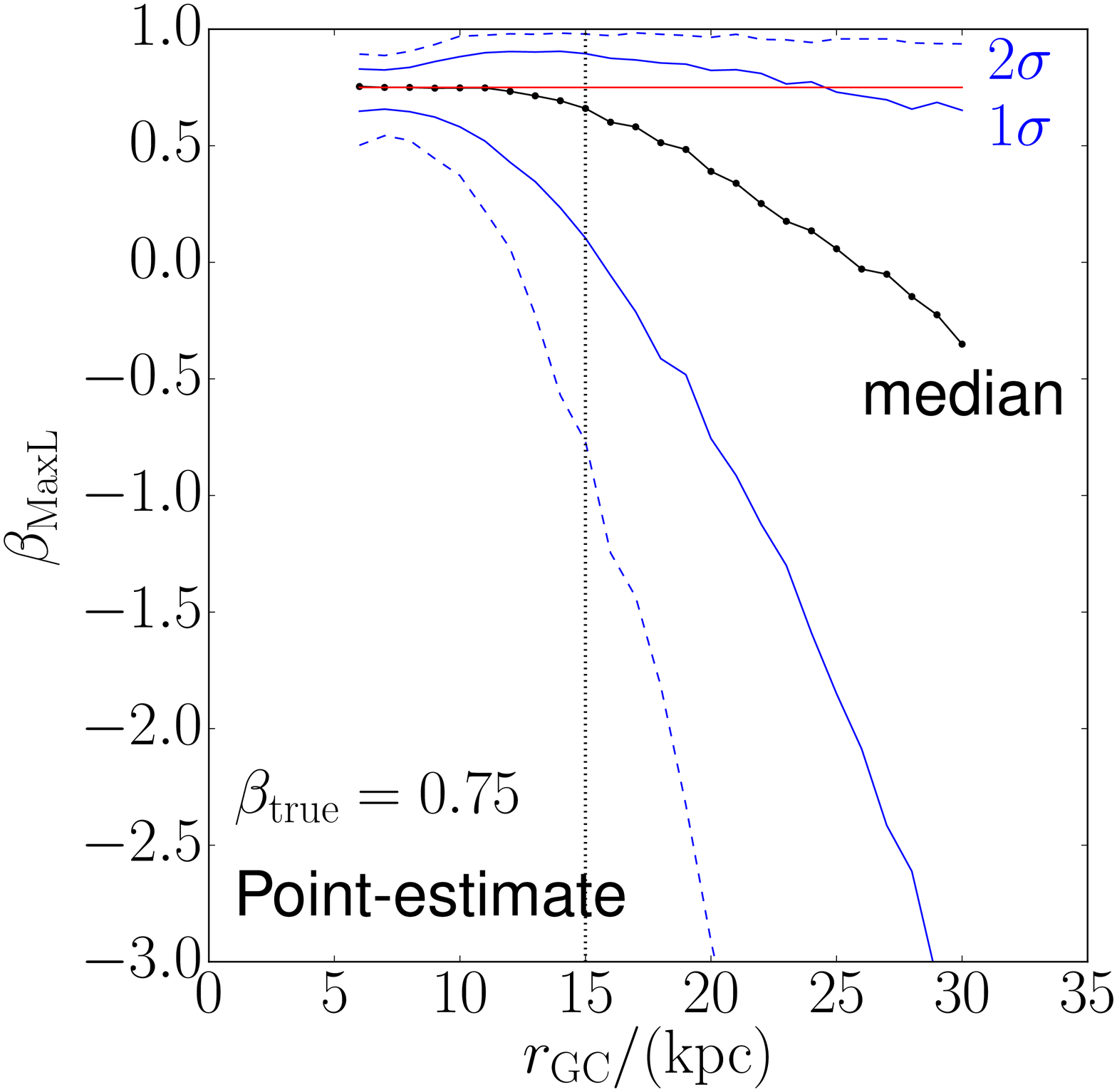} 
	\includegraphics[angle=0,width=0.475\columnwidth]{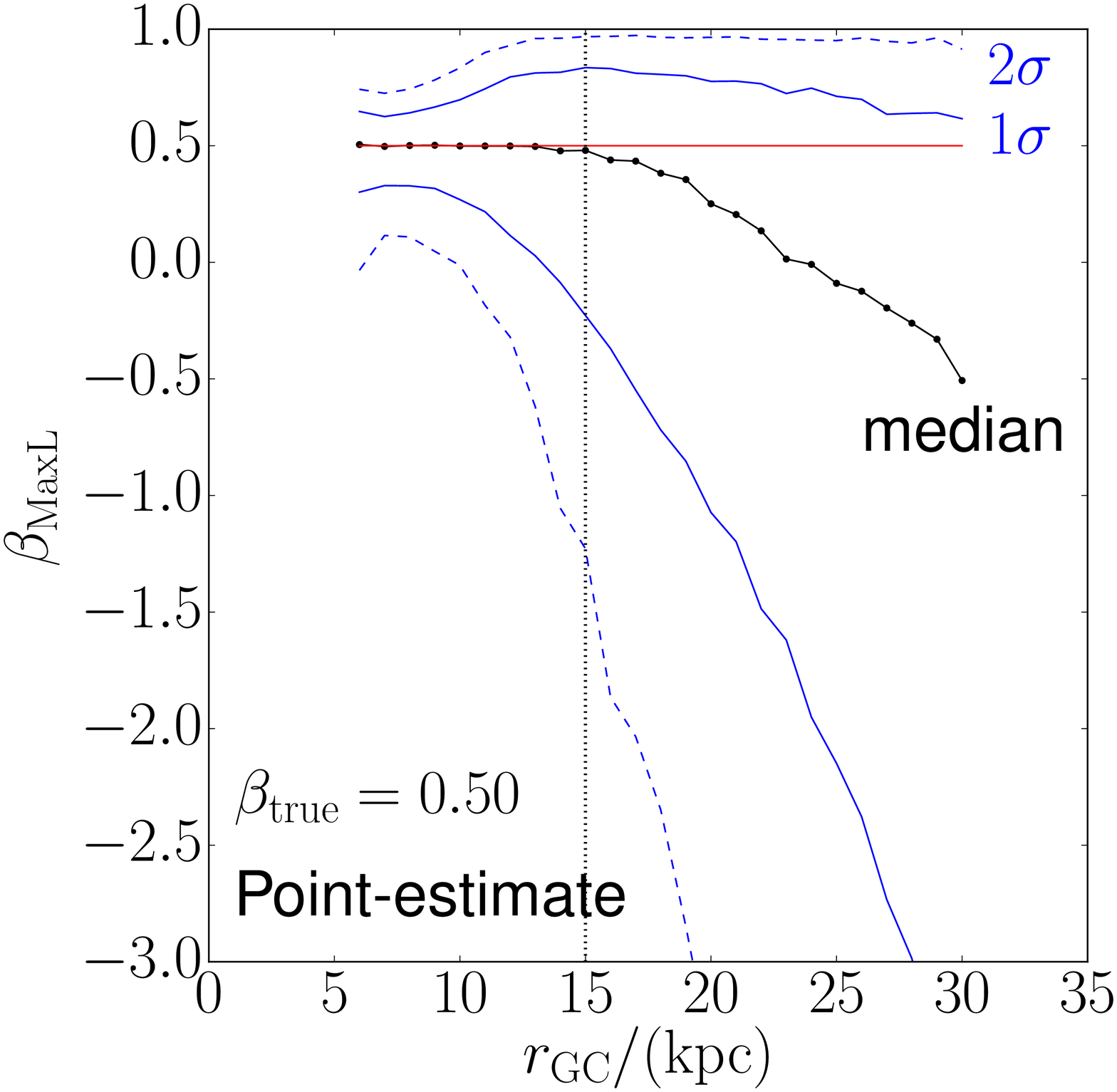} 
	\includegraphics[angle=0,width=0.475\columnwidth]{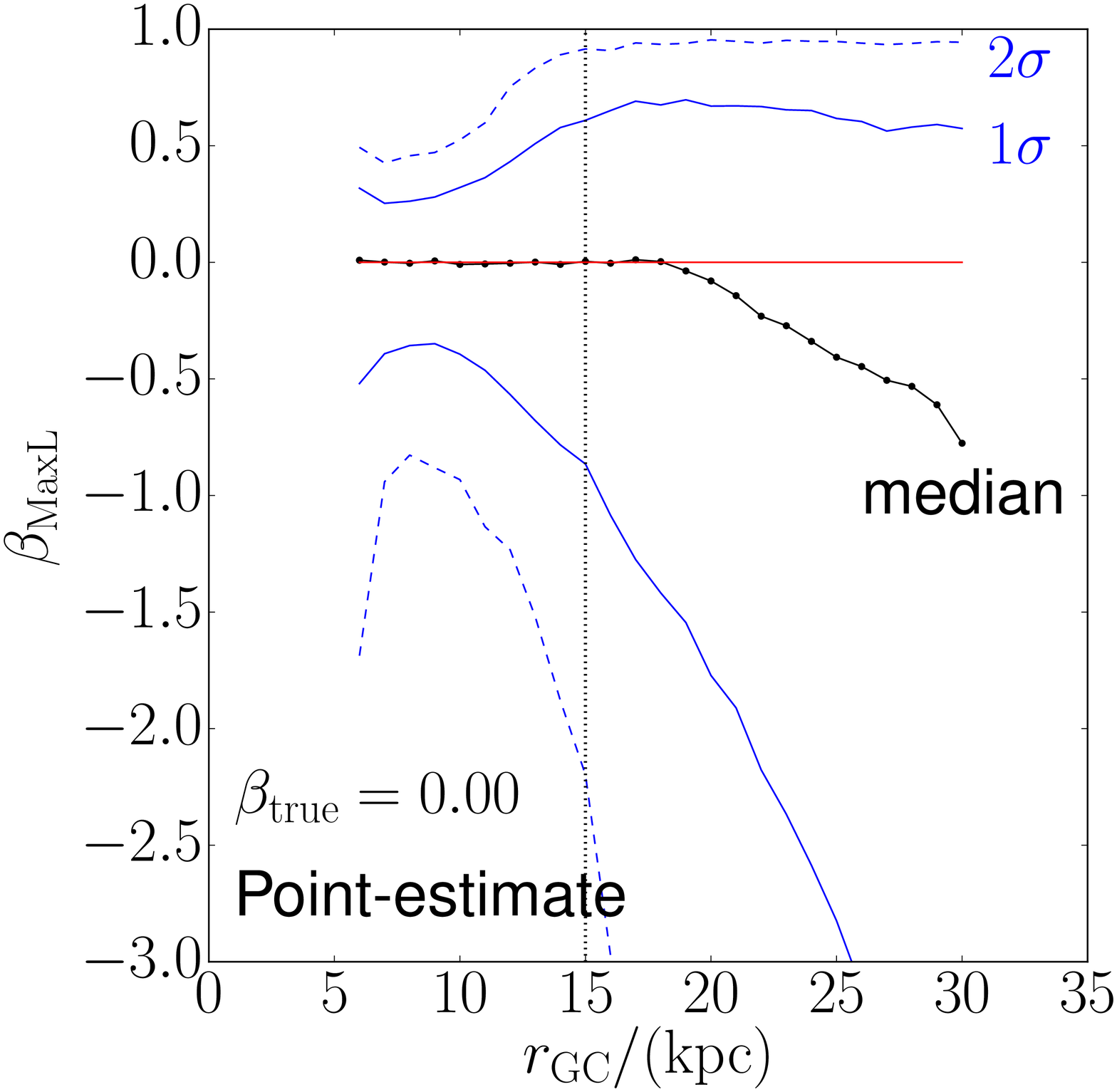} 
	\includegraphics[angle=0,width=0.475\columnwidth]{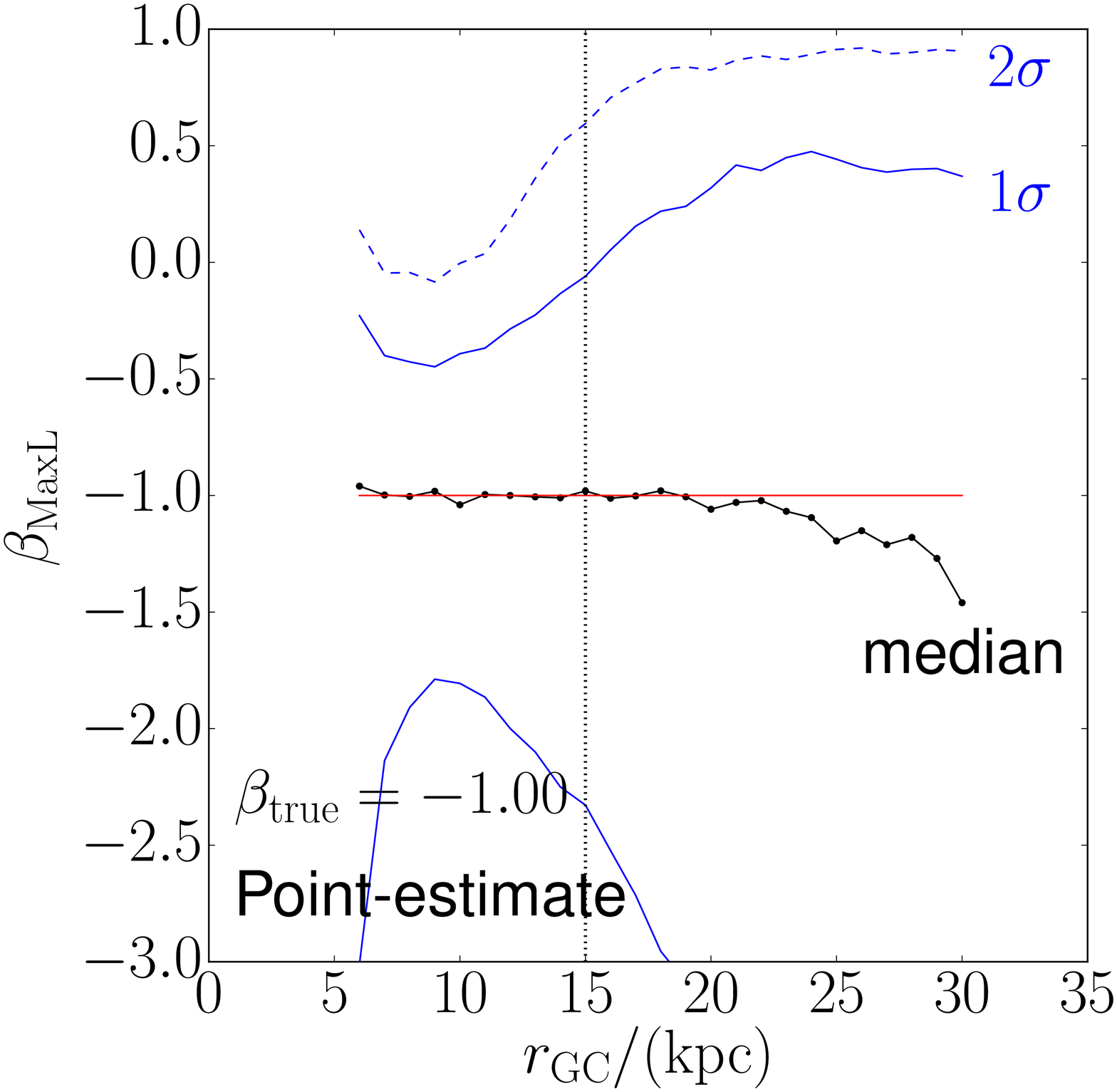} \\
\end{center}
\caption{
The distribution of velocity anisotropy $\beta_{\rm MaxL}$ estimated from the maximum-likelihood method 
as a function of Galactocentric radius $r$. 
From left panel to right, the assumed $\beta_{\rm true}$ is $0.75, 0.5, 0$, and $-1$ 
(as shown by the horizontal solid red line). 
The black solid curve indicates the median value of $\beta_{\rm MaxL}$. 
The solid and dashed blue curves cover 68\% and 95\% of the distribution of $\beta_{\rm MaxL}$.
The vertical dotted lines at $r=15 \kpc$ are added to guide the eye. 
The distributions of $\beta_{\rm MaxL}$ on these panels are also used in Figure \ref{fig:fixedBeta_data}.
}
\label{fig:fixedBeta_bootstrap_MaxL}
\end{figure*}

Figure \ref{fig:fixedBeta_bootstrap_MaxL} shows the 
distribution of $\beta_{\rm MaxL}$ as a function of $r$. 
In each panel, the value of $\beta_{\rm true}$ is fixed, and it is shown by the horizontal red line. 
In order to understand the systematic error on $\beta_{\rm MaxL}$, 
let us first focus on the behavior of the median value of $\beta_{\rm MaxL}$. 
From  
Figure \ref{fig:fixedBeta_bootstrap_MaxL}, we can see that the median value of $\beta_{\rm MaxL}$ 
matches the value of $\beta_{\rm true}$ within a certain radius $r_{\rm reliable}$ (marked by the vertical black line). 
For example, when $\beta_{\rm true}=0.5$, 
the median value of $\beta_{\rm MaxL}$ coincides with $\beta_{\rm true}$ 
at $r<r_{\rm reliable}=15 \kpc$. 
At $r>r_{\rm reliable}$, 
the median value of $\beta_{\rm MaxL}$ is systematically smaller than $\beta_{\rm true}$, 
and this deviation from $\beta_{\rm true}$ grows larger with increasing $r$. 
These properties are also true for other values of $\beta_{\rm true}$. 
Although the value of $r_{\rm reliable}$ is slightly smaller than $15 \kpc$ for $\beta_{\rm true}=0.75$, 
and larger for $\beta_{\rm true}=0$ and $-1$, 
it is safe to say that the maximum-likelihood solutions are reliable within $r \leq 15 \kpc$ 
given the small offset between $\beta_{\rm true}$ and the median value of $\beta_{\rm MaxL}$.

We now focus on the spread in the distributions of $\beta_{\rm MaxL}$. 
The blue solid curves in each panel of Figure \ref{fig:fixedBeta_bootstrap_MaxL} 
show the 16 and 84 percentiles of the distributions of $\beta_{\rm MaxL}$. 
Similarly, the blue dashed curves in this figure show the 2.5 and 97.5 percentiles. 
We hereafter refer to these percentile ranges bracketing 68\% and 95 \% of the distribution as one- and two-$\sigma$ ranges, respectively. 
We can see from these panels that the spread of these distributions grows rapidly as a function of $r$ 
and that its growth is especially prominent at $r\gtrsim15 \kpc$. 
This finding can be understood in a following manner. 
As $r$ becomes larger, the line-of-sight direction $\vector{e}_{\rm los}$ 
becomes closer to the radial direction $\vector{e}_r$. 
This means that the line-of-sight velocity $v_{\rm los}$ is more dominated by the radial velocity $v_r$. 
Therefore, when $r$ is large enough compared to the Galactocentric radius of the Sun ($8 \kpc$), 
the contribution of $\sigma_\theta$ or $\sigma_\phi$ to $\sigma_{\rm los}$ 
becomes less significant, making it harder to reliably extract information on 
the tangential velocity components. 
A geometrical explanation for this result is given in Appendix \ref{appendix:geometry}.

To summarize, 
the maximum-likelihood method tends to underestimate the value of $\beta$ beyond a certain radius $r_{\rm reliable}$, 
and that this systematic error on $\beta$ increases 
as the Galactocentric radius $r$ of the sample increases, and 
as the true anisotropy $\beta_{\rm true}$ becomes larger. 
Also, the random error on $\beta$ increases with increasing $r$ at $r\gtrsim15 \kpc$. 
These systematic and random errors make the estimated values of $\beta$ unreliable at large $r$. 
Based on these findings, 
it is fair to conclude that the maximum-likelihood method can in principle reliably estimate $\beta$ at $r \leq 15 \kpc$, 
but is unable to estimate $\beta$ at $r>15 \kpc$ 
if one uses only 4-dimensional information for $1000$ stars at a given radius.

\subsubsection{Results for fixed $r$} \label{section:MaxL_fixedrGC}

Here we shall view our results from a different perspective and 
investigate how the performance of the maximum-likelihood method depends on $\beta_{\rm true}$.

\begin{figure*}
\begin{center}
	\includegraphics[angle=0,width=0.475\columnwidth]{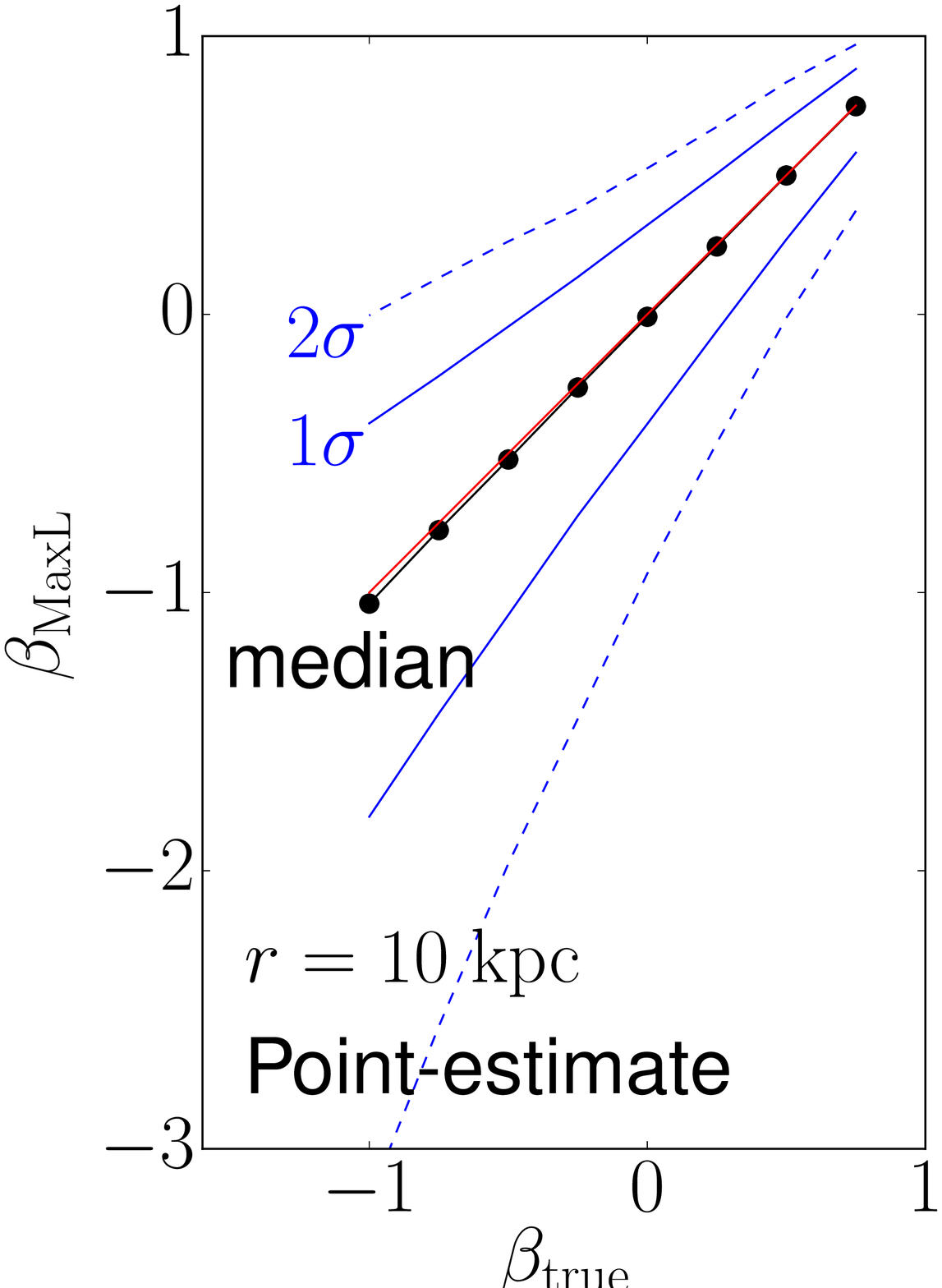}
	\includegraphics[angle=0,width=0.475\columnwidth]{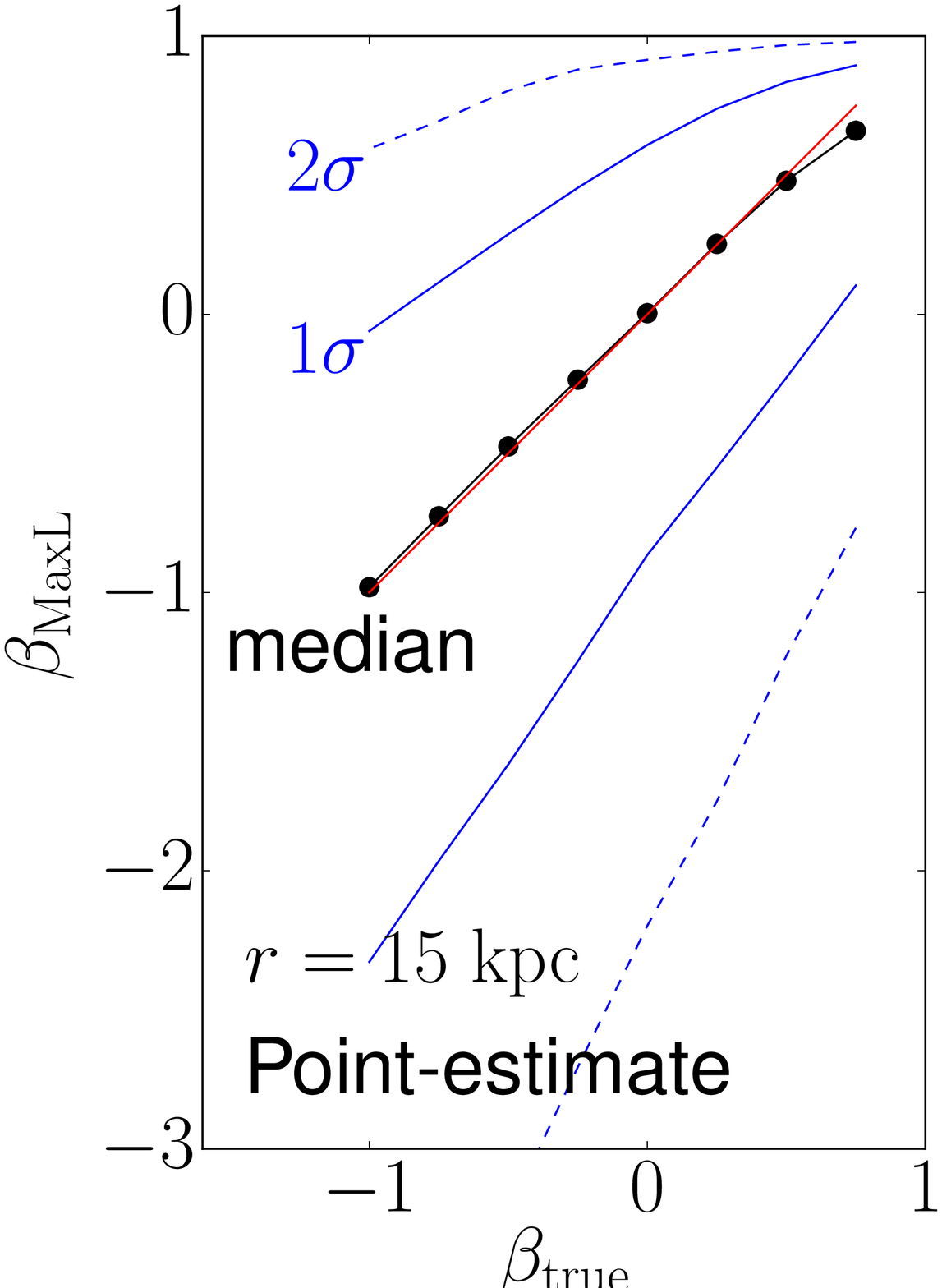}
	\includegraphics[angle=0,width=0.475\columnwidth]{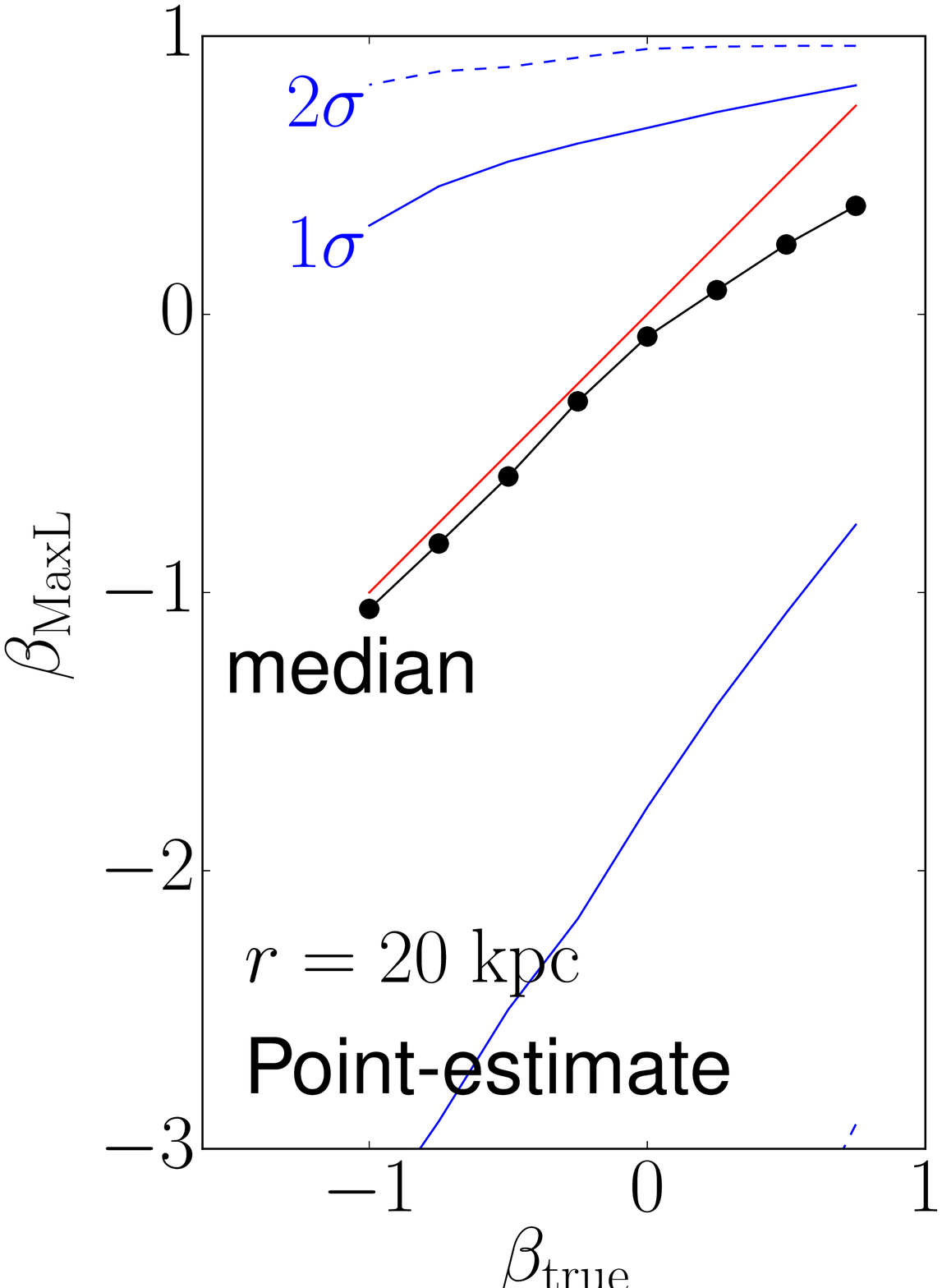}
	\includegraphics[angle=0,width=0.475\columnwidth]{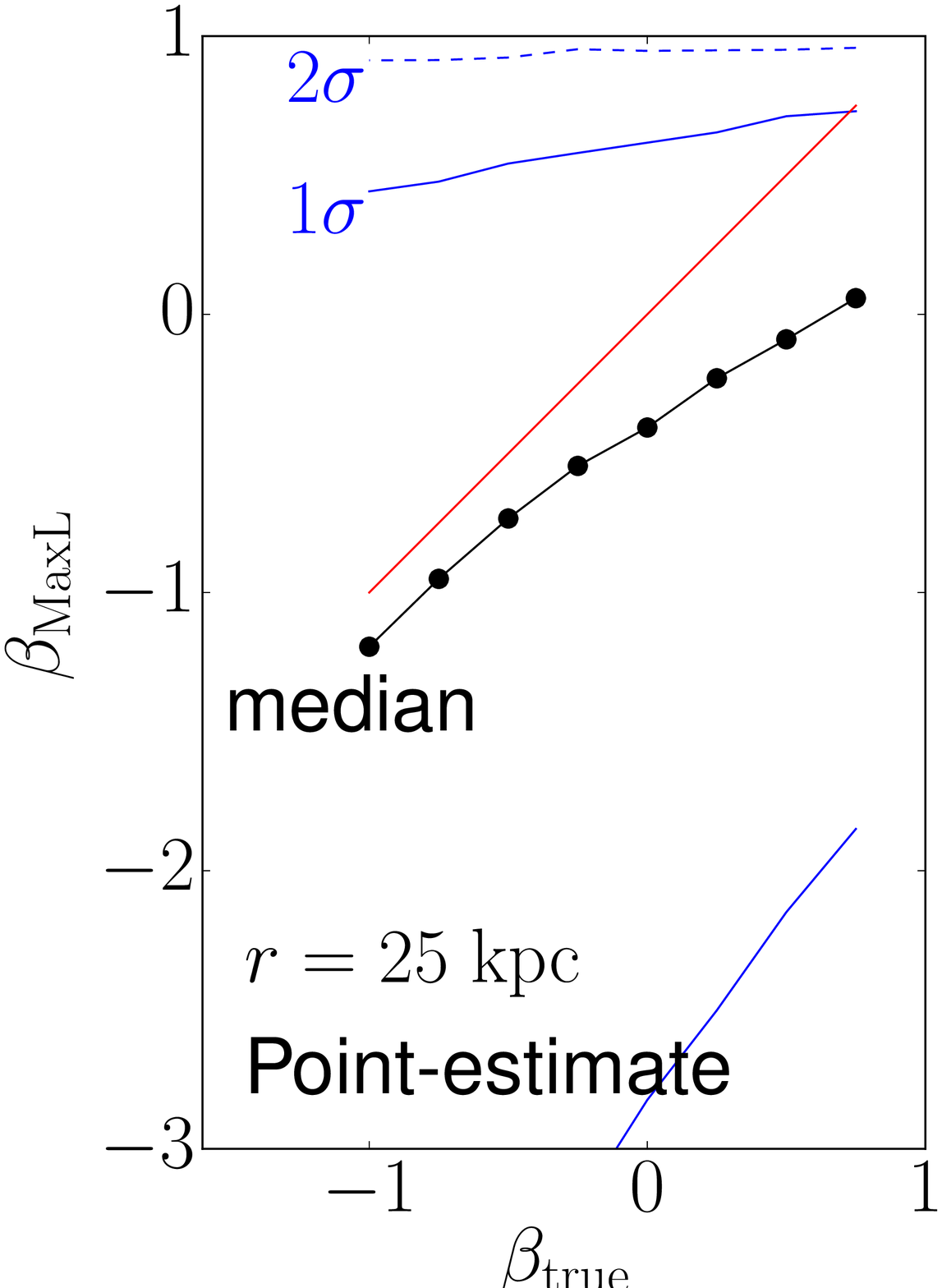} \\
\end{center}
\caption{
The distribution of the velocity anisotropy $\beta_{\rm MaxL}$ as a function of the input value of $\beta_{\rm true}$. 
From left panel to right, the assumed Galactocentric radius is $r/\kpc=10, 15, 20$, and $25$.
The red diagonal line corresponds to $\beta_{\rm MaxL}=\beta_{\rm true}$. 
The solid black curve indicates the median value of $\beta_{\rm MaxL}$. 
The solid and dashed blue curves cover 68\% and 95\% of the distribution of $\beta_{\rm MaxL}$.
}
\label{fig:fixedrGC_bootstrap_MaxL}
\end{figure*}

Figure \ref{fig:fixedrGC_bootstrap_MaxL} shows the 
distribution of $\beta_{\rm MaxL}$ as a function of the input anisotropy $\beta_{\rm true}$. 
In each panel, the Galactocentric radius $r$ is fixed, 
and the diagonal red line indicates the line of $\beta_{\rm MaxL}=\beta_{\rm true}$. 
From Figure \ref{fig:fixedrGC_bootstrap_MaxL}, we can see the growth of  the systematic and random errors as a function of $r$. 
At $r=10 \kpc$, we see that the maximum-likelihood method on average returns the correct velocity anisotropy, 
independent of $\beta_{\rm true}$.
At $r=15 \kpc$, the one-$\sigma$ (68\%) range of $\beta_{\rm MaxL}$ becomes wider (larger random error), 
but the median value of $\beta_{\rm MaxL}$ is still very close to $\beta_{\rm true}$. 
However, at $r > 15 \kpc$, the systematic error on $\beta_{\rm MaxL}$ becomes prominent. 
Especially, if $\beta_{\rm true}>0$, the median value of $\beta_{\rm MaxL}$ is systematically smaller than $\beta_{\rm true}$. 
This systematic offset as well as the larger random error on $\beta_{\rm MaxL}$ 
indicates that there is a large probability that 
the maximum-likelihood method mistakenly returns a highly negative value of $\beta_{\rm MaxL}$ at $r > 15 \kpc$ 
even if the true value of $\beta_{\rm true}$ is positive. 
Although the median value of $\beta_{\rm MaxL}$ is much closer to $\beta_{\rm true}$ if $\beta_{\rm true} <0$, 
the one-$\sigma$ range is so large at $r > 15 \kpc$ 
that it is practically impossible to determine if a measured negative  $\beta_{\rm MaxL}$ results 
from a positive or a negative value of $\beta_{\rm true}$.

\section{Result 2: Bayesian method} \label{section:resultBayes}

In Section \ref{section:resultMaxL}, 
we have found that the performance of the maximum-likelihood method deteriorates at $r > 15 \kpc$.
In this Section, we shall confirm this result by using a Bayesian method. 

Due to the relatively large computational cost of Bayesian analyses, 
we apply this method only to a fraction of our mock catalogues. 
By using 100 mock catalogs for each pair of $(r, \beta_{\rm true})$, 
we derive the posterior distributions of $(\sigma_r,\sigma_\theta,\sigma_\phi,V_{\rm rot})$ as well as $\beta_{\rm Bayes}$ 
(hereafter $\beta_{\rm Bayes}$ denotes the velocity anisotropy obtained from Bayesian analyses). 
We used three types of priors, A, B, and C (see Section \ref{section:method_Bayes}), 
but it turned out that the use of priors A and B results in almost identical posterior distributions. 
Since our main aim here is to quantify the systematic error in this method,
we combine these 100 posterior distributions for each pair of $(r, \beta_{\rm true})$ for each prior. 
Then we calculate the 2.5, 16, 50, 84, and 97.5 percentiles of the posterior distribution of $\beta_{\rm Bayes}$. 
These calculations were performed by using a publicly available Python package \emcee $\;$ \citep{Foreman-Mackey2013}.

\subsection{Results for fixed $\beta_{\rm true}$} \label{section:resultBayesFixedBeta}

Figure \ref{fig:fixedBeta_BayesAB} shows the posterior distribution of $\beta_{\rm Bayes}$ as a function of $r$. 
In each panel, the value of $\beta_{\rm true}$ is fixed and shown by the red horizontal line. 
The black solid line shows the median value of the posterior distribution of $\beta_{\rm MaxL}$, 
and blue solid and blue dashed lines respectively cover 68\% and 95\% of the posterior distribution.

The results of the Bayesian analyses for priors A and B are almost identical to each other, 
while the results for priors B and C look distinctly different 
(hence we do not show results for prior A in Figure \ref{fig:fixedBeta_BayesAB}). 
This prior-dependence of the resultant posterior distributions can be explained in a following manner. 
On the one hand, the uncertainty in  $\sigma_r$ is relatively small even at large Galactocentric radius $r$ (see Section \ref{section:showCase}). 
This small uncertainty means that the likelihood function is strongly peaked near the true value of $\sigma_r$. 
Thus, although the priors A and B have very different  $\sigma_r$-dependence, 
the resultant posterior distributions of $\sigma_r$ are almost the same. 
On the other hand, errors in $\sigma_\theta$ and $\sigma_\phi$ are large at large $r$ (see Section \ref{section:showCase}), 
and the error in $\beta_{\rm Bayes}$ is dominated by such errors. 
This large uncertainty means that the likelihood function only weakly depends on $(\sigma_\theta, \sigma_\phi)$, 
so the posterior distribution is sensitive to the $(\sigma_\theta, \sigma_\phi)$-dependence of the prior. 
Since prior C 
is weighted more heavily towards
smaller values of $(\sigma_\theta, \sigma_\phi)$, 
adopting prior C is equivalent to adopting a strong prior on $\beta$. 
As a result, the resultant posterior distribution is strongly peaked near $\beta_{\rm Bayes} \simeq 1$, 
making the median value of $\beta_{\rm Bayes}$ biased toward large values. 
Since prior B does not depend on $(\sigma_\theta, \sigma_\phi)$, 
adopting prior B is equivalent to adopting a flat prior on $\beta$. 
As a result, the posterior distribution traces the  $(\sigma_\theta, \sigma_\phi)$-dependence of the likelihood function, 
making the posterior distribution of $\beta_{\rm Bayes}$ for prior B
more or less similar to the distribution of $\beta_{\rm MaxL}$.

Next we 
compare the posterior distribution $\beta_{\rm Bayes}$ for prior B with the distribution of $\beta_{\rm MaxL}$. 
As seen in Figure \ref{fig:fixedBeta_BayesAB}, 
the median value of $\beta_{\rm Bayes}$ is close to $\beta_{\rm true}$ 
at small Galactocentric radii, 
but it gradually deviates from  $\beta_{\rm true}$ at large $r$. 
When $\beta_{\rm true}>0$ and $r \gtrsim 15 \kpc$, 
the median value of $\beta_{\rm Bayes}$ is systematically lower than $\beta_{\rm true}$, similar to the results of maximum-likelihood method. 
However, when $\beta_{\rm true} < 0$ and $r \gtrsim 15 \kpc$, 
the median value of $\beta_{\rm Bayes}$ is larger than $\beta_{\rm true}$, unlike in the case of the maximum-likelihood method. 
In any case, the one- and two-$\sigma$ ranges in the $\beta_{\rm Bayes}$-distribution 
rapidly increases at $r \gtrsim 15 \kpc$, making the estimation of velocity anisotropy 
as difficult and uncertain as 
with the maximum-likelihood method. 

\subsection{Results for fixed $r$}

\begin{figure*}
\begin{flushleft}
	\includegraphics[angle=0,width=0.45\columnwidth]{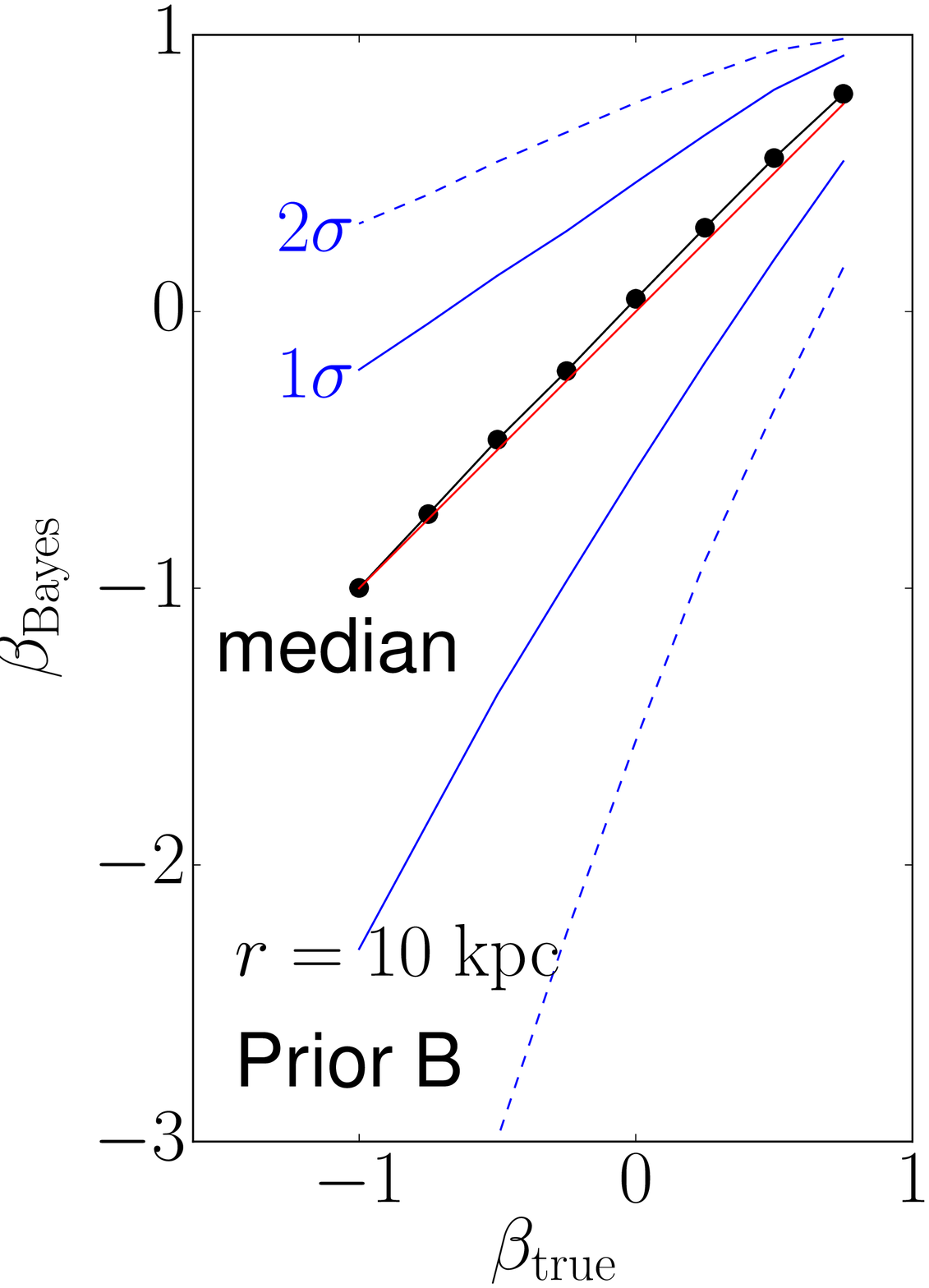} 
	\includegraphics[angle=0,width=0.45\columnwidth]{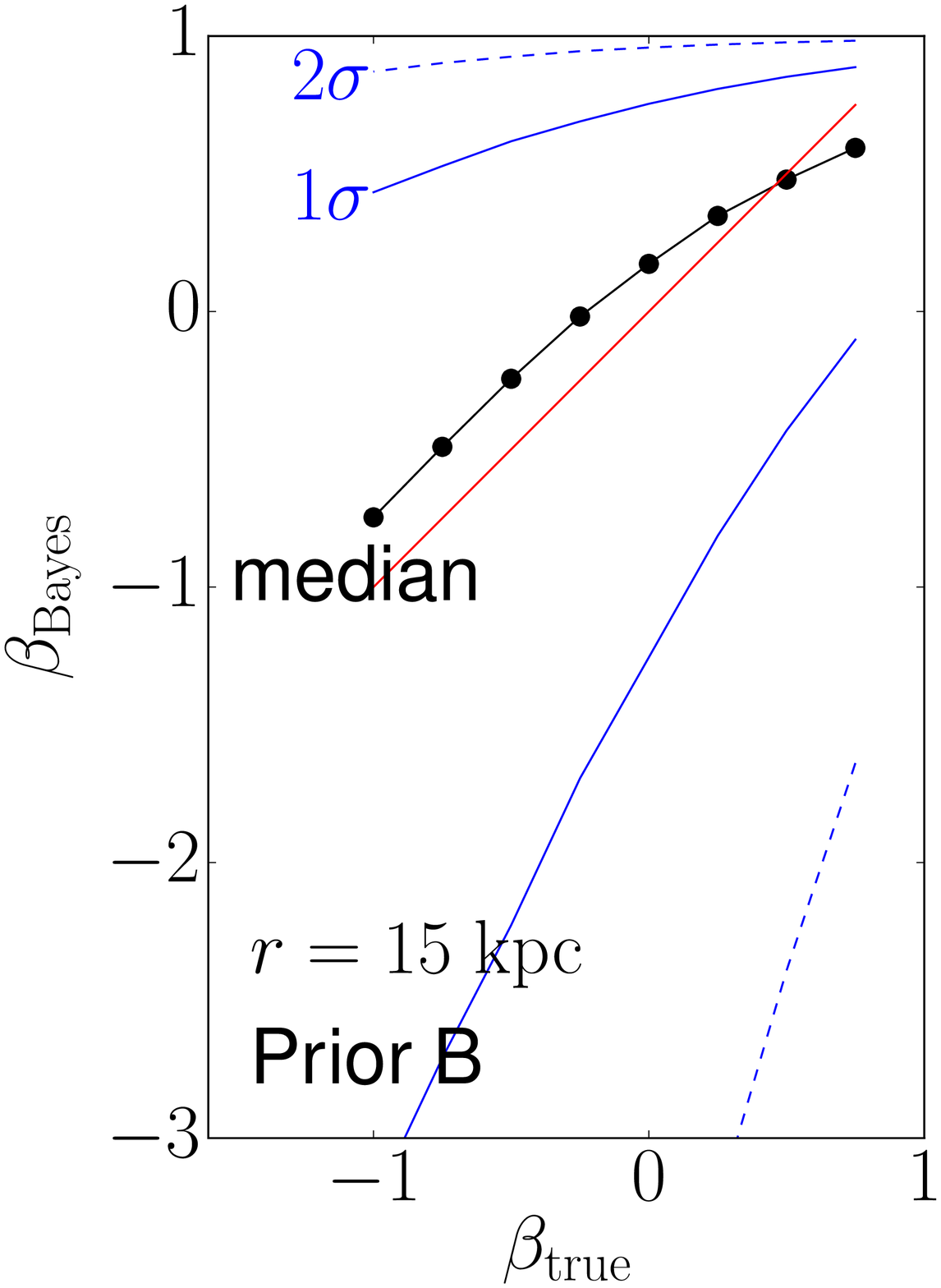} 
	\includegraphics[angle=0,width=0.45\columnwidth]{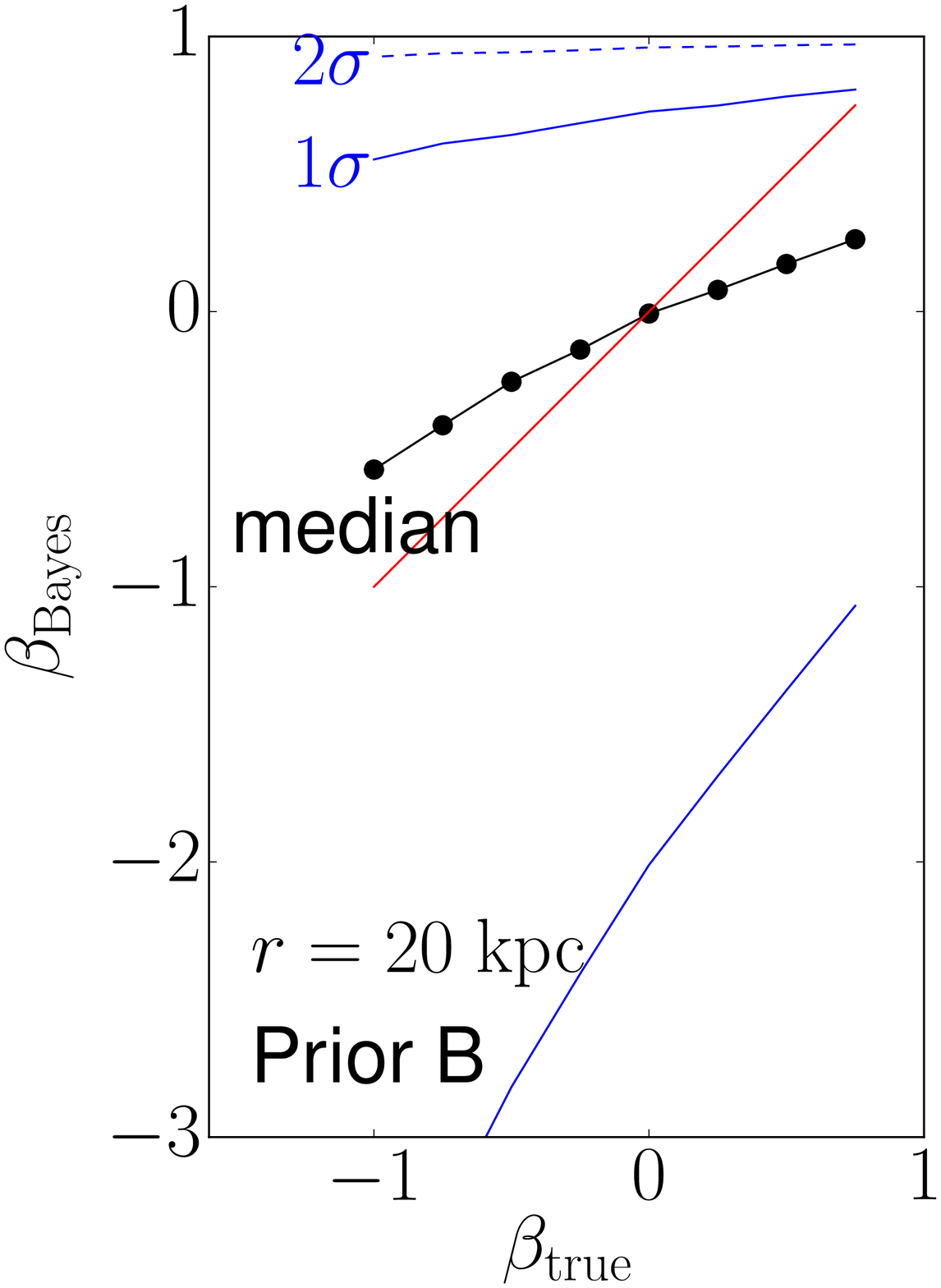} 
	\includegraphics[angle=0,width=0.45\columnwidth]{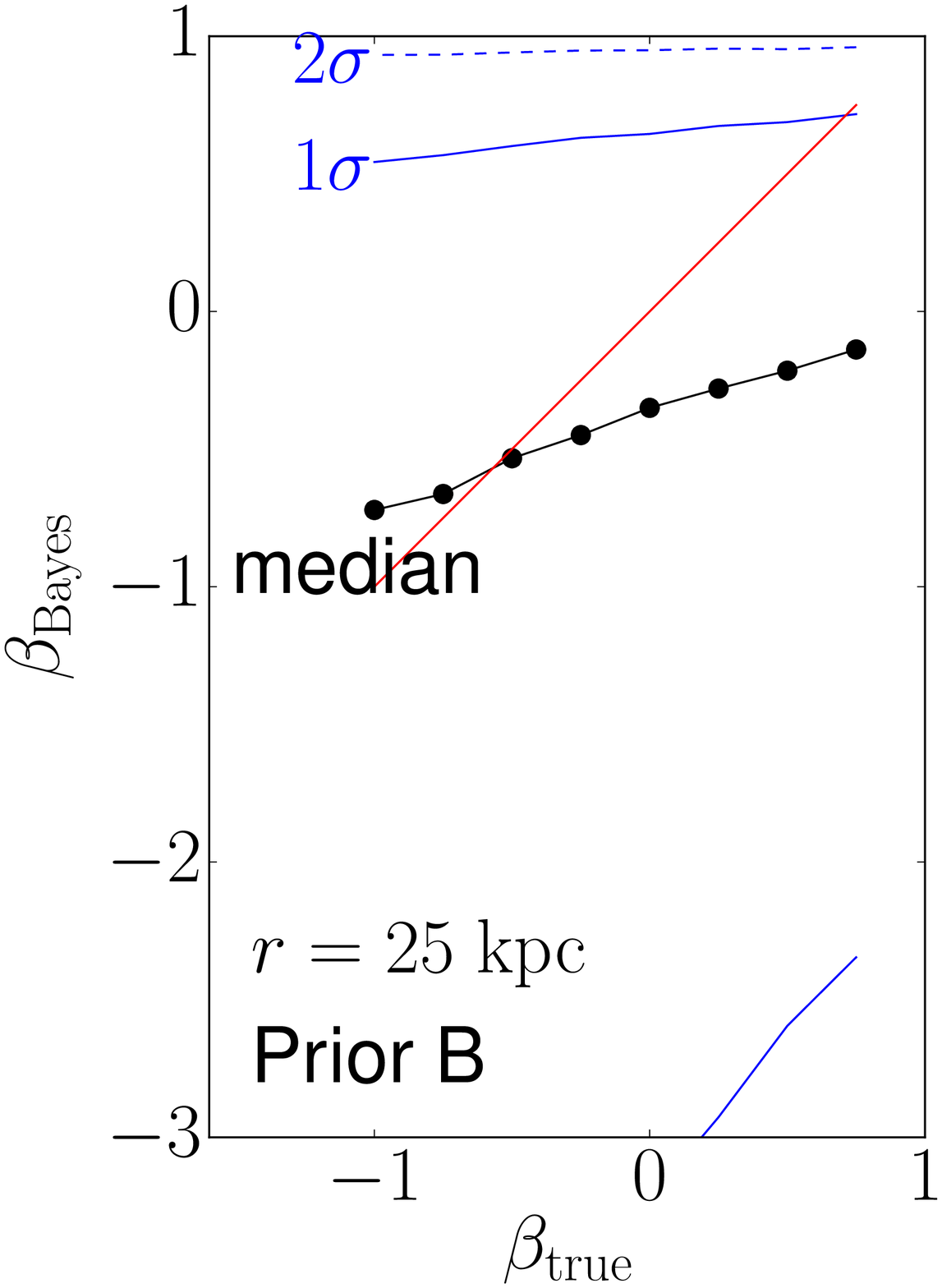} \\
	\includegraphics[angle=0,width=0.45\columnwidth]{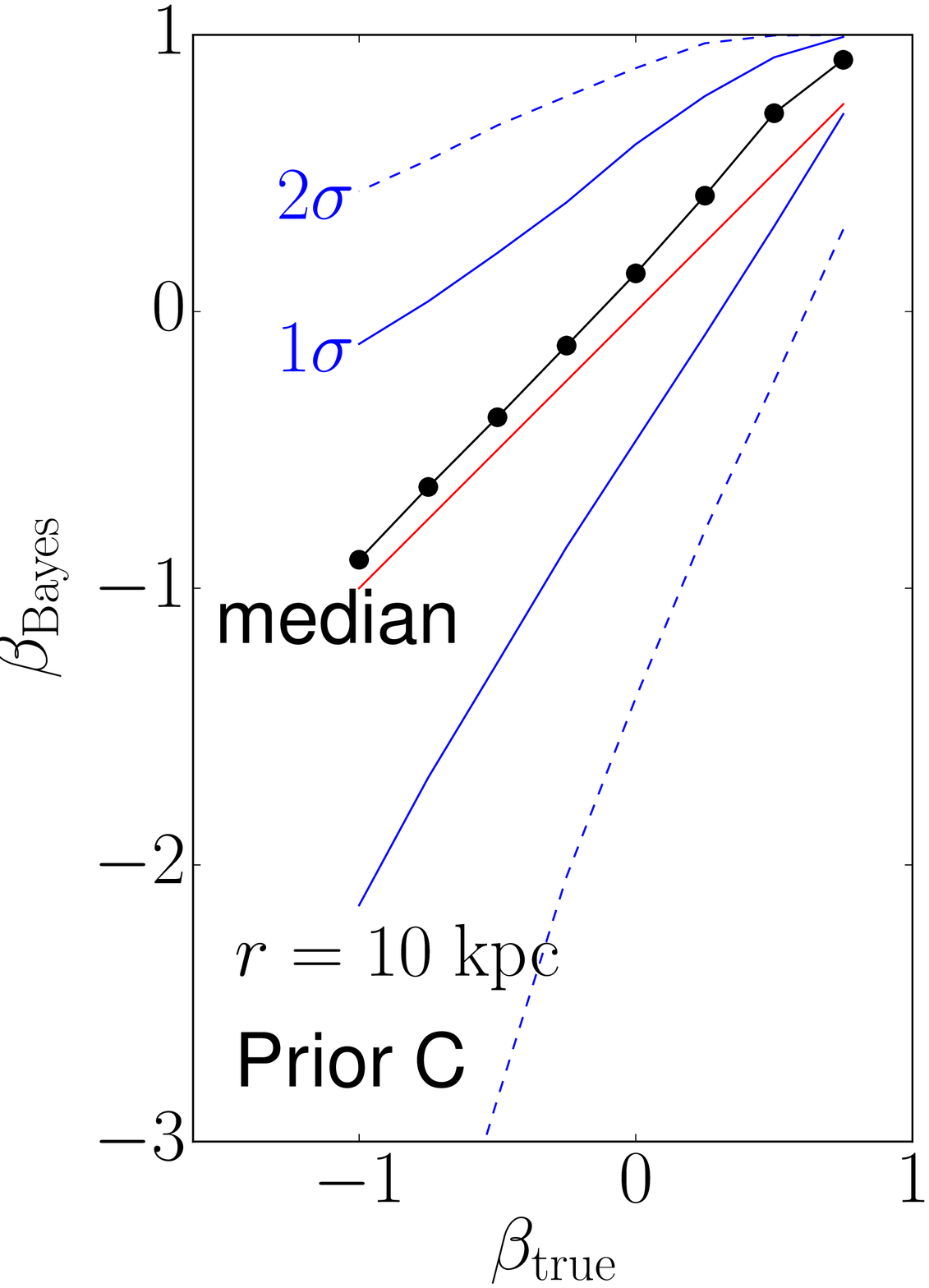} 
	\includegraphics[angle=0,width=0.45\columnwidth]{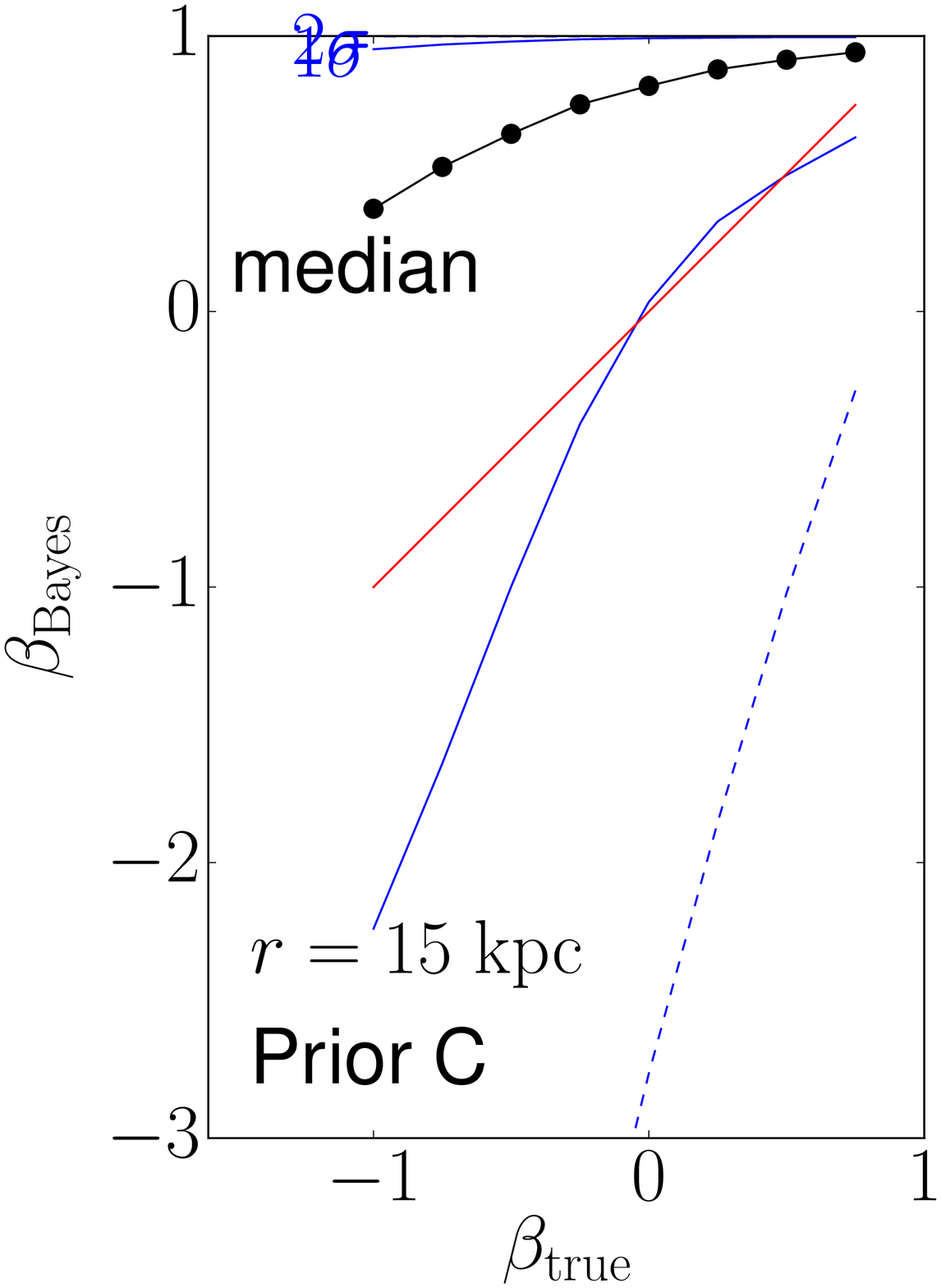} 
	\includegraphics[angle=0,width=0.45\columnwidth]{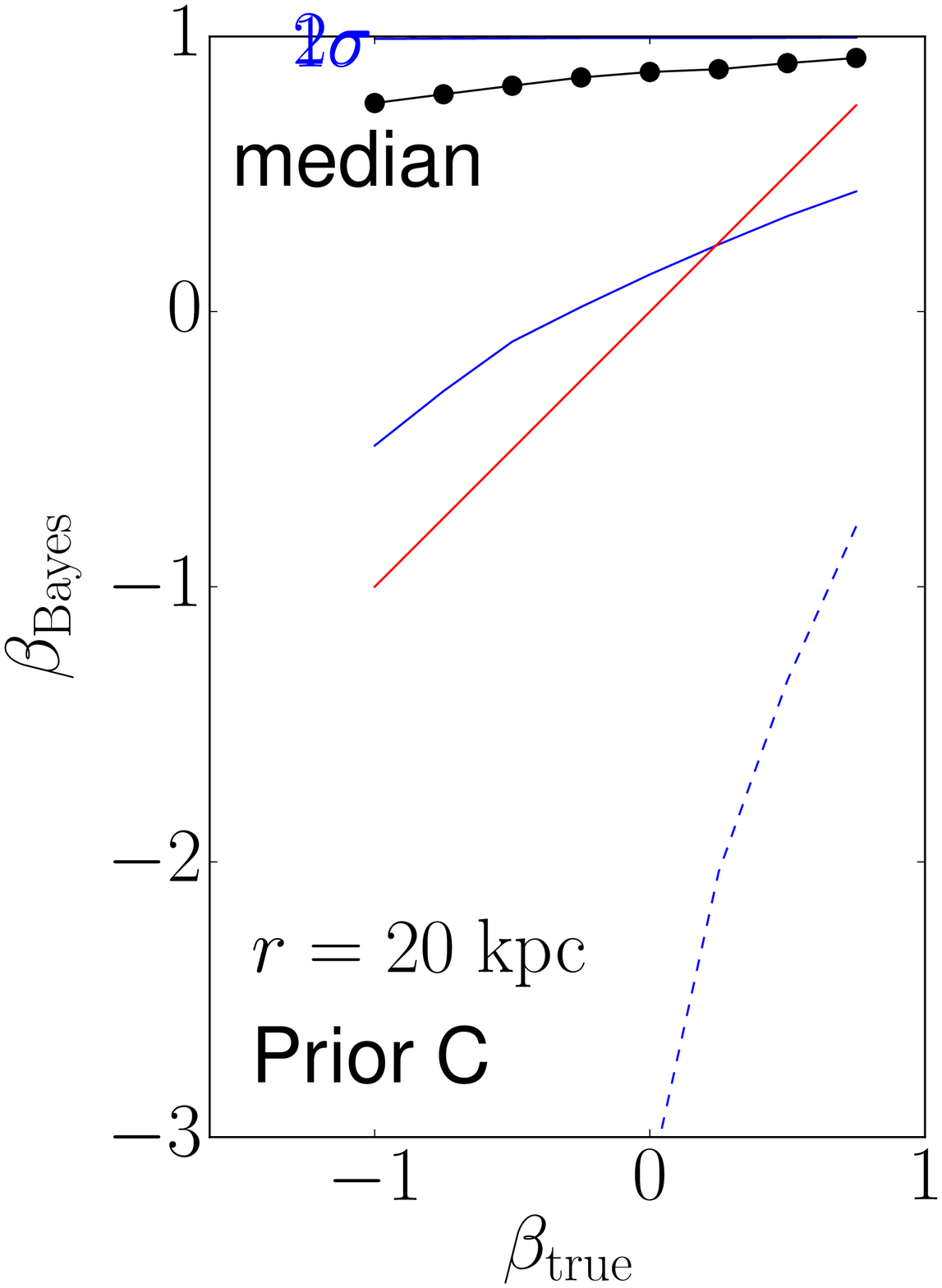} 
	\includegraphics[angle=0,width=0.45\columnwidth]{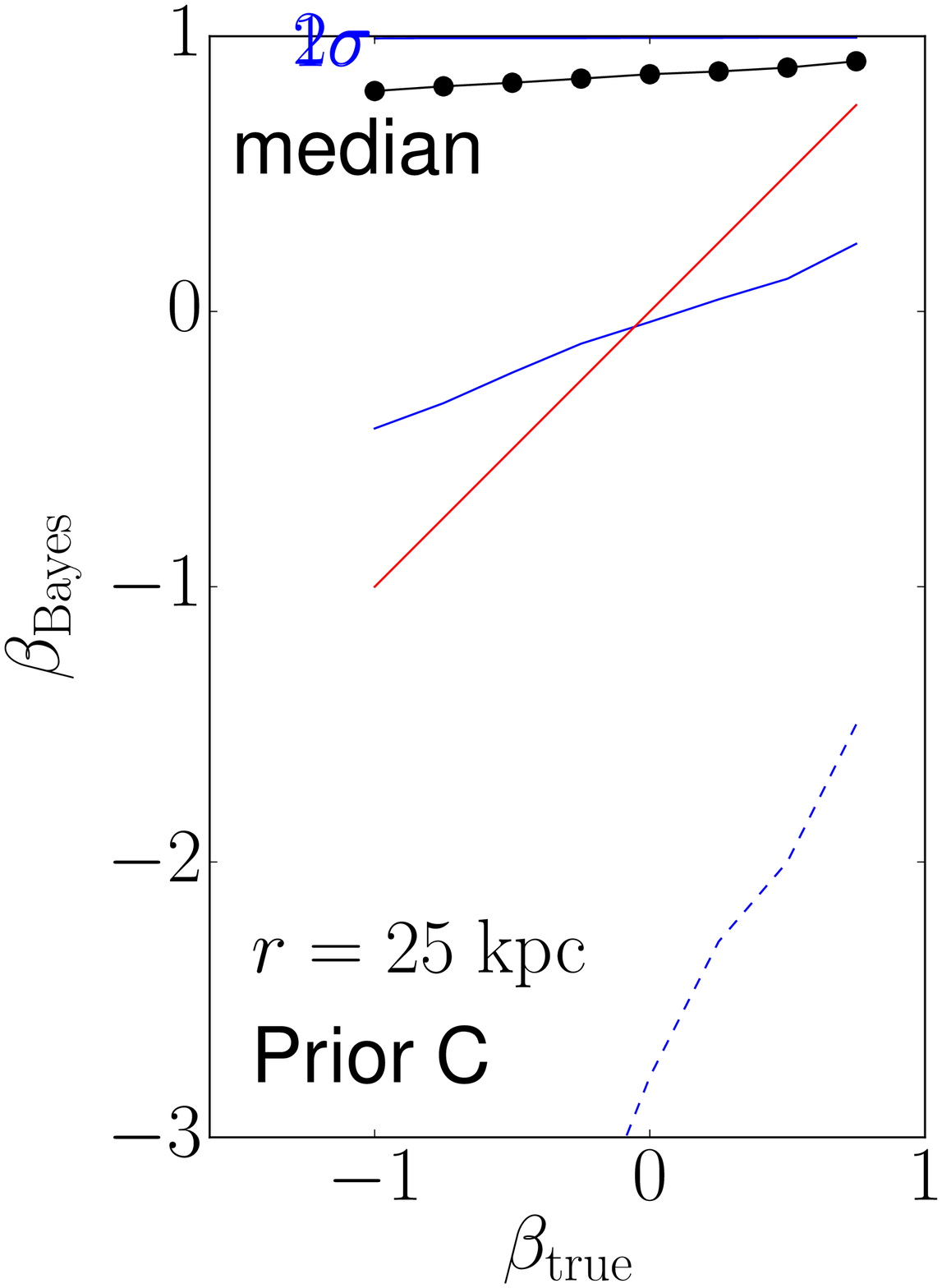} 
\end{flushleft}
\caption{
The posterior distribution of velocity anisotropy $\beta_{\rm Bayes}$ for our mock catalogues 
as a function of true anisotropy $\beta_{\rm true}$. 
The top and bottom rows correspond to the results for priors B and C, respectively, 
assumed in the Bayesian analyses.
From left panel to right, the assumed Galactocentric radius is $r/\kpc=10, 15$, and $20$. 
In each panel, the diagonal red solid line indicates $\beta_{\rm  Bayes} = \beta_{\rm true}$. 
The black solid line shows the median value of the posterior distribution of $\beta_{\rm  Bayes}$, 
and blue solid and blue dashed lines respectively cover 68\% and 95\% of the posterior distribution. 
}
\label{fig:fixedrGC_BayesAB}
\end{figure*}

Figure \ref{fig:fixedrGC_BayesAB} 
is similar to Figure \ref{fig:fixedrGC_bootstrap_MaxL} and 
shows 
the distribution of $\beta_{\rm Bayes}$ as a function of $\beta_{\rm true}$. 
In each panel, the Galactocentric radius $r$ of the sample is fixed, 
and the diagonal red line indicates the line of $\beta_{\rm Bayes}=\beta_{\rm true}$. 

\begin{figure*}
\begin{flushleft}
	\includegraphics[angle=0,width=0.47\columnwidth]{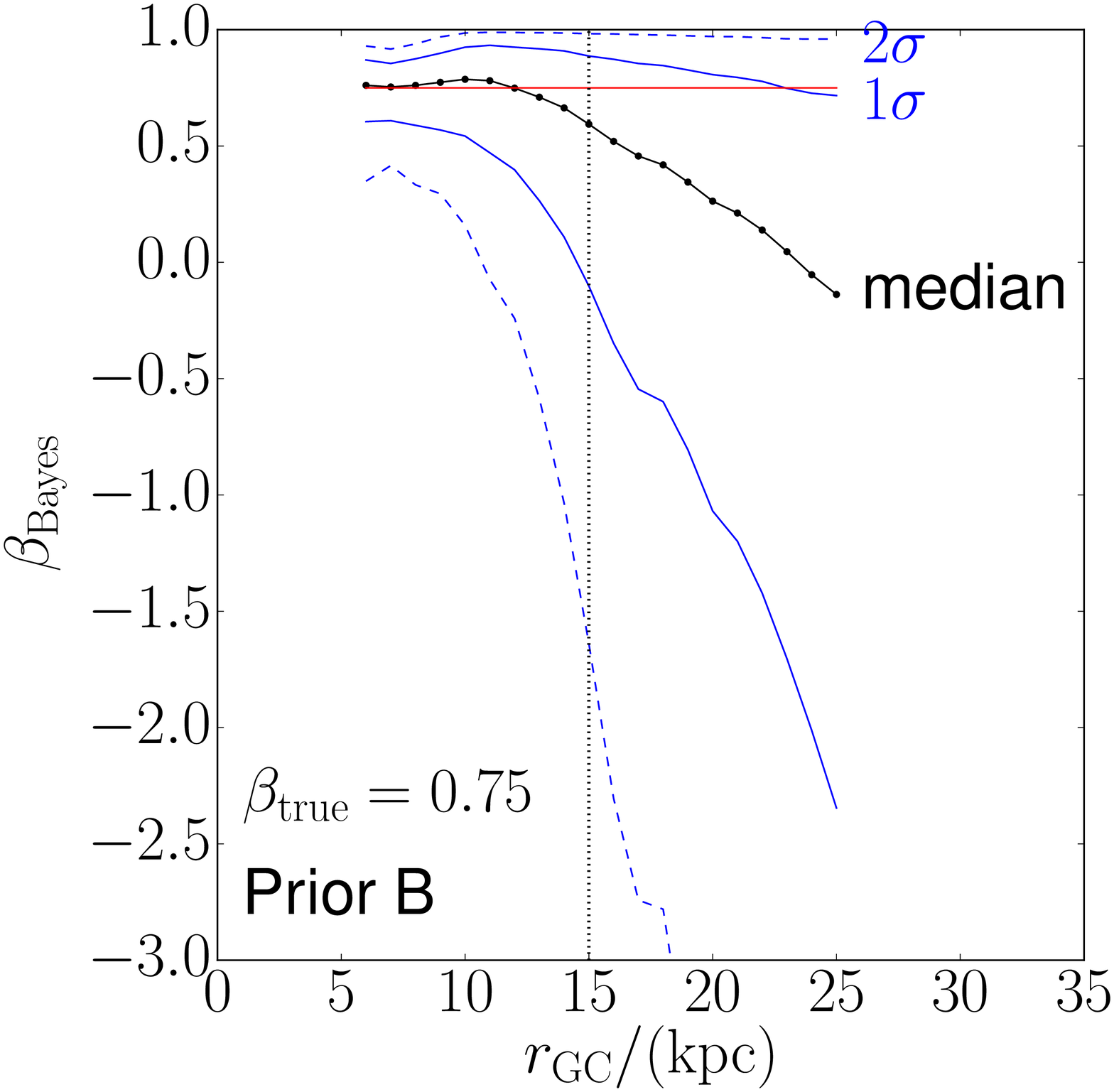} 
	\includegraphics[angle=0,width=0.47\columnwidth]{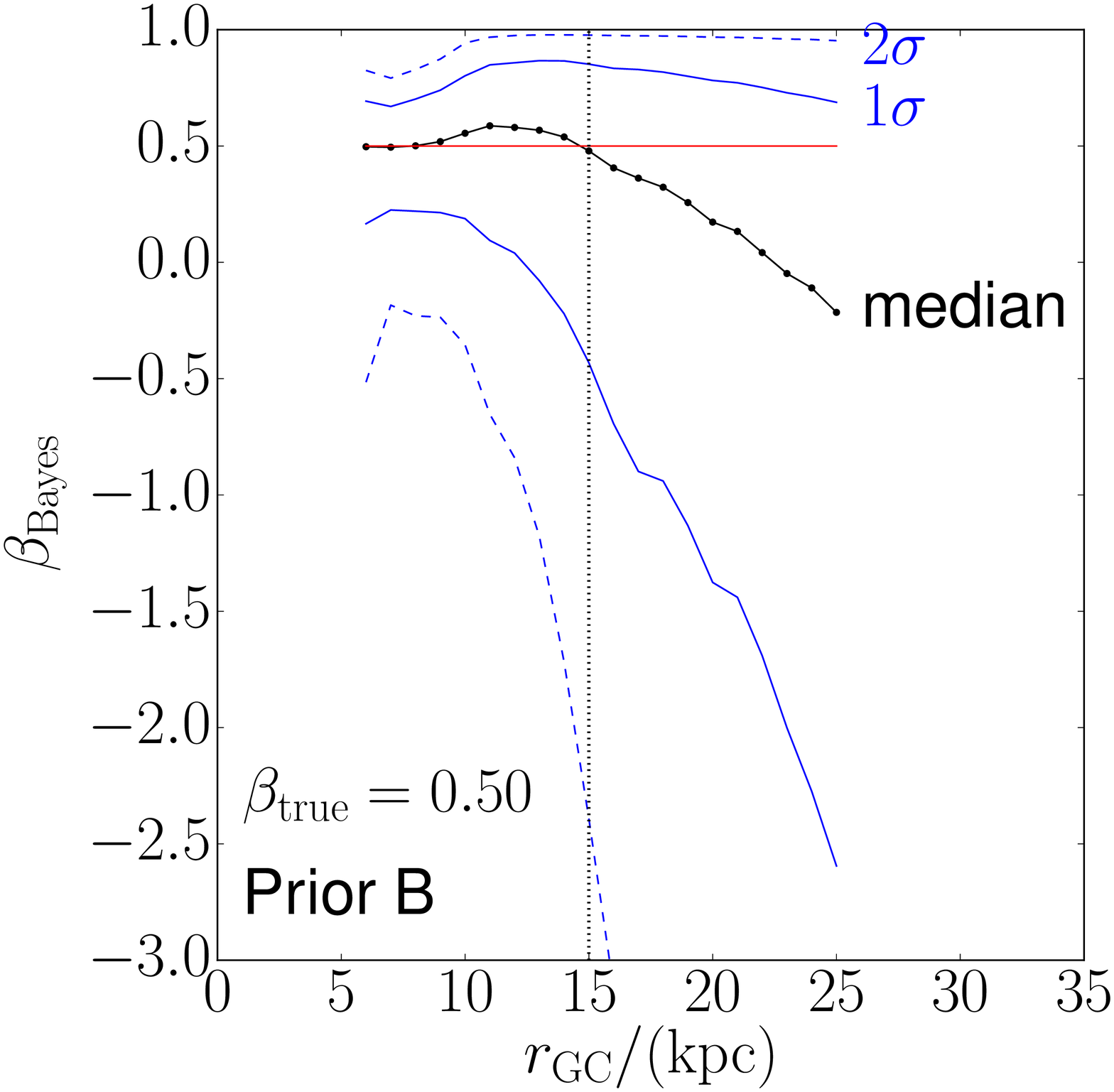} 
	\includegraphics[angle=0,width=0.47\columnwidth]{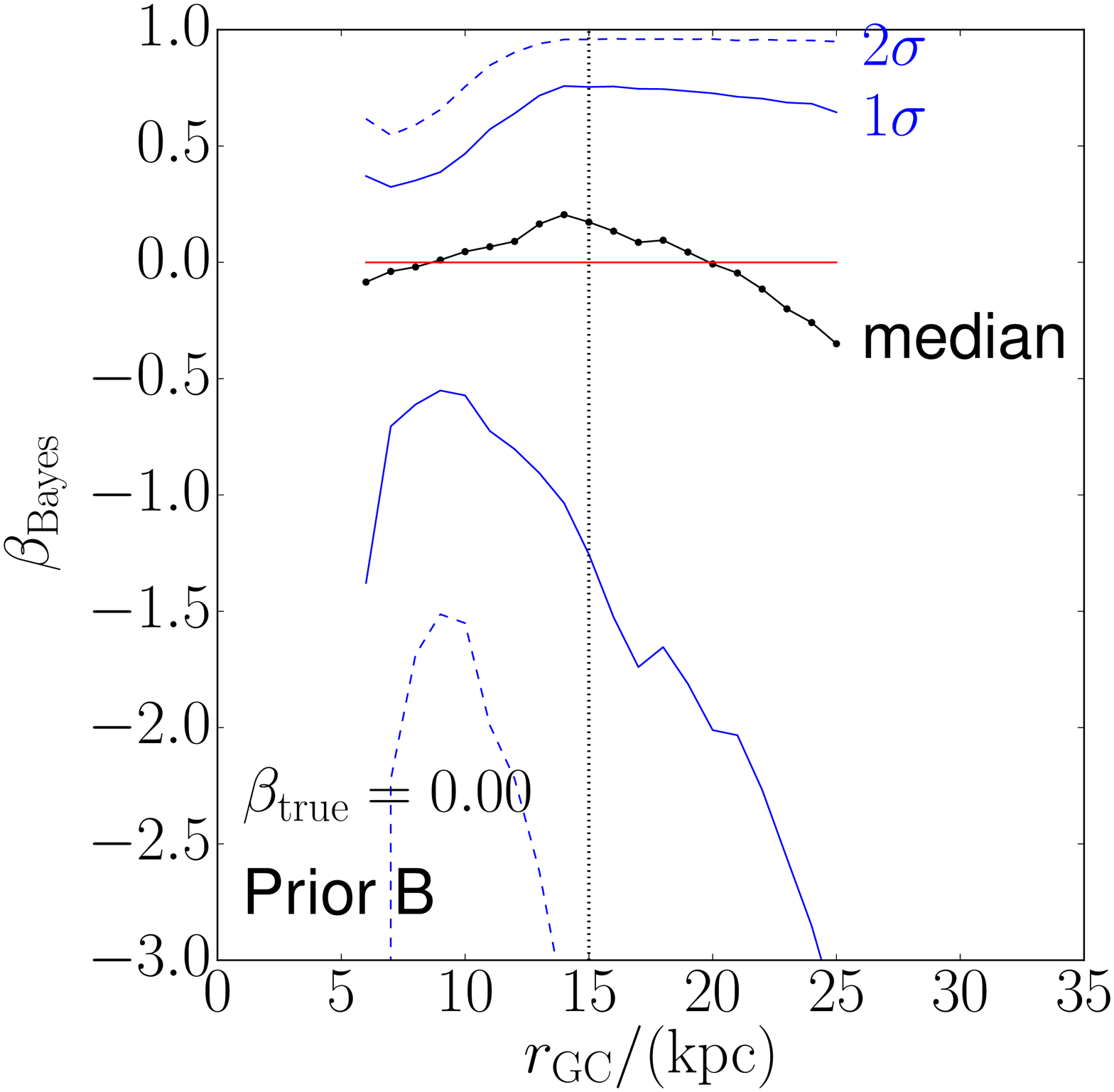} 
	\includegraphics[angle=0,width=0.47\columnwidth]{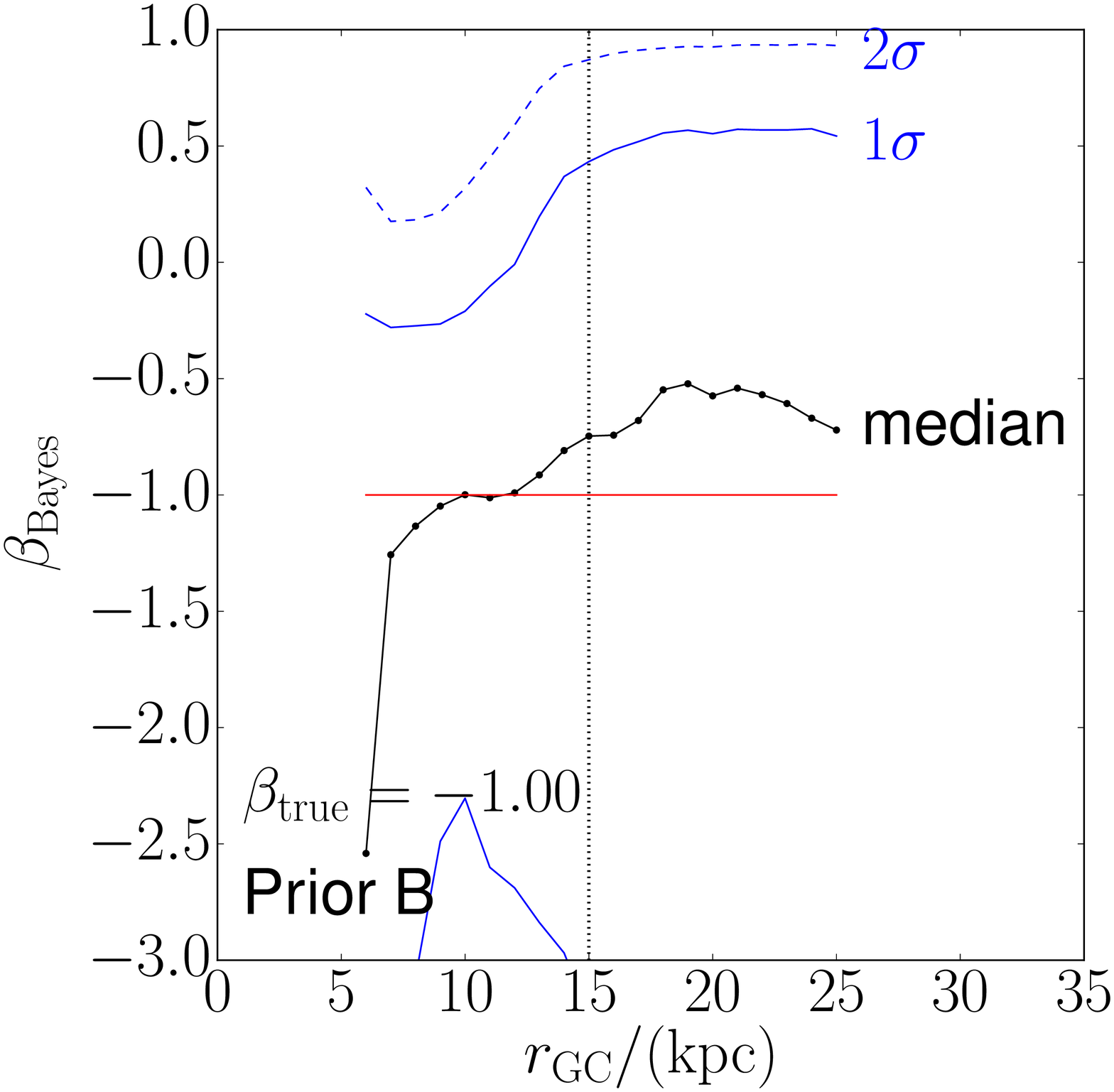} \\
	\includegraphics[angle=0,width=0.47\columnwidth]{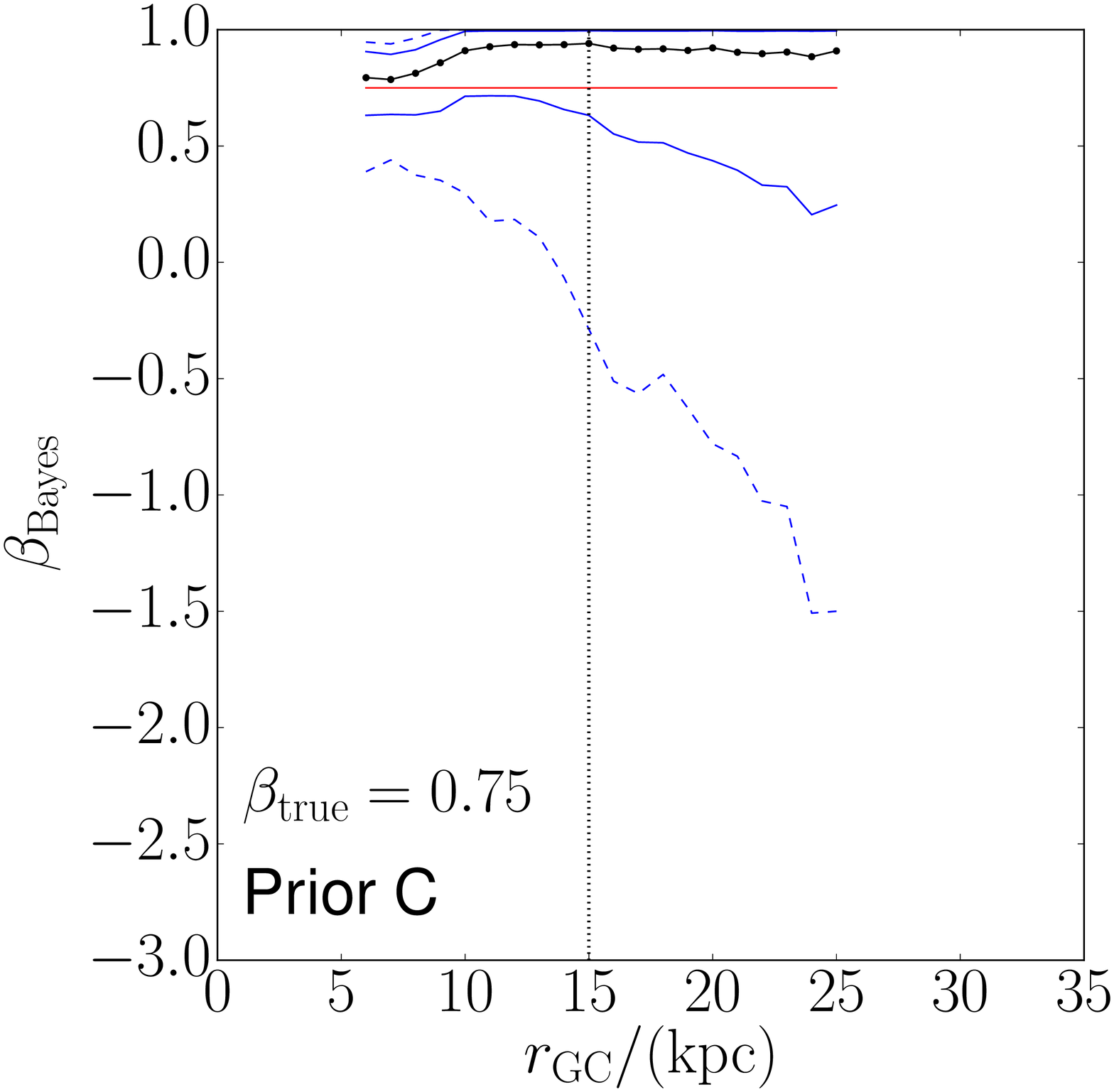} 
	\includegraphics[angle=0,width=0.47\columnwidth]{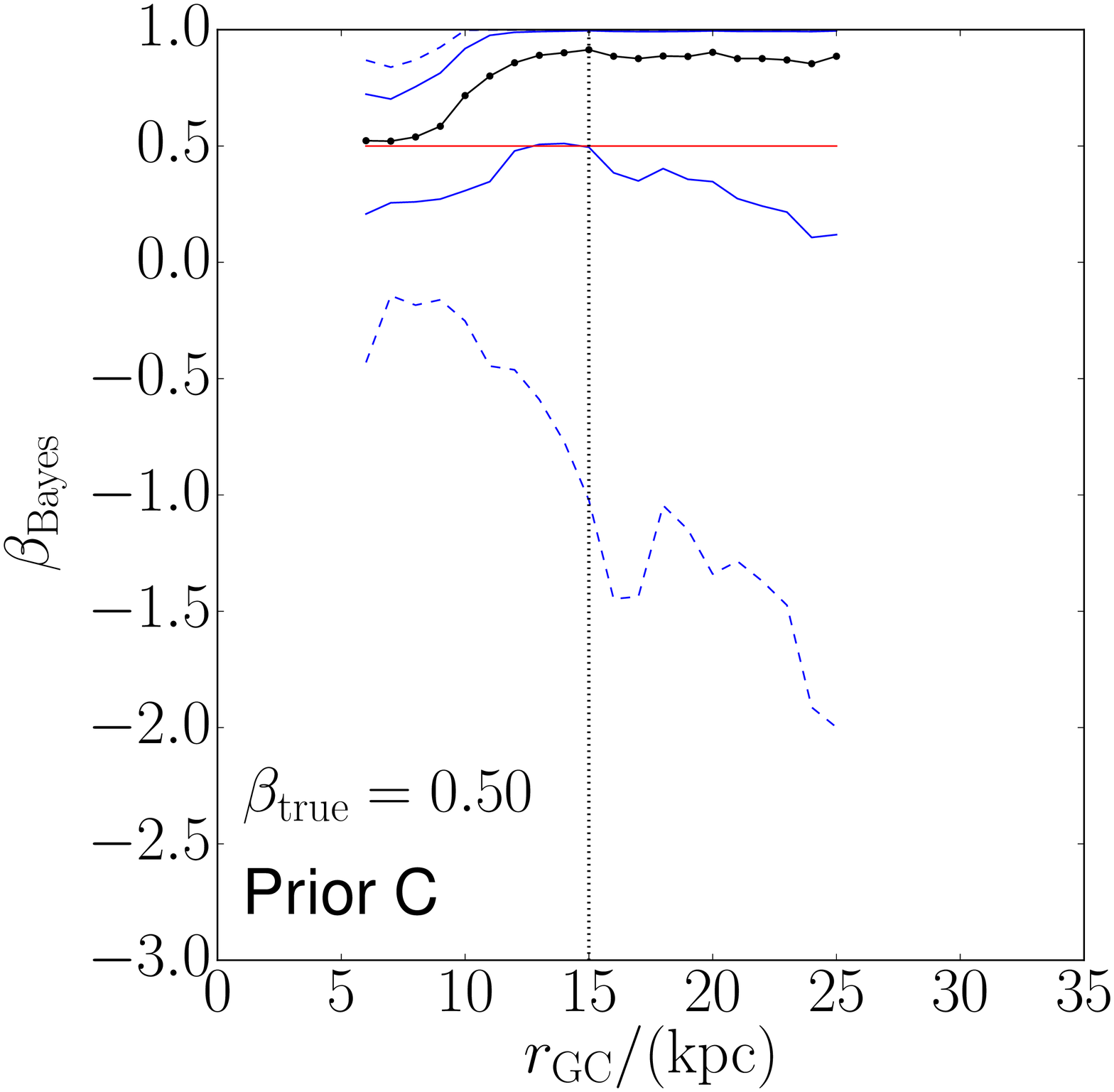} 
	\includegraphics[angle=0,width=0.47\columnwidth]{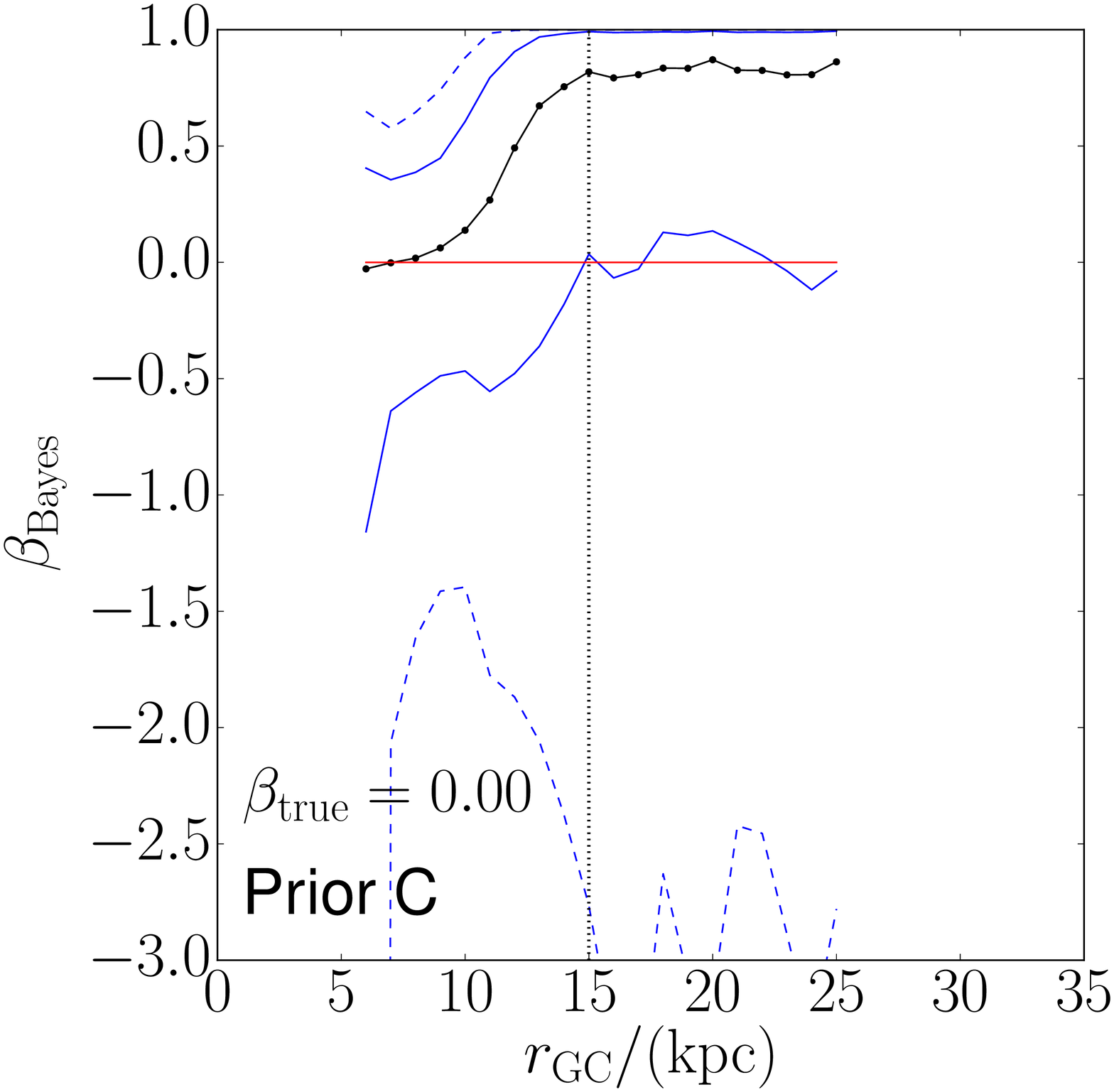} 
	\includegraphics[angle=0,width=0.47\columnwidth]{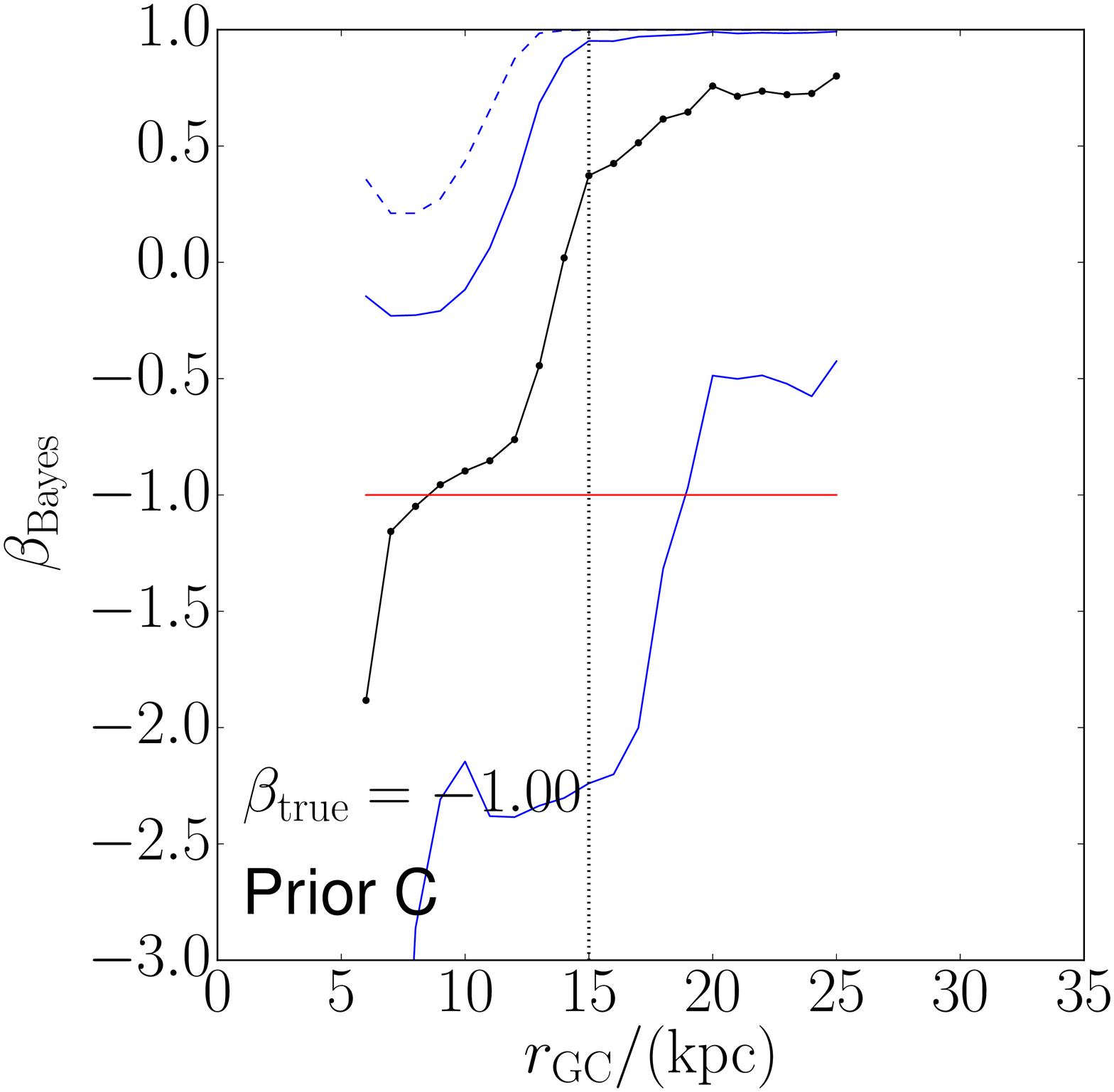} 
\end{flushleft}
\caption{
The posterior distribution of $\beta_{\rm Bayes}$ as a function of Galactocentric radius $r$ for our mock catalogues. 
The top and bottom rows correspond to the results for priors B and C, respectively, 
assumed in the Bayesian analyses. 
From left panel to right, the assumed $\beta_{\rm true}$ is $0.75, 0.5, 0$, and $-1$ 
(as shown by the horizontal solid red line). 
The black solid line shows the median value of the posterior distribution of $\beta_{\rm  Bayes}$, 
and blue solid and blue dashed lines respectively cover 68\% and 95\% of the posterior distribution. 
The vertical dotted lines at $r=15 \kpc$ are added to guide the eye. 
}
\label{fig:fixedBeta_BayesAB}
\end{figure*}

From the top row of 
this figure, we see that the Bayesian analysis on average 
returns a nearly unbiased estimate of velocity anisotropy independent of $\beta_{\rm true}$, 
for $r \leq 15 \kpc$ and 
for prior B (nearly identical plots were obtained for prior A and are not shown). 
However,
the spread in the posterior distribution becomes quite large at $r > 15 \kpc$, 
making it hard to infer the true anisotropy. 
On the other hand, the bottom row of Figure \ref{fig:fixedrGC_BayesAB} 
(as well as the bottom row of Figure \ref{fig:fixedBeta_BayesAB}) 
shows that a prior that is biased towards large $\beta$ 
like prior C results in 
an overestimate of $\beta$ $r > 10 \kpc$.

\section{Discussion}

In Sections \ref{section:resultMaxL} and \ref{section:resultBayes}, 
we have explored the systematic and random errors inherent in the 
local fitting methods with maximum-likelihood and Bayesian formulations. 
We found that these methods can reliably estimate $\beta$ only at $r \leq 15 \kpc$. 
In order to better understand the errors inherent in these methods, 
we discuss how the performance improves if we increase the sample size in Section \ref{section:N1e4}. 
Then we discuss how our results in this paper can be used in interpreting the 
measured values of $\beta$ in Section \ref{section:interpret_observed_beta}.
We also comment on the use of different (non-Gaussian) functions in the local fitting methods in Section \ref{section:other_estimates}.

\subsection{Effects of the increased sample size} \label{section:N1e4}

From a mathematical point of view, 
we expect that we can recover the value of $\beta$ from mock data accurately 
if we have a large enough sample of stars (and we know the functional form of the distribution function). 
In order to confirm this expectation, 
we did the same analyses as in Sections \ref{section:showCase} and \ref{section:MaxL_fixedrGC} 
with sample size of $N=10^4$ (at a given radius) instead of $N=10^3$. 
We found that we can recover $\beta$ accurately out to a significantly larger radius of $r = 25 \kpc$ (see Appendix \ref{appendix:N1e4} for details). 
From these experiments, 
we conclude that the systematic bias in $\beta_{\rm MaxL}$ seen in Figures \ref{fig:showCaseBetap0.5_MaxL}-\ref{fig:fixedrGC_bootstrap_MaxL} arises from the small sample size of $N=10^3$. 
Also, from a geometric argument in Appendix \ref{appendix:geometry}, 
we found that the random error on $\beta_{\rm{MaxL}}$ at $r>R_0$ ($R_0$ is the Galactocentric radius of the Sun) is approximately given by 
\eq{
|\Delta \beta_{\rm{MaxL}}| = 2 \sqrt{\frac{8}{N}} \left( \frac{r^2}{R_0^2} - \beta_{\rm true} \right).
}
This expression indicates that the performance of the maximum-likelihood method 
deteriorates as $r$ increases and improves as $N$ increases. 
We expect the similar results would be obtained if we use Bayesian method and prior B, based on our results in Sections \ref{section:resultMaxL} and \ref{section:resultBayes}.

We warn that these experiments do not guarantee 
that a reliable estimation of $\beta$ can always be obtained with with just 4-dimensional information 
for an adequately large number of sample stars. 
Local fitting 
measurements of $\beta$ presented in this paper make use of the fact that $\sigma_{\rm los}$ depends on the heliocentric direction on the sky when the velocity distribution is anisotropic ($\beta \neq 0$) but not if it is isotropic ($\beta=0$) [see equation (\ref{eq:LOSVD})]. 
Our analyses use this direction-dependence of $\sigma_{\rm los}$ across the sky to estimate $\beta$ by assuming that the principal axes of the velocity ellipsoid are perfectly aligned with the Galactocentric spherical coordinates and that $(\sigma_r,\sigma_\theta,\sigma_\phi,V_{\rm rot})$ are functions of $r$ only.
In the Solar neighborhood, these assumptions are approximately valid \citep{Bond2010,Evans2016}, but there is no guarantee that they are valid outside the Solar neighborhood. 
Therefore, we expect that in order to obtain a reliable determination of $\beta$ at $r \gtrsim 20 \kpc$  proper motion data for halo stars are required.

\subsection{Literature values of $\beta$ based on the local fitting methods} \label{section:interpret_observed_beta}

In this paper, we found that the use of 4-dimensional information for $\sim 1000$ halo stars at a given radius is not sufficient to reliably estimate $\beta$ for Galactocentric radii $r>15 \kpc$ (even with error-free, noise-free data). 
In this subsection we use our results to better understand recent observational determinations of $\beta$ that were based on similar 
local fitting methods. 
In particular, we focus on \cite{Kafle2012} and \cite{King2015} 
as the typical studies in which the Bayesian and the maximum-likelihood methods are applied to a halo sample without any metallicity cuts. 
Also, we consider the results from \cite{Hattori2013}, 
who used a slightly different method based on solving the matrix equation to 4-dimensional information for a halo sample that was split into two different metallicity ranges. 
(We note \citealt{Kafle2013} did essentially the same analyses as in \citealt{Hattori2013}, 
but  \citealt{Kafle2013}  did not discard data points at large $r$ which, as we have shown, are likely to be biased.)

\begin{figure*}
\begin{flushright}
	\includegraphics[angle=0,width=0.65\columnwidth]{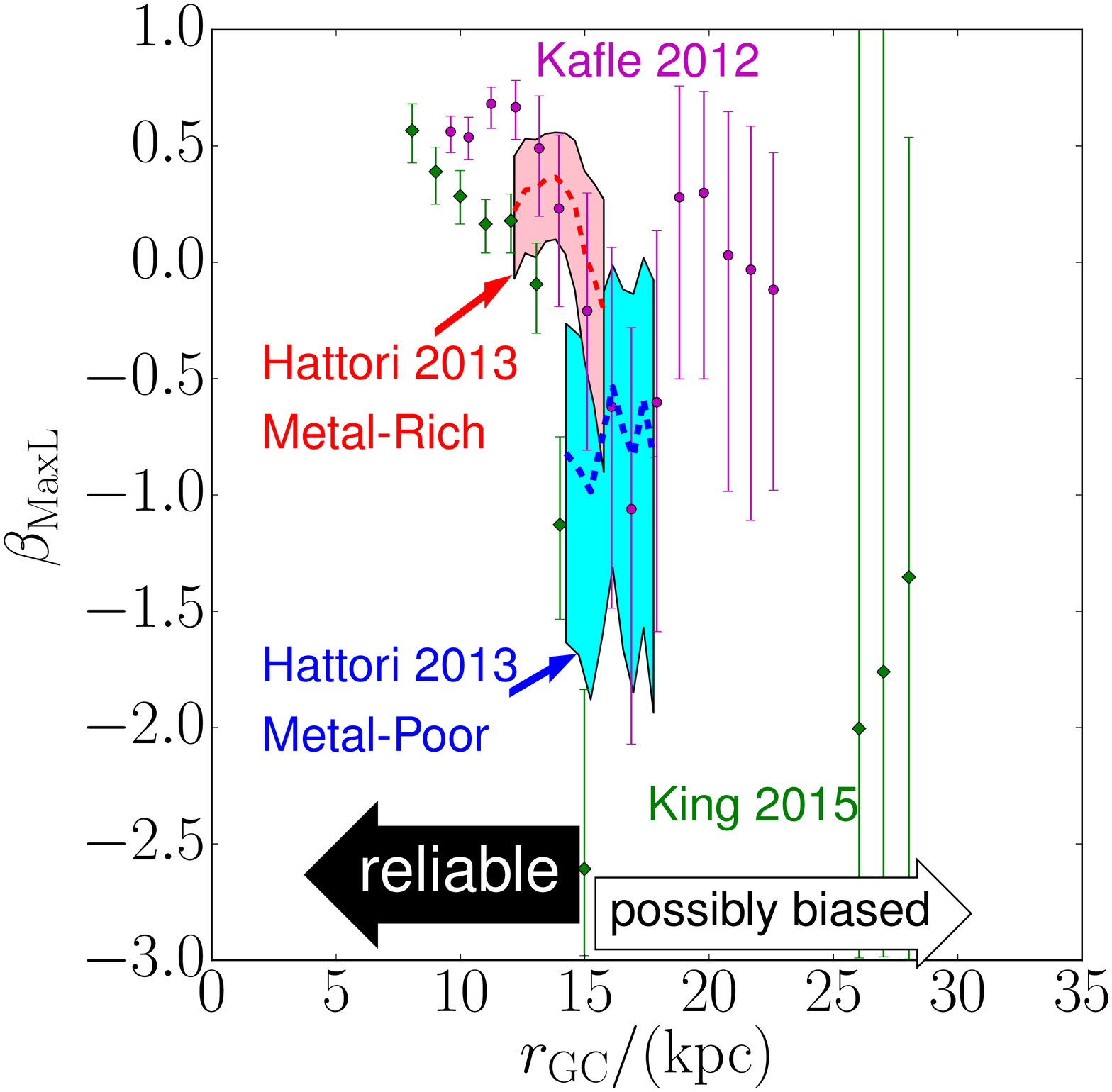} 
	\includegraphics[angle=0,width=0.65\columnwidth]{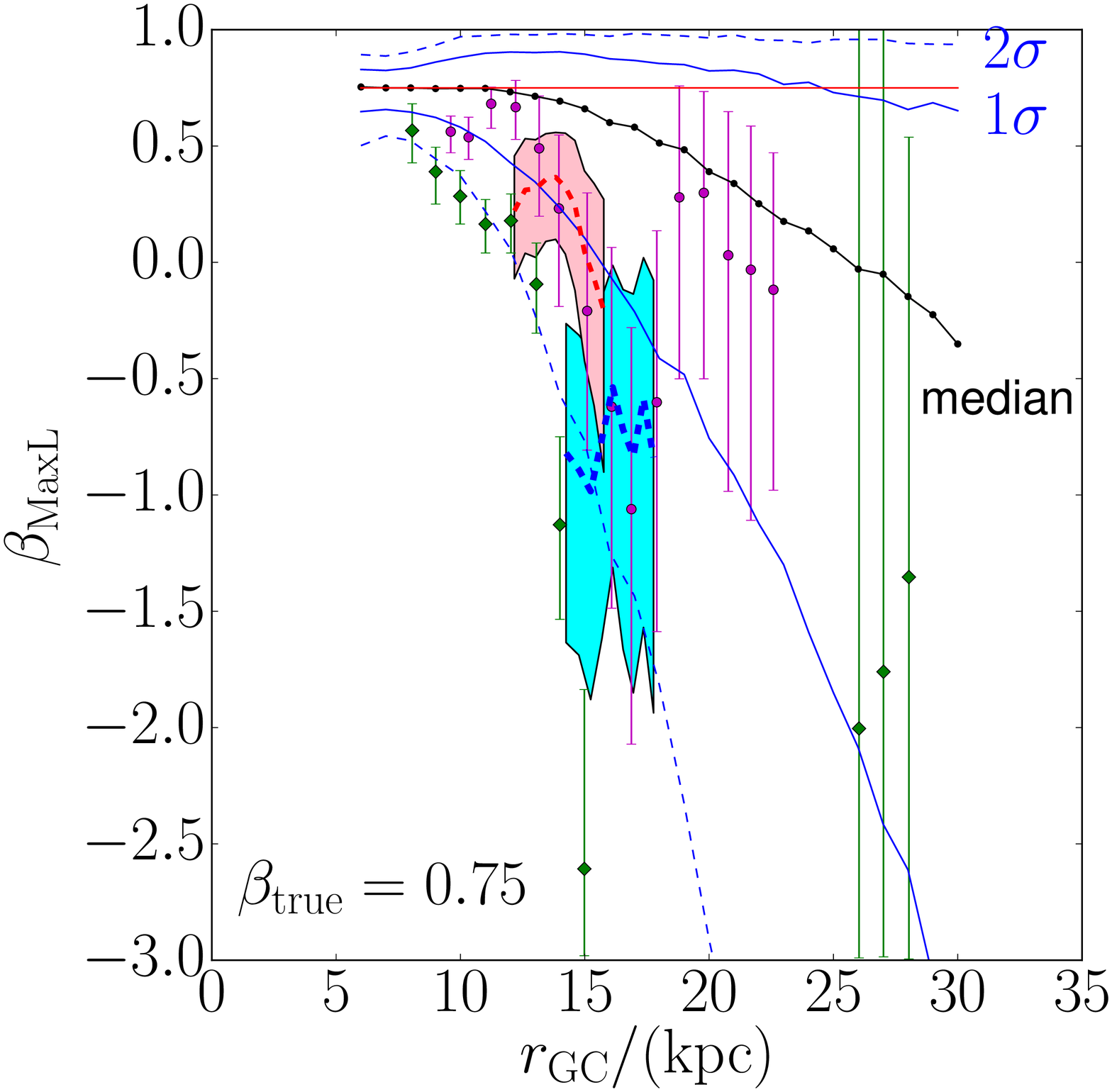} 
	\includegraphics[angle=0,width=0.65\columnwidth]{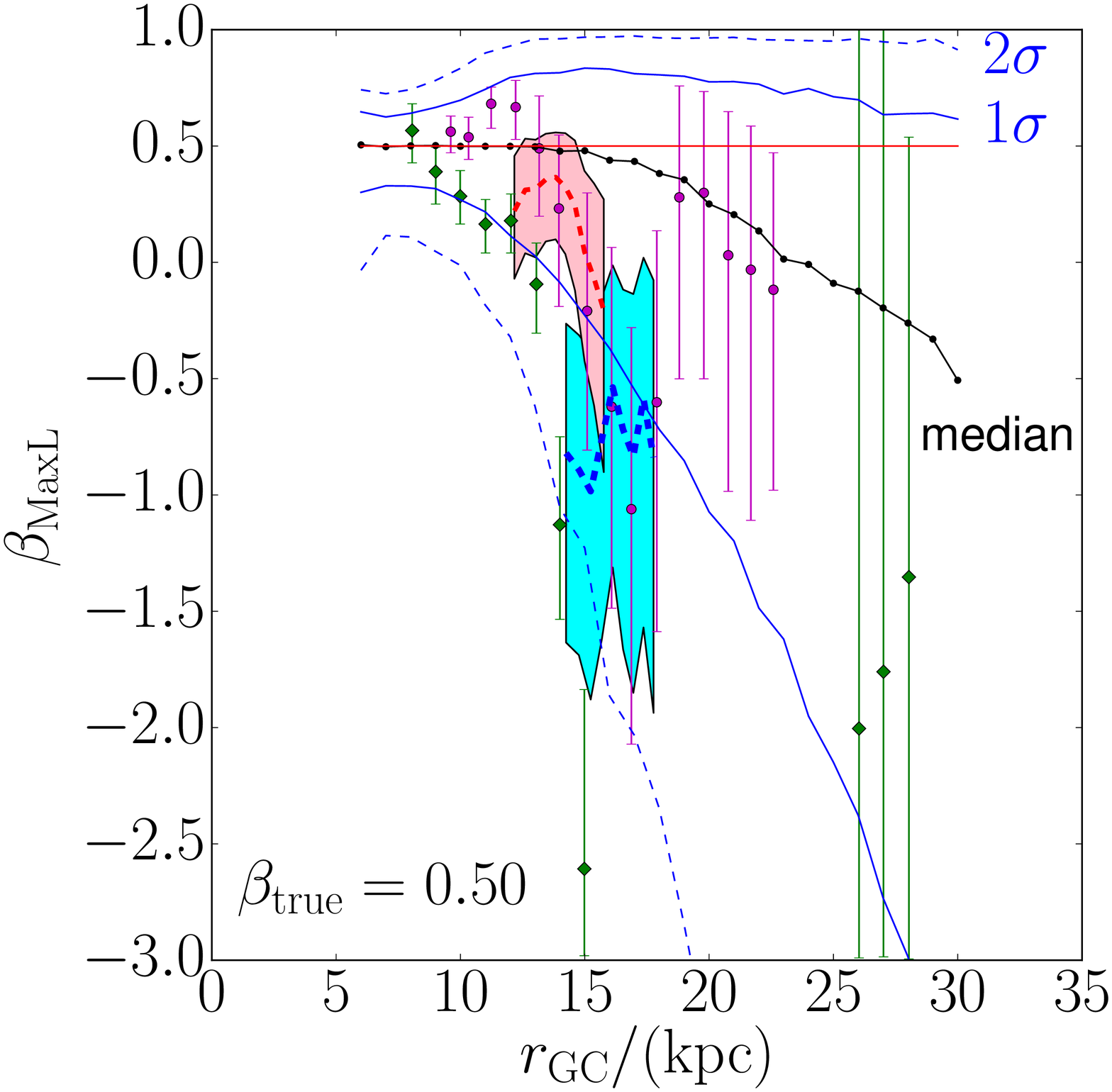} \\
	\includegraphics[angle=0,width=0.65\columnwidth]{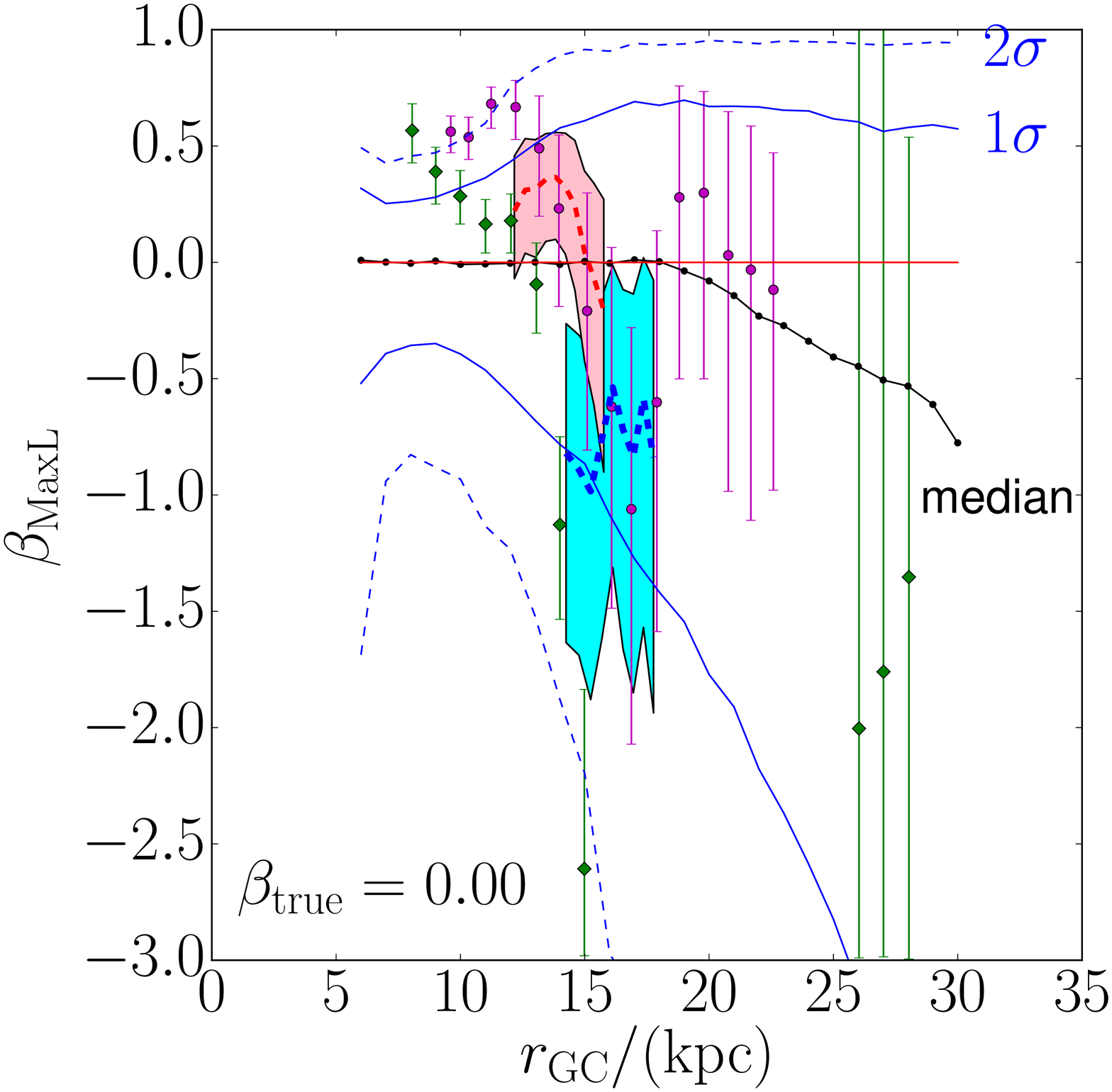} 
	\includegraphics[angle=0,width=0.65\columnwidth]{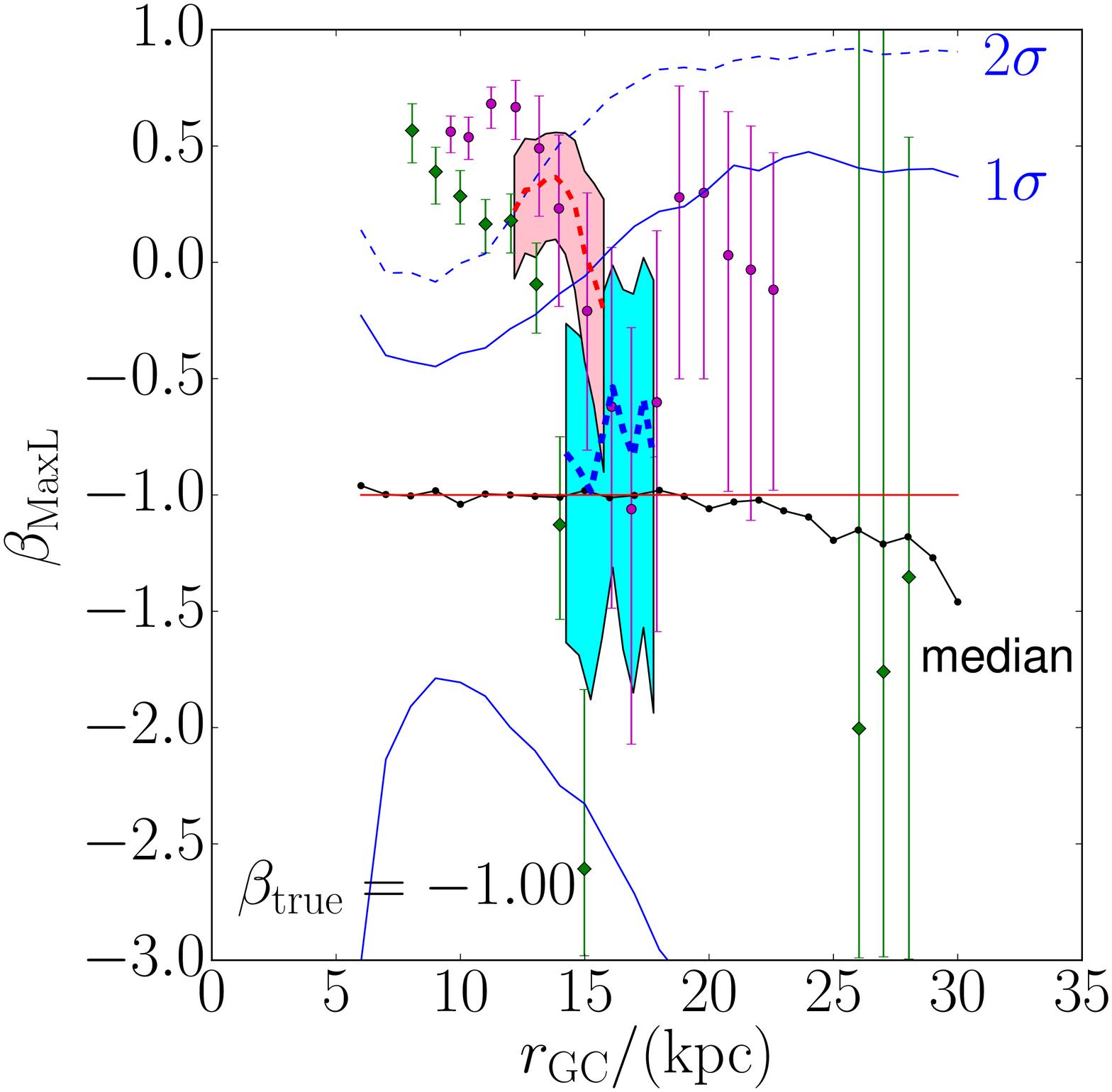} 
\end{flushright}
\caption{
A comparison of 
the literature values of $\beta$ of halo stars based on local fitting methods  
and our results of mock-catalogue analyses. 
In the top-left panel, we show only the results from observational estimates of $\beta$. 
The pink filled range of $\beta$ at $\beta\simeq 0.3$ and 
the light-blue filled range of $\beta$ at $\beta\simeq -1$ 
show the results of metal-rich and metal-poor halo stars, respectively, taken from \cite{Hattori2013}. 
The magenta circles and green diamonds with error bars show the results from \cite{Kafle2012} and \cite{King2015}, respectively.
The radial range of $r<15 \kpc$ shown at the bottom the top-left panel 
indicates the radial range where the estimate of $\beta$ is reliable, based on our results in this paper. 
In the middle and rightmost columns, 
we also show the distribution of $\beta_{\rm MaxL}$ as a function of $r$ for a fixed value of $\beta_{\rm true}$, 
taken from Figure \ref{fig:fixedBeta_bootstrap_MaxL}. 
}
\label{fig:fixedBeta_data}
\end{figure*}

In Figure \ref{fig:fixedBeta_data},
we show the point-estimate distribution of $\beta_{\rm MaxL}$ taken from Figure \ref{fig:fixedBeta_bootstrap_MaxL} 
as well as the measurements of $\beta$ in the above-mentioned papers.
It is worth noting that all published measurements of $\beta$ from line-of-sight velocities 
rely on the maximum-likelihood, Bayesian, or other similar local fitting methods 
and that the results of these methods are quite similar (as shown in this paper). 
Hence a comparison of $\beta$ profiles obtained from maximum-likelihood analyses of our mock data with observationally determined $\beta$ profiles provides 
a useful way to evaluate the accuracy of previous observational measurements. 
The distribution of $\beta_{\rm MaxL}$ in Figure \ref{fig:fixedBeta_data} 
represents the expected distribution of the maximum-likelihood solution for a given value of $\beta_{\rm true}$. 
Thus, if a given data point (not the error bar) in Figure \ref{fig:fixedBeta_data} 
is located outside the two-$\sigma$ range of the distribution of $\beta_{\rm MaxL}$ for a certain value of $\beta_{\rm true}$, 
then that value of $\beta_{\rm true}$ is disfavored by the data point.

First, we focus on the panels with $\beta_{\rm true}=0$ and $0.5$. 
These panels suggest that
even if the Milky Way stellar halo has a constant profile of, say, $\beta(r)=0.5$, 
most of the  
measured values of $\beta$ from observations (data points)
shown here lie within two-$\sigma$ 
of the expected deviation (using the maximum-likelihood method).

Second, let us focus on data points of \cite{Kafle2012} and \cite{King2015} at $r>13 \kpc$. 
For these data points, we see a declining profile of $\beta(r)$ as a function of $r$, although the error bars are quite large at large radii. 
This declining profile is a reminiscent of the declining profile of the median curve of $\beta_{\rm MaxL}$. 
Therefore, even if the measured $\beta(r)$ profile is mildly declining, it does not necessarily mean that the true $\beta(r)$ is declining. 
The apparent dip of $\beta(r)$ at $r\simeq $(15-17)$ \kpc$ is intriguing, and this dip might be a true signal. 
Indeed, \cite{Loebman2016} demonstrate that 
dips in $\beta(r)$ can 
arise due to substructure in the stellar halo. 
However, it is worthwhile to point out that 
the measurement of a dip in $\beta$ profile at $r=17 \kpc$ 
reported by \cite{Kafle2012} is consistent with $-1 \leq \beta_{\rm true} \leq 0.5$. 
Also, the data points at $r \gtrsim 20 \kpc$ do not seem to help our understanding of $\beta(r)$ profile, 
since both the systematic and random errors grows rapidly as a function of $r$. 
For example, all the data points at $r > 20 \kpc$ of \cite{Kafle2012} and \cite{King2015} are consistent with $-1 \leq \beta_{\rm true} \leq 0.75$, 
which is the full range of $\beta_{\rm true}$ we have explored in this paper.

Third, let us focus on data points of \cite{Kafle2012} and \cite{King2015} at $r\simeq 12 \kpc$. 
Although their sample stars are partially overlapping (both of their samples include SDSS blue-horizontal branch stars), 
their estimated values of $\beta$ are inconsistent with each other if their error bars are correct. 
Currently we do not know the origin of this discrepancy, 
but the published error bars might be too small. 
For example, most of their data points at $r\simeq 12 \kpc$ are consistent with our results of $\beta_{\rm true}=0.5$ within one-$\sigma$ 
range. 
On the other hand, the error bars in \cite{Kafle2012} and \cite{King2015} are $\sim $30\% of the one-$\sigma$ range of our mock catalogue analyses, 
while the error bars in \cite{Hattori2013} are $\sim$ 65\% of the one-$\sigma$ range in this paper. 
Since our mock catalogues do not include observational errors
and yet the error bars in our mock catalogues are larger than the above-mentioned papers \citep{Kafle2012, Hattori2013, King2015}, 
the published error bars might not represent the uncertainty in $\beta$. 
However, it is premature to conclude that the published error bars are incorrect, 
since our mock catalogues are not realistic enough (e.g., we do not consider the spread in $r$ of sample stars).

Lastly, let us now focus on data points of \cite{Hattori2013}. 
Based on the apparent difference of $\beta$ for the metal-poor and metal-rich samples, 
they claimed that the kinematics of the halo stars depends on the metallicity 
(for stars with $r < 18 \kpc$). 
Although they carefully avoid possible systematic errors on $\beta$ 
by discarding the results at $r>16 \kpc$ for metal-rich halo stars 
and at $r>18 \kpc$ for metal-poor halo stars, their cut may not be sufficient. 
Based on the analyses in this paper, we argue that a safer approach is to discard data points at $r>15 \kpc$. 
(Even with this spatial cut, a metallicity-dependence in $\beta$ can still  be seen in the surviving data points.)
Admittedly, there is a possibility that these differences appeared by mere chance. 
For example, if the Milky Way stellar halo has a constant profile of $\beta(r)=\beta_{\rm true}=0$, 
then the estimated $\beta(r)$ profiles of both the metal-rich and metal-poor samples are inside the one-$\sigma$ range expected from our model. Although the current 4-dimensional data are not good enough to conclusively assert that $\beta(r)$ depends on stellar metallicity, the current data within $15 \kpc$ do hint at such a possibility. 
For example, let us suppose that the $\beta$ profile of metal-rich halo is $\beta(r)=\beta_{\rm true}=0.5$. In this case, the data points for the metal-poor sample at $r\leq 15 \kpc$ are marginally outside the one-$\sigma$ range of the model prediction. Conversely, let us suppose that the metal-poor halo has $\beta(r)=\beta_{\rm true}=-1$. 
In this case, most of the data points of the metal-rich sample at $r\leq 15 \kpc$ are outside the one-$\sigma$ range. 
A robust determination of a metallicity dependence in $\beta(r)$  awaits confirmation with kinematical data from Gaia and chemical information from ground-based surveys. 

\subsection{Local fitting methods with non-Gaussian functions} \label{section:other_estimates}

In previous sections, we generated mock catalogues based on a Gaussian velocity distribution 
and performed local fitting analyses by assuming that the underlying velocity distribution is also a Gaussian function. 
However, in reality we do not know the correct functional form of the velocity distribution of the halo stars. 
In order to investigate the reliability of our local fitting methods, we perform some additional tests.

First, we introduce two simple distribution function models that are both functions of energy $E$ and total angular momentum $L$ 
(see Appendix \ref{appendix:twoDF} for details).  
One model, $f_{\rm const}(E,L)$,  has a constant profile of $\beta(r) = \beta_{\rm const}$, 
and another model, $f_{\rm OM}(E,L)$, is an Osipkov-Merritt model with a rising $\beta$ profile of $\beta(r) = r^2 / ( r_a^2 + r^2)$ with $r_a$ a constant.
Then we assume that the potential of the Milky Way is a spherical singular isothermal potential (with a flat rotation curve) 
and generate three sets of mock catalogues in the same manner as in Section \ref{section:mock_catalogues}.  
For two sets of mock catalogues, we adopt $f_{\rm const}(E,L)$ with $\beta_{\rm const} = 0.25$ and $-0.42$. 
For the other set of mock catalogues, we adopt $f_{\rm OM}(E,L)$ with $r_a = 10 \kpc$.

We fit each of these three sets of mock catalogues with local fitting methods with the maximum-likelihood formulation. 
Specifically, at each Galactocentric radius $r$, 
we fit the data by assuming that the underlying velocity distribution is described by either 
$f_{\rm Gauss}(\vector{v}|\vector{x})$, 
$f_{\rm const}(E,L|\vector{x})$, or $f_{\rm OM}(E,L|\vector{x})$. 
Here, $f(E,L|\vector{x})$ denotes the {\it velocity distribution} at a given location $\vector{x}$ 
of a system obeying a distribution function $f(E,L)$, and it is different from $f(E,L)$ itself. 
In the following, however, we omit the arguments of $f_{\rm const}$ and $f_{\rm OM}$ for brevity.

\begin{figure*}
\begin{flushright}
	\includegraphics[angle=0,width=0.65\columnwidth]{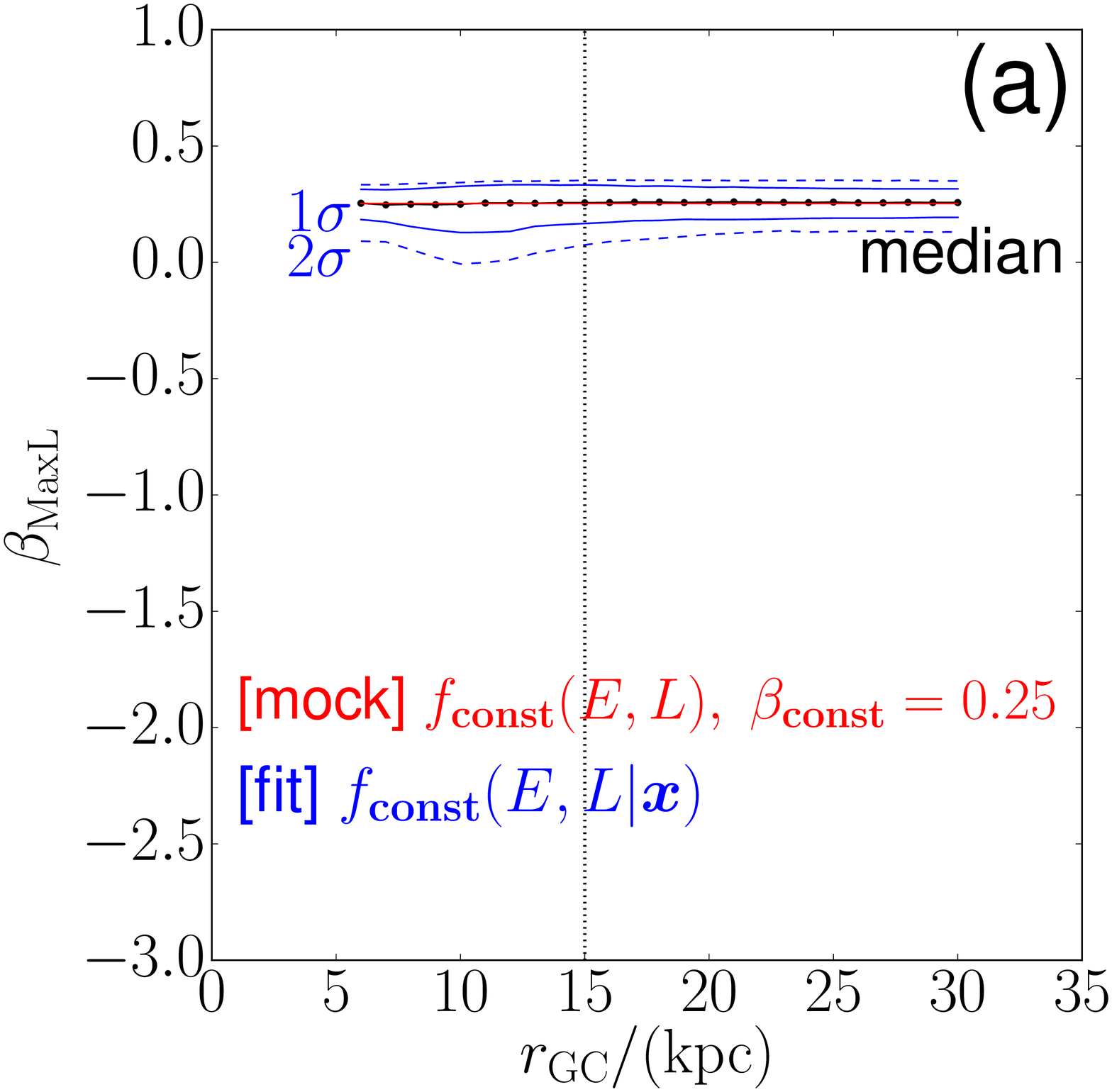} 
	\includegraphics[angle=0,width=0.65\columnwidth]{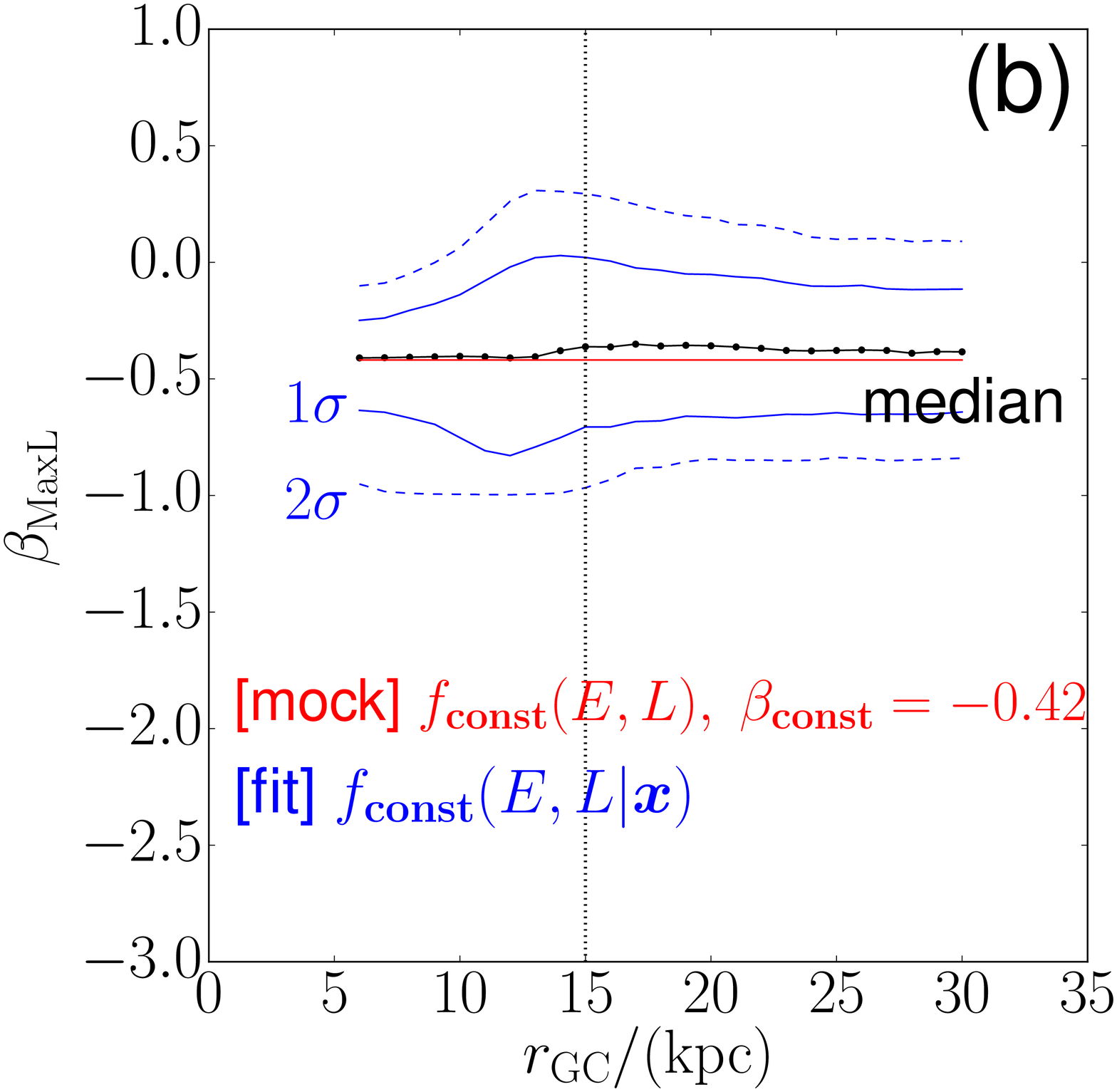} 
	\includegraphics[angle=0,width=0.65\columnwidth]{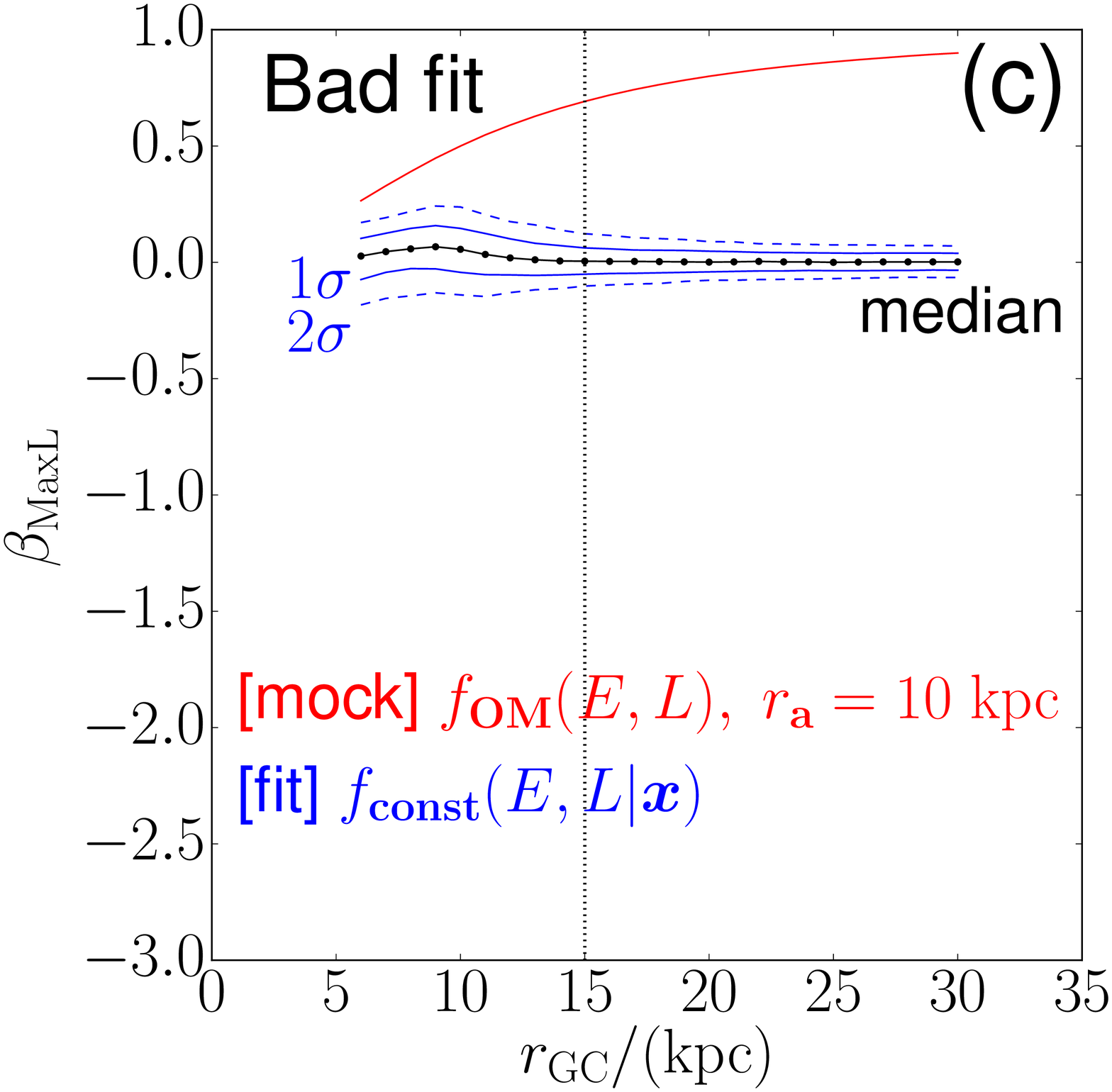} \\
	\includegraphics[angle=0,width=0.65\columnwidth]{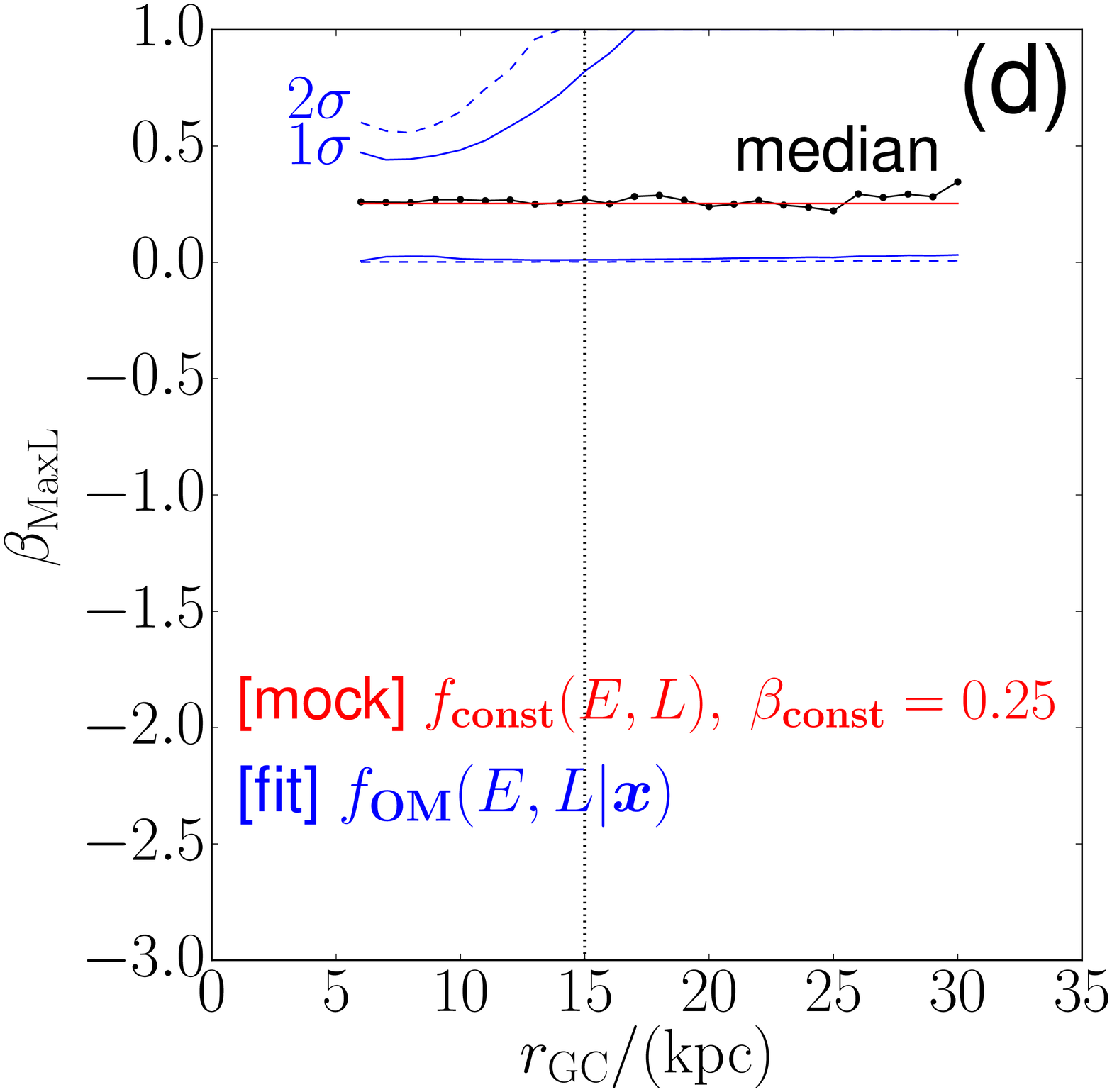} 
	\includegraphics[angle=0,width=0.65\columnwidth]{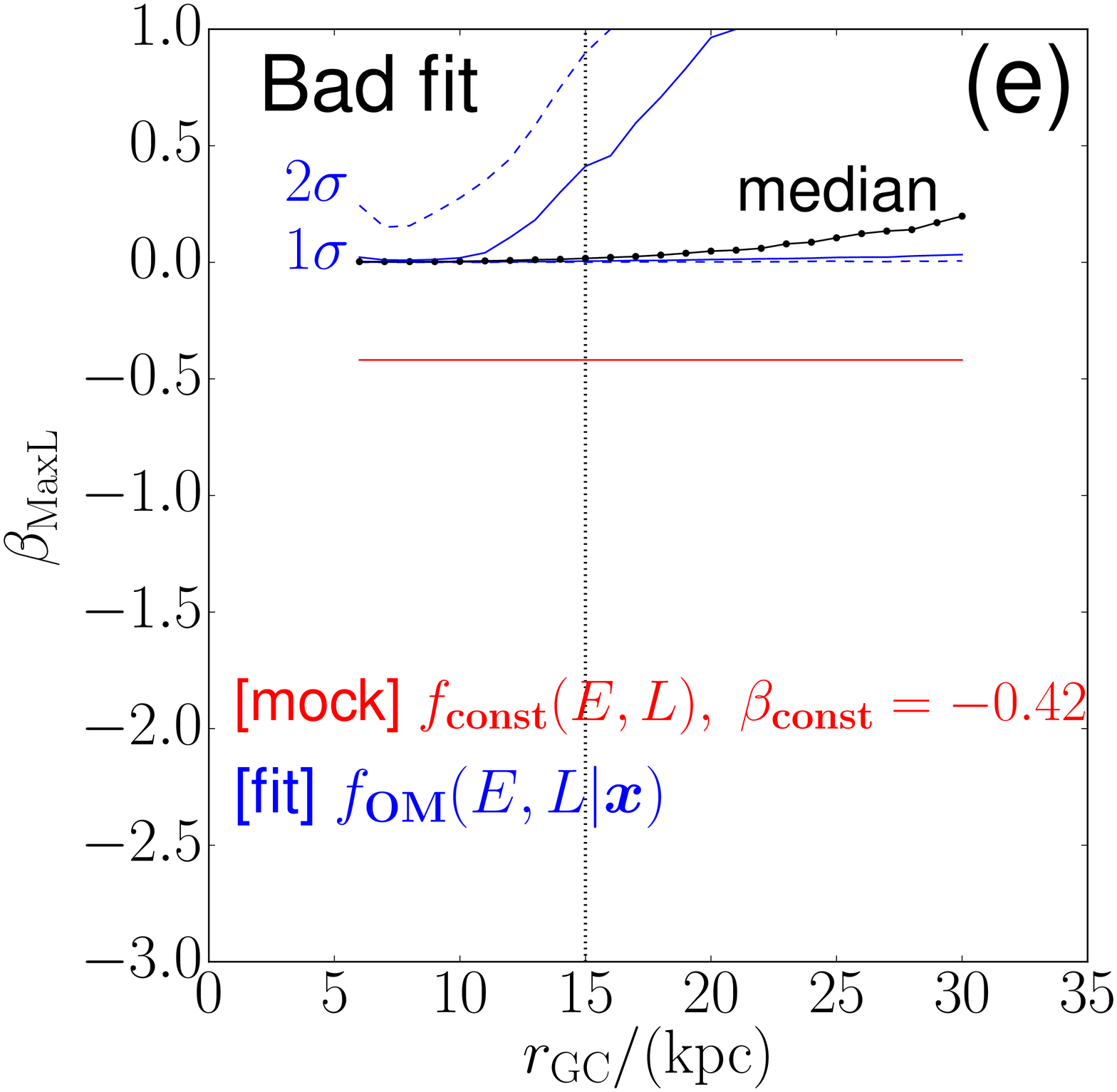} 
	\includegraphics[angle=0,width=0.65\columnwidth]{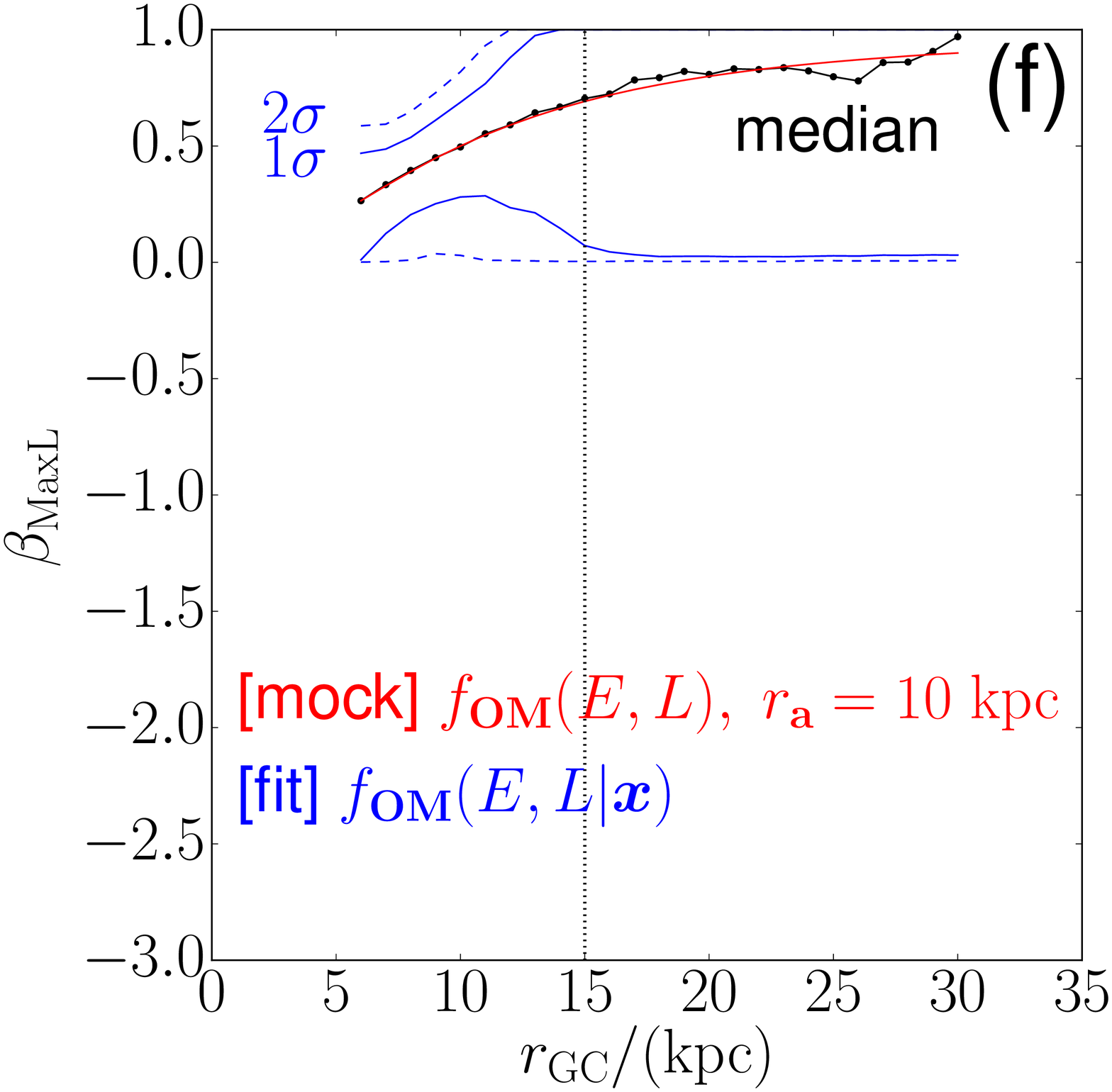} \\
	\includegraphics[angle=0,width=0.65\columnwidth]{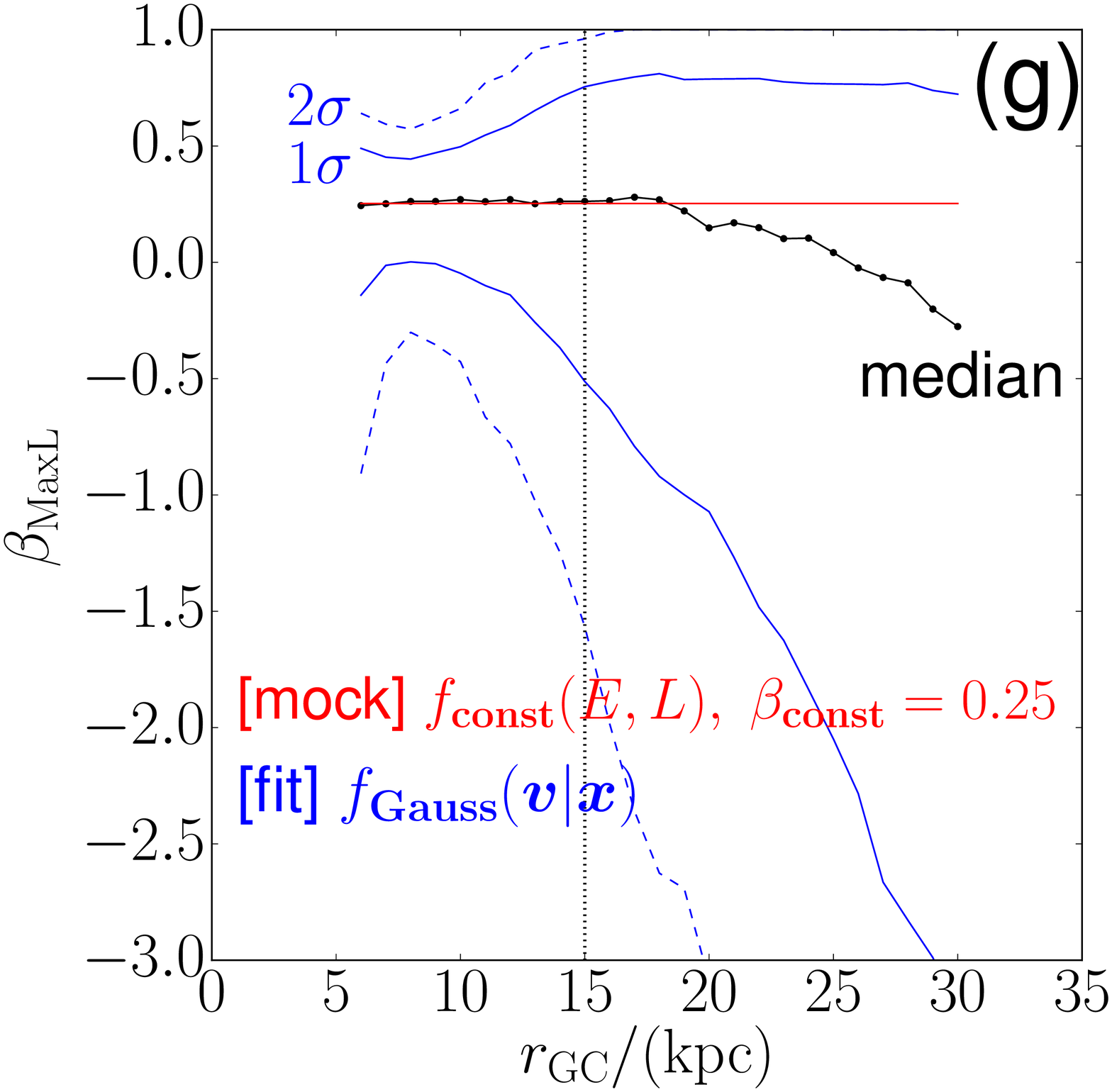} 
	\includegraphics[angle=0,width=0.65\columnwidth]{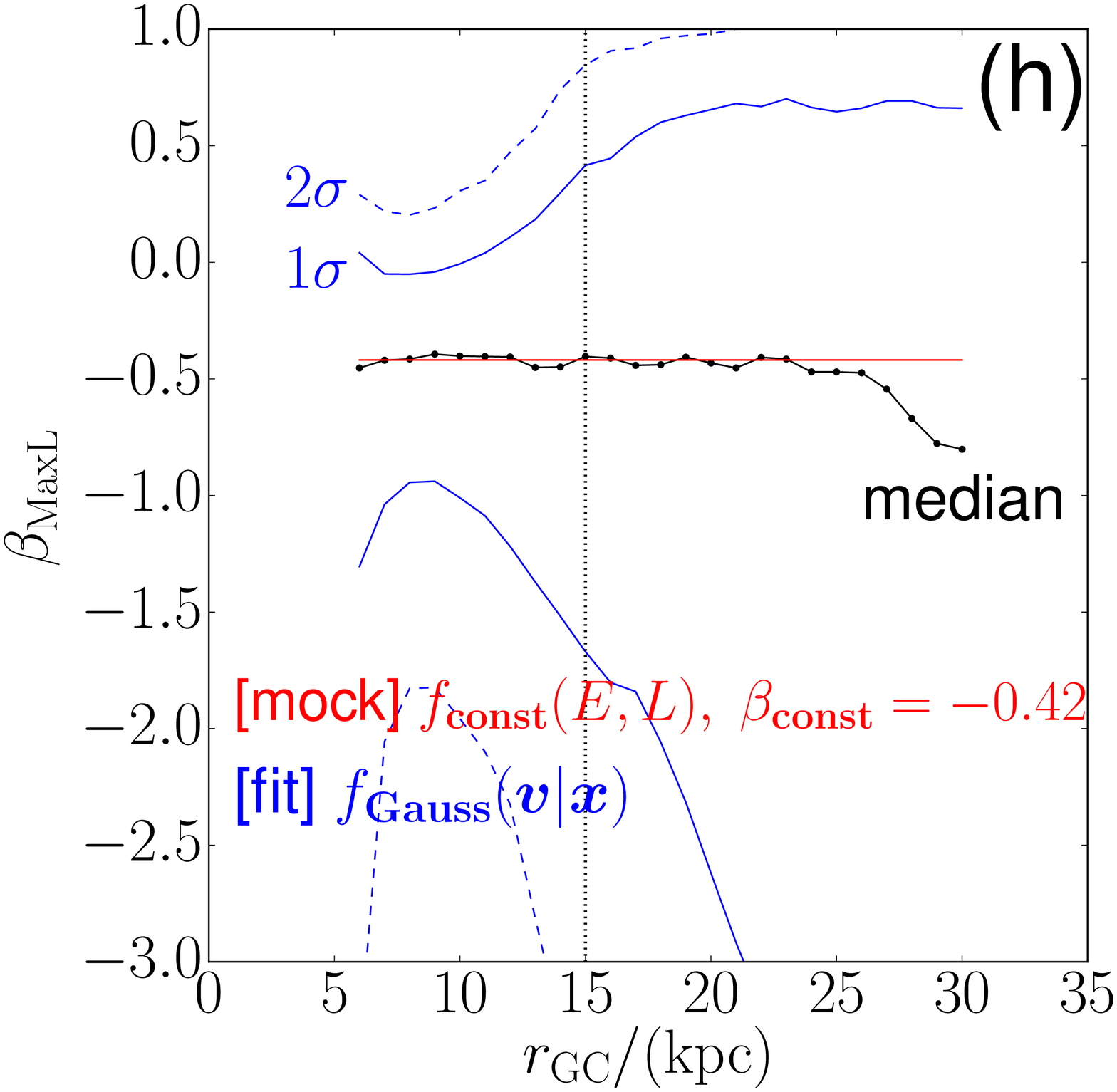} 
	\includegraphics[angle=0,width=0.65\columnwidth]{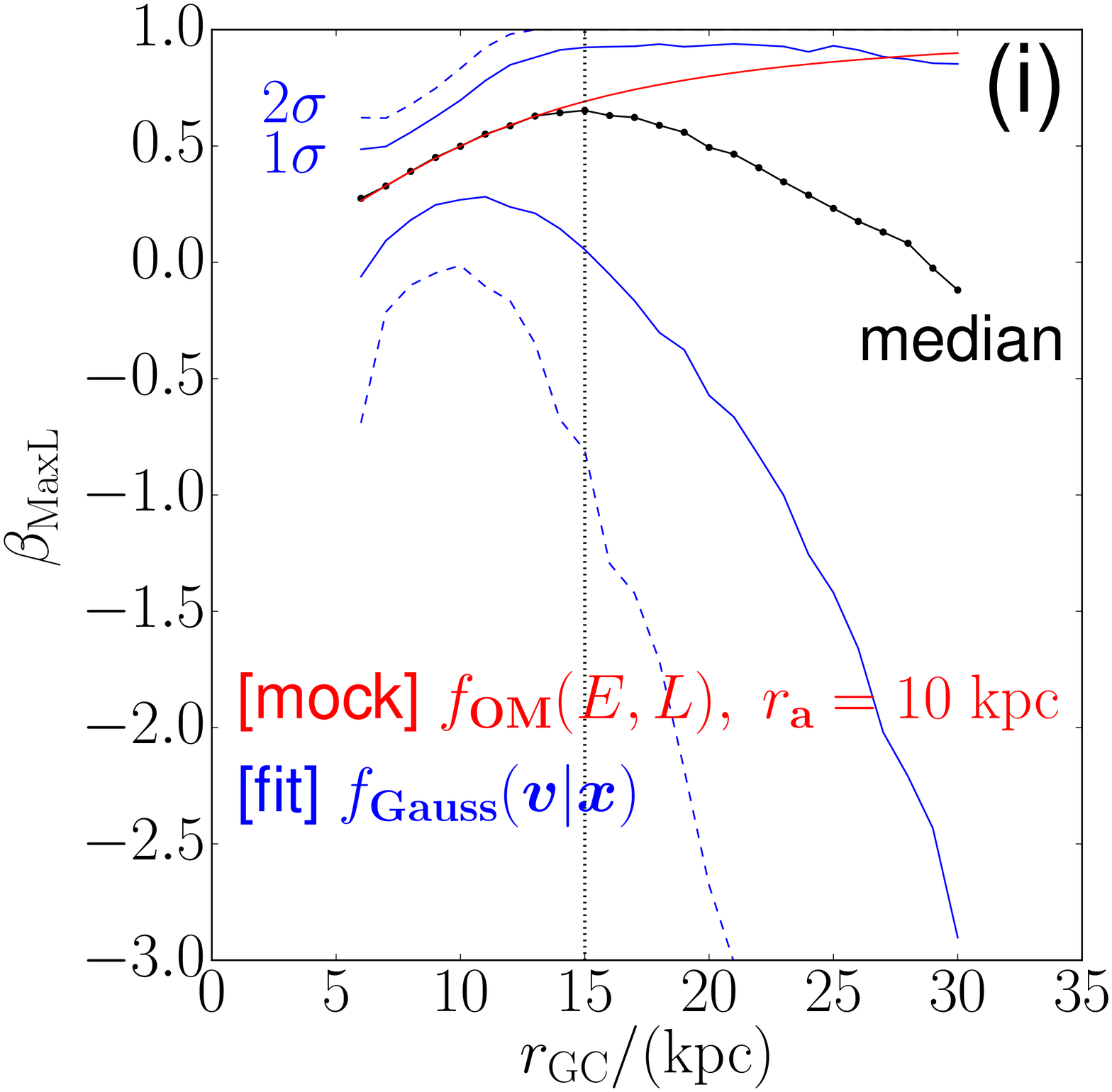} \\
\end{flushright}
\caption{
A comparison of the anisotropy $\beta$ measured by local fitting methods 
and the correct value of $\beta$ in the additional mock catalogues described in Section \ref{section:other_estimates}. 
On each panel, mock catalogues generated from either $f_{\rm Gauss}(\vector{v} | \vector{x})$, $f_{\rm const}(E,L)$, or $f_{\rm OM}(E,L)$ 
are locally fitted with a velocity distribution model, which is either $f_{\rm Gauss}(\vector{v} | \vector{x})$, $f_{\rm const}(E,L | \vector{x})$, or $f_{\rm OM}(E,L | \vector{x})$.
As seen in panels (c) and (e), 
when an incorrect velocity distribution is used to fit the data, 
the estimated value of $\beta$ may suffer a significant systematic error even at $r \leq 10 \kpc$.  
Also, as seen in panels (g)-(i), 
the estimated value of $\beta$ is unbiased at $r \leq 15 \kpc$ independent of the type of the mock catalogues, 
when we use the $f_{\rm Gauss}$ model to fit the data. 
}
\label{fig:twoDF}
\end{figure*}

Figure \ref{fig:twoDF} shows the results of these analyses. 
As seen in Figure \ref{fig:twoDF}(f),
when the mock catalogues generated from $f_{\rm OM}$ are locally fitted with $f_{\rm OM}$, 
the median value of the $\beta_{\rm MaxL}$ for the mock catalogues 
almost overlaps the true profile of $\beta(r)$ at $6 \kpc \leq r \leq 30 \kpc$. 
However, at $r>15 \kpc$, 
the one-$\sigma$ range fills essentially the entire range of $0\leq \beta \leq 1$, 
which is the entire range of the allowed anisotropy in Osipkov-Merritt models 
(recall that the Osipkov-Merritt model does not permit models with negative $\beta$). 
Thus, the recovered value of $\beta$ is informative only at $r \leq 15 \kpc$, 
just as in the results in Figure \ref{fig:fixedBeta_bootstrap_MaxL}.
On the other hand, 
as seen in Figures \ref{fig:twoDF}(a) and \ref{fig:twoDF}(b),
when the mock catalogues generated from $f_{\rm const}$ are locally fitted with $f_{\rm const}$,  
the median value of the $\beta_{\rm MaxL}$ for the mock catalogues 
is very close to the correct value $\beta_{\rm const}$ 
at $6 \kpc \leq r \leq 30 \kpc$ for both cases of $\beta_{\rm const}=0.25$ and $-0.42$. 
Interestingly, the uncertainty in $\beta$ does not change a lot as a function of $r$ for this model. 
This result suggests that choosing the correct functional form is beneficial in estimating $\beta$.

When an incorrect velocity distribution is assumed, on the other hand, 
the resultant $\beta$ profiles are sometimes not reliable. 
For example, 
as seen in Figure \ref{fig:twoDF}(c),
when mock catalogues generated from $f_{\rm OM}$
are locally fitted with $f_{\rm const}$, 
the $\beta(r)$ profile can not be recovered even at $r \leq 10 \kpc$. 
What is interesting in these fits is that the formal uncertainties associated with the recovered value of $\beta$ 
are very small despite the fact that the true profile of $\beta(r) = r^2/( (10 \kpc)^2 + r^2 )$ is well outside the two-$\sigma$ range. 
Also,
as seen in Figure \ref{fig:twoDF}(e),
when mock catalogues generated from $f_{\rm const}$ with $\beta_{\rm const}=-0.42$ 
are locally fitted with $f_{\rm OM}$, 
the estimated profile of $\beta$ is far from the true profile at all the radii explored,
since $\beta<0$ can not be attained with $f_{\rm OM}$. 
These examples suggest that 
when local fitting methods are applied to 4D data, 
the use of wrong functional forms for the velocity distribution 
can result in a significant systematic error on the recovered value of $\beta$. 
On the other hand,
as seen in Figures \ref{fig:twoDF}(g)-(i),
when we use $f_{\rm Gauss}$ to fit the mock catalogues generated from $f_{\rm const}$ and $f_{\rm OM}$, 
the $\beta$ profile is reliably estimated at $r < 15 \kpc$. 
This result is intriguing, since the velocity distribution for $f_{\rm const}$ is non-Gaussian.\footnote{
We note that the velocity distribution for $f_{\rm OM}$ is always Gaussian in our current example.
}

These results, combined with the findings in Section \ref{section:resultMaxL}, can be summarized in the following manner. 
When local fitting methods are used to estimate $\beta$, 
\begin{itemize}
\item 
if the correct functional form of the velocity distribution is assumed, 
the estimated value of $\beta$ and the associated error bars are reliable at least at $r<15 \kpc$, 
and even at $r=30 \kpc$ if the halo obeys $f_{\rm const}$ [see Figures \ref{fig:twoDF}(a),(b),(f)]; 
\item 
if a wrong functional form of the velocity distribution is assumed,
the estimated value of $\beta$ may be significantly biased (even at $r<10 \kpc$), 
and the associated formal error may not be reliable [see Figures \ref{fig:twoDF}(c),(e)]; 
\item 
if the functional form of the velocity distribution is unknown, 
the assumption of a Gaussian function $f_{\rm Gauss}$ is a reasonable choice, 
since this choice allows unbiased estimation of $\beta$ at $r \leq 15 \kpc$ 
for the various kinds of mock catalogues explored in this paper  [see Figures \ref{fig:twoDF}(g)-(i)].
\end{itemize}

Recently, many authors have developed sophisticated distribution function models 
that fit the observed positions and velocities of halo stars in the Milky Way 
\citep{Deason2012, Williams2015, DasBinney2016, Das2016}. 
With these global fitting methods, 
authors use a sample of halo stars that are not restricted to a single Galactocentric radius $r$ (as with the local fitting methods), 
but are distributed over a wide range of $r$. 
However, given that our local fitting methods fail to recover $\beta$ (sometimes even at $r<10 \kpc$) when a wrong functional form is assumed, 
it may be worthwhile checking the performance of these global fitting methods 
when a wrong functional form of the stellar halo distribution function is assumed.

\section{Conclusion}

In the past 10 years, many authors have tried to infer the velocity distribution of distant halo stars 
from stellar samples without reliable proper motion measurements (see references in Section \ref{section:introduction}). 
A common way of inferring the 3-dimensional velocity dispersion of halo stars 
from 4-dimensional position and line-of-sight velocity measurements is the local fitting methods. 
In these methods, they estimate 3-dimensional velocity dispersion by using information on 
how the line-of-sight velocity dispersion $\sigma_{\rm los}$ varies across the sky, 
which reflects the different line-of-sight projections of the velocity ellipsoid. 
However, as stars get farther away from us and from the Galactic center, 
$v_{\rm los}$ (corrected for the Solar reflex motion) 
becomes increasingly closer to $v_r$.
As a result, $\sigma_{\rm los}$ becomes closer to $\sigma_r$, 
and the variation of $\sigma_{\rm los}$ across the sky becomes harder to evaluate, 
making it difficult to estimate the tangential components of the velocity dispersion (see Appendix \ref{appendix:geometry}). 
Thus it is important to explore the random and systematic uncertainties inherent in such methods using mock datasets, 
in order to build intuition about how far out in the stellar halo we can reliably recover the velocity anisotropy. 
In this paper, we tackled this problem by performing a series of mock analyses. 
The main messages of this paper can be summarized as follows.

\begin{enumerate}

\item  
As shown in Figure \ref{fig:fixedrGC_bootstrap_MaxL}, 
the local fitting methods with the maximum-likelihood formulation can in principle 
reliably estimate the velocity anisotropy $\beta$ of the stellar halo at $r \leq 15 \kpc$ 
but is unable to reliably estimate $\beta$ at larger $r$ 
if 4-dimensional data only (position and line-of-sight velocity) are used
for $N=10^3$ halo stars at a given radius. 

\item 
If a flat prior on $\beta$ is adopted, 
the Bayesian formulation yields 
similar results to those obtained with the maximum-likelihood formulation
[see Figure \ref{fig:fixedrGC_bootstrap_MaxL} and the top row (prior B) of Figure \ref{fig:fixedrGC_BayesAB}]. 

\item  
Previous local fitting analyses to measure $\beta$ from 4-dimensional information 
used a few hundred halo stars (in each radial bin). 
Our results suggest that these measurements of $\beta$ are very likely to be biased to low/negative values at $r > 15 \kpc$. 

\item 
The performance of the local fitting methods 
to estimate $\beta$ with 4-dimensional information 
can be improved if we enlarge the sample size (see Section \ref{section:N1e4} and Appendix \ref{appendix:N1e4}). 
However, we expect that a more direct way of better estimating $\beta$ at large $r$ is to use accurate proper motion data. 

\item 
If the correct functional form of the velocity distribution of the stellar halo is assumed,
the performance of the local fitting methods to estimate $\beta$ can be improved;
otherwise assuming a Gaussian velocity distribution is a reasonable choice
(see Section \ref{section:other_estimates} and Figure \ref{fig:twoDF}). 

\item
It is important to point out that 
\cite{Deason2013} and 
\cite{Cunningham2016}
used halo stars at $18 \kpc < r < 30 \kpc$ with reliable proper motion data 
and reported $\beta \simeq 0$, although their sample size was small ($N=13$). 
Their results combined with the results in this paper suggest 
that the negative values of $\beta$ at $r > 20 \kpc$ reported by \cite{Kafle2012} and \cite{King2015} 
likely resulted from the large systematic and random errors inherent in the maximum-likelihood method (see Figure \ref{fig:fixedBeta_data}). 

\item 
In this paper we have used idealized mock catalogues 
for which the stellar heliocentric distances and line-of-sight velocities are measured with infinite precision. 
Also, all the stars in each mock catalogue are assigned the same Galactocentric radius $r$. 
The errors and noise in real data will only make the task of measuring $\beta$ from 4-dimensional data even harder.
However, in the next 3 to 5 years, Gaia \citep{Perryman2001, Lindegren2016} 
will provide  proper motion data for a large number of halo stars, 
opening new avenues for measuring the velocity distribution of halo stars at $r>20 \kpc$. 
These new data will yield important insights into our understanding of 
the structure and the merger history of the Milky Way.

\end{enumerate}

\acknowledgments{
The authors thank the stellar halo group at the Department of Astronomy, University of Michigan for stimulating discussions.
MV and KH are supported by NASA-ATP award NNX15AK79G.
SRL acknowledges support from the Michigan Society of Fellows. 
}       


\clearpage
\appendix

\section{Geometry of local fitting methods} \label{appendix:geometry}

Here we use a geometrical argument to
explain why the estimation of $\beta$ deteriorates as $r$ increases and improves as $N$ increases. 
We assume that the velocity distribution obeys a Gaussian distribution described in equation (\ref{eq:DF}). 
Also, for brevity, we assume that the sample stars have an identical Galactocentric radius $r>R_0$, 
where $R_0$ is the Galactocentric radius of the Sun, 
and that they are distributed along Galactic longitude $\ell = 0^\circ$ or $180^\circ$ (with any value of Galactic latitude $b$). 
In this case, the line-of-sight velocity dispersion $\sigma_{\rm los}$ at a given position in the Milky Way is given by 
\eq{
\sigma_{\rm los}^2 = Q_r^2 \sigma_r^2 + (1-Q_r^2) \sigma_\theta^2, \label{eq:sig_los}
}
where $Q_r = \vector{e}_{\rm los} \cdot \vector{e}_r$ [see equation (\ref{eq:LOSVD})].
We note that this linear dependence of $\sigma_{\rm los}^2$ on $Q_r^2$ suggests that 
$\sigma_{\rm los}^2 = \sigma_\theta^2$ at $Q_r^2 = 0$, 
while $\sigma_{\rm los}^2 = \sigma_\theta^2$ at $Q_r^2 = 1$. 
Also, we note that at a given Galactocentric radius $r(>R_0)$, 
the maximum value of $Q_r^2$ is $Q^2_{r, \rm{max}}(r)=1$ ($b=0^\circ$) and 
the minimum value of $Q_r^2$ is $Q^2_{r, \rm{min}} (r) = 1- {R_0^2}/{r^2}$ ($b=\pm90^\circ$).

\begin{figure*}
\begin{flushleft}
	\includegraphics[angle=0,width=0.95\columnwidth]{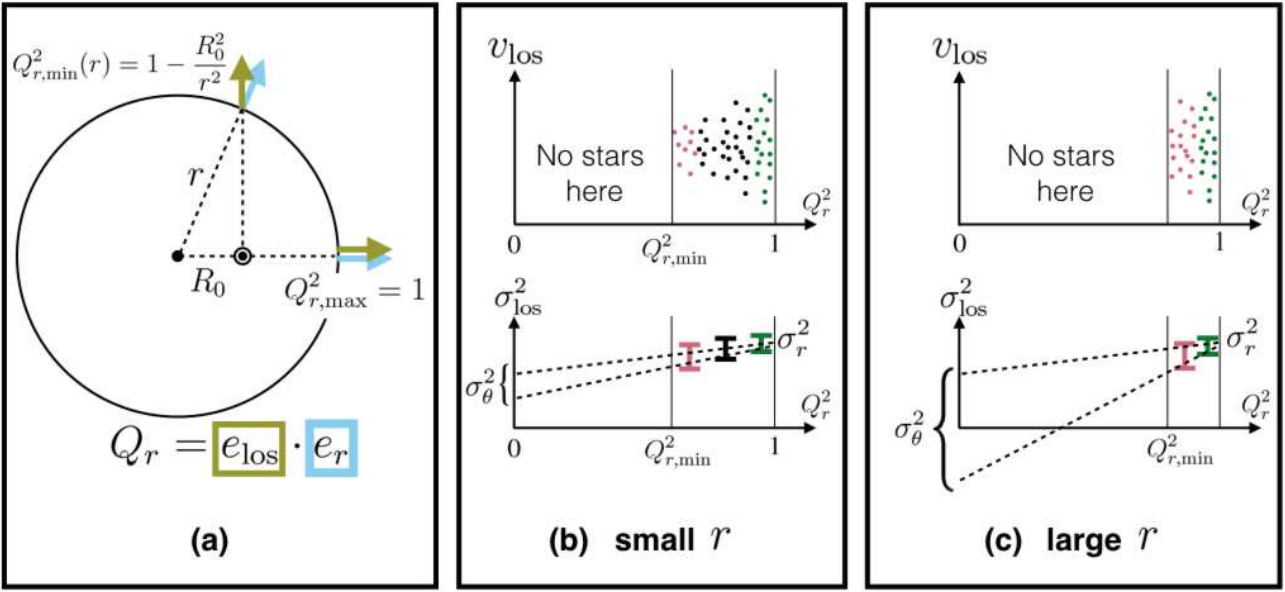}
\end{flushleft}
\caption{
A geometrical explanation for the local fitting method. 
(a) 
Illustration of how $Q_r$ depends on the line-of-sight direction. 
(b) 
At a fixed radius $r$, the observational constraints on 
$\sigma_{\rm los}^2$ as a function of $Q_r^2$ can be only obtained at 
$Q^2_{r, {\rm min}}(r) \leq Q_r^2 \leq 1$. 
The observed value of $\sigma_{\rm los}^2$ near $Q_r^2 \simeq 1$ is a good measure for $\sigma_r^2$, 
but estimation of $\sigma_\theta^2$ requires an extrapolation of $\sigma_{\rm los}^2 (Q_r^2)$ towards $Q_r^2 = 0$. 
This is why $\sigma_\theta^2$ is associated with larger error than $\sigma_r^2$. 
(c) 
At large $r$, it is hard to estimate the slope ${\rm d} \sigma_{\rm los}^2 / {\rm d} Q_r^2$ 
since $Q^2_{r, {\rm min}}$ becomes very close to $1$. 
This is why at large $r$ the estimation of $\sigma_\theta^2$ is hard 
and sometimes the best-fit value of $\sigma_\theta^2$ is unphysical ($0$ or negative; see Appendix \ref{appendix:unphysical}). 
}
\label{fig:appendix_geometry}
\end{figure*}

If we have 4D information of $N$ stars, these data can be transformed into 
$\{ (Q_r, v_{\rm los})_i | i = 1,\cdots N \}$. 
The local fitting method fits the distribution of $(Q_r, v_{\rm los})$ with a model 
in which $\sigma_{\rm los}^2$ varies linearly as a function of $Q_r^2$ as described in equation (\ref{eq:sig_los}). 
To put it differently, if we bin the data according to $Q_r^2$ and derive $\sigma_{\rm los}^2$ for each bin, 
the derived $\sigma_{\rm los}^2$  profile is fitted with a line. 
The values of $\sigma_{\rm los}^2$ at $Q_r^2 = 1$ and $Q_r^2 = 0$ 
correspond to the best-fit values of $\sigma_r^2$ and $\sigma_\theta^2$, respectively. 
The data points are distributed only at $Q^2_{r, {\rm min}}(r) \leq Q_r^2 \leq 1$, 
so the derivation of $\sigma_\theta^2$ requires an extrapolation of the linear relationship 
inferred at $Q^2_{r, {\rm min}}(r) \leq Q_r^2 \leq 1$ to $Q_r^2=0$ [see Figure \ref{fig:appendix_geometry}(b)]. 
At larger $r$, $Q_{r,\rm{min}}^2$ becomes increasingly closer to $1$, 
so that the estimation of the slope ${\rm d} \sigma_{\rm los}^2 / {\rm d} Q_r^2$ 
becomes increasingly more difficult. 
This difficulty results in large uncertainty in $\sigma_\theta^2$ at large $r$ [see Figure \ref{fig:appendix_geometry}(c)],
and thus sometimes the maximum-likelihood routine 
finds unphysical solutions of $\sigma_\theta^2 = 0$ (if a constraint of $\sigma_\theta^2 \geq 0$ is imposed), 
or even $\sigma_\theta^2 < 0$ (if such a constraint is not imposed; see Appendix \ref{appendix:unphysical}). 
On the other hand, the estimation of $\sigma_r^2$ is not difficult, 
since $\sigma_r^2$ is approximately the observed value of $\sigma_{\rm los}^2$ at $Q_r^2 \simeq 1$.

\subsection{Ideal distribution of sample stars}

Let us consider a case where we have $N (\gg 1)$ sample stars at a Galactocentric radius $r$, 
and half of them ($N/2$ stars) are observed in the direction of $Q_r^2=1$ (hereafter `QMAX direction')
and the other half of them are observed in the direction of $Q_r^2=Q_{r,\rm{min}}^2(r)$ (hereafter `QMIN direction'). 
This spatial distribution of stars is not realistic, 
but is ideal for inferring the slope ${\rm d} \sigma_{\rm los}^2 / {\rm d} Q_r^2$ with local fitting method.

In this case, the true values of $\sigma_{\rm los}^2$ in the QMAX and QMIN directions are given by 
\eq{
&\sigma_{\rm los}^2 (QMAX) = \sigma_r^2 , \\
&\sigma_{\rm los}^2 (QMIN) = \sigma_r^2 + \frac{R_0^2}{r^2} (\sigma_\theta^2 - \sigma_r^2) = \sigma_r^2 \left[ 1 - \frac{R_0^2}{r^2} \beta_{\rm true} \right], 
}
respectively. 
By using observed values of $\sigma_{\rm los}^2 (QMIN)$ and $\sigma_{\rm los}^2 (QMAX)$, 
the value of $\beta$ can be expressed as
\eq{
\beta = \frac{r^2}{R_0^2} \left( 1- \frac{ \sigma_{\rm los}^2 (QMIN) }{ \sigma_{\rm los}^2 (QMAX) } \right) .
\label{eq:Beta_QmaxQmin}
}

Since the distribution of $v_{\rm los}$ follows a Gaussian distribution [see equation (\ref{eq:GaussLOS})], 
the observed values of $\sigma_{\rm los}^2$ in the QMAX and QMIN directions are associated with uncertainties of 
\eq{
\Delta \sigma_{\rm los}^2 (QMAX) = \frac{\sqrt{2} }{\sqrt{(N/2)}} \sigma_r^2 ,
}
\eq{
\Delta \sigma_{\rm los}^2 (QMIN) = \frac{\sqrt{2}  }{\sqrt{(N/2)}} 
\sigma_r^2 \left[ 1 - \frac{R_0^2}{r^2} \beta_{\rm true} \right],
}
respectively. 
By using equation (\ref{eq:Beta_QmaxQmin}) 
and by assuming that the uncertainties $\Delta \sigma_{\rm los}^2 (QMAX)$ and  $\Delta \sigma_{\rm los}^2 (QMIN)$ are not correlated, 
we can express the uncertainty in $\beta$ as follows:
\eq{
|\Delta \beta|_{\mathrm{ideal}} 
= 
\left(\frac{r^2}{R_0^2}\right) 
\left\{ 
\left( \frac{ \sigma_{\rm los}^2 (QMIN) }{  \sigma_{\rm los}^4 (QMAX) } \Delta \sigma_{\rm los}^2 (QMAX)  \right)^2 
+\left( \frac{ \Delta \sigma_{\rm los}^2 (QMIN)  }{ \sigma_{\rm los}^2 (QMAX) }  \right)^2 
\right\}^{1/2} 
= \sqrt{\frac{8}{N}} \left( \frac{r^2}{R_0^2} - \beta_{\rm true} \right) .
}

\subsection{Realistic distribution of sample stars}

In reality, the sample stars are distributed in a wide area in $(\ell, b)$-space, 
and are not confined around the QMAX and QMIN directions. 
Therefore, we need to rescale the value of $|\Delta \beta|_{\mathrm{ideal}}$. 
From our results in Section \ref{section:resultMaxL}, we find that 
\eq{
|\Delta \beta| = 2 \sqrt{\frac{8}{N}} \left( \frac{r^2}{R_0^2} - \beta_{\rm true} \right) 
}
is a good approximation to the random error on $\beta$. 
This expression clearly illustrates 
how the performance of the maximum-likelihood method 
deteriorates when $r$ increases and improves when $N$ increases. 
For example, with $N=10^3$ and $\beta_{\rm true}=0.5$, 
we see that $\Delta \beta$ is smaller than $(1- \beta_{\rm true})$ only at $r < 14.5 \kpc$ 
and hence estimation of $\beta$ is not reliable beyond this radius (as discussed in Sections \ref{section:resultMaxL} and \ref{section:resultBayes}). 
However, when $N=10^4$ and $\beta_{\rm true}=0.5$ are assumed, 
$\Delta \beta$ is smaller than $(1- \beta_{\rm true})$ at $r<24.4 \kpc$.

\section{Origin of the unrealistic solutions in maximum-likelihood analyses} \label{appendix:unphysical}

In Section \ref{section:showCase}, we showed that the maximum-likelihood method 
returns unrealistic solutions with $\sigma_\theta=0$ or $\sigma_\phi=0$ for a fraction of mock catalogues at $r \gtrsim 15 \kpc$. 
In Appendix \ref{appendix:geometry}, we explained with an geometrical argument that 
it becomes increasingly more difficult to estimate the tangential components of the velocity dispersion at larger $r$, 
and that unphysical solutions arise due to this difficulty.
Here we investigate the origin of the unrealistic solutions from a different perspective, 
by performing maximum-likelihood analyses for 100 mock catalogues with $\beta_{\rm true}=0.5$ and $r=15 \kpc$ in two ways.

The first set of analyses are done with the same formulation 
as in Section \ref{section:formulationMaxL}. 
To be specific, we search for a set of parameters 
$(\sigma_r, \sigma_\theta, \sigma_\phi, V_{\rm rot})$ 
that maximize the log-likelihood $\ln L$ [see equation (\ref{eq:lnL})]
under the condition that 
$0 \leq \sigma_k < \infty$ $(k=r, \theta, \phi)$ and $-\infty<V_{\rm rot}<\infty$. 
Since $0\leq \sigma_k$ is a physical requirement, 
we refer to these solutions as {\it physical solutions}.

The second set of analyses are done with a different parametrization.
Here, 
we define new variables 
$(s_r, s_\theta, s_\phi) = (\sigma^2_r, \sigma^2_\theta, \sigma^2_\phi)$ 
and we search for a set of parameters 
$(s_r, s_\theta, s_\phi, V_{\rm rot})$ 
that maximize $\ln L$ under the conditions of 
$-\infty <s_k < \infty$ $(k=r, \theta, \phi)$ and $-\infty<V_{\rm rot}<\infty$. 
Obviously, when $s_k <0$ for any of $k=r, \theta, \phi$, 
there is no physical distribution function given by equation (\ref{eq:DF}). 
However, since $\ln L$ is a function of 
$(\sigma^2_r, \sigma^2_\theta, \sigma^2_\phi, V_{\rm rot})$, 
it is mathematically justified to search for the solutions 
with negative values of $s_k$. 
Hereafter, we refer to these solutions as {\it mathematical solutions}.

\begin{figure}
\begin{flushleft}
	\includegraphics[angle=0,width=0.47\columnwidth]{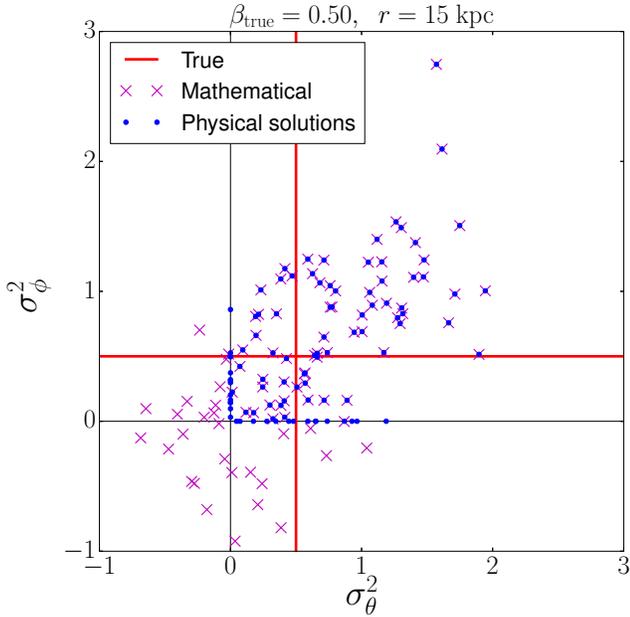}
\end{flushleft}
\caption{
A comparison of the mathematical and physical solutions of the maximum-likelihood analyses. 
Here we use 100 mock catalogues with $\beta_{\rm true}=0.5$ and $r=15 \kpc$. 
Note that both $\sigma_\theta^2$ and  $\sigma_\phi^2$ are normalized by a constant $(100 \; \kms)^2$. 
}
\label{fig:appendix_zero_sigma}
\end{figure}

Figure \ref{fig:appendix_zero_sigma} shows the distribution of the 
physical and mathematical solutions 
in $(\sigma^2_\theta, \sigma^2_\phi)$-space 
[or equivalently, $(s_\theta, s_\phi)$-space]. 
The blue dots and magenta crosses indicate 
the physical and mathematical solutions, respectively. 
We see that the mathematical solutions are 
more or less distributed around the exact location of 
$(\sigma^2_{\theta, {\rm true}}, \sigma^2_{\phi, {\rm true}})$
with a rather large scatter. 
Since $(\sigma^2_{\theta, {\rm true}}, \sigma^2_{\phi, {\rm true}})$ is 
located within the first quadrant, 
mathematical solutions for a large fraction of mock catalogues 
are located within the first quadrant. 
In such cases, the physical and mathematical solutions are identical. 
However, for a fraction of mock catalogues,
the mathematical solutions are located outside the first quadrant.
In such cases, the corresponding physical solutions 
are located along the axes of $\sigma_\theta^2=0$ or $\sigma_\phi^2=0$.

The origin of these unrealistic solutions can be explained in a following manner. 
At $r\gtrsim 15 \kpc$, the log-likelihood $\ln L$  depends weakly on 
$(s_\theta, s_\phi)$ (as mentioned in Section \ref{section:resultBayesFixedBeta}). 
Therefore, for a fraction of mock catalogues, 
depending on the spatial distribution or velocity distribution of the sample stars, 
$\ln L$ happens to attain its maximum at $s_\theta<0$ or $s_\phi<0$. 
In such situations, 
$\ln L$ increases as $s_\theta$ or $s_\phi$ decreases within the first quadrant, 
so that the physical solutions are distributed along the axes.

Since the scatter in the mathematical solutions of $s_\theta$ and $s_\phi$ 
becomes larger with increasing $r$ 
(due to the enhanced difficulties in extracting 
the information regarding tangential velocity components), 
a larger fraction of mathematical solutions are located outside the 
first quadrant in $(\sigma^2_\theta, \sigma^2_\phi)$-space. 
This is why the fraction of unrealistic physical solutions with $\sigma_\theta=0$ or $\sigma_\phi=0$ 
increases with increasing $r$, as seen in Figure \ref{fig:showCaseBetap0.5_MaxL}.

\section{Experiments with larger sample size} \label{appendix:N1e4}

In Sections \ref{section:resultMaxL} and \ref{section:resultBayes}, 
we use $N=10^3$ stars at a given radius.  Here we briefly explore whether the results improve if we use $N=10^4$ stars instead. 

To this end, we generated 1000 mock catalogues containing $N=10^4$ stars for each pair of $(r, \beta_{\rm true})$. 
We adopted four values of $r/\kpc=10,15,20,$ and $25$ and $\beta_{\rm true}=0.75, 0.5, 0,$ and $-1$. 

\begin{figure}
\begin{center}
	\includegraphics[angle=0,width=0.9\columnwidth]{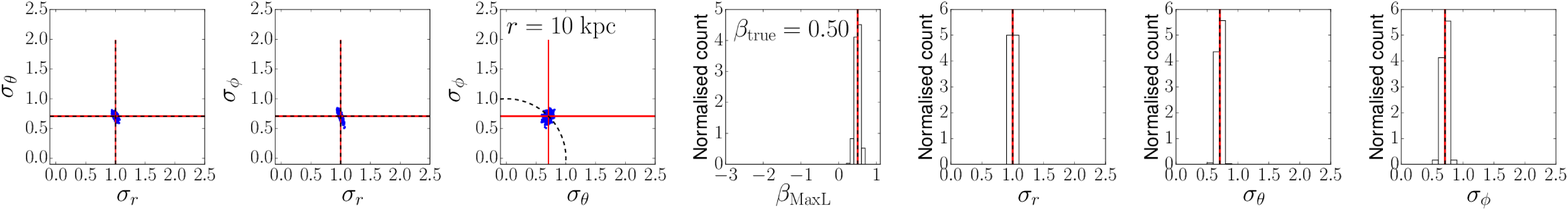} \\
	\includegraphics[angle=0,width=0.9\columnwidth]{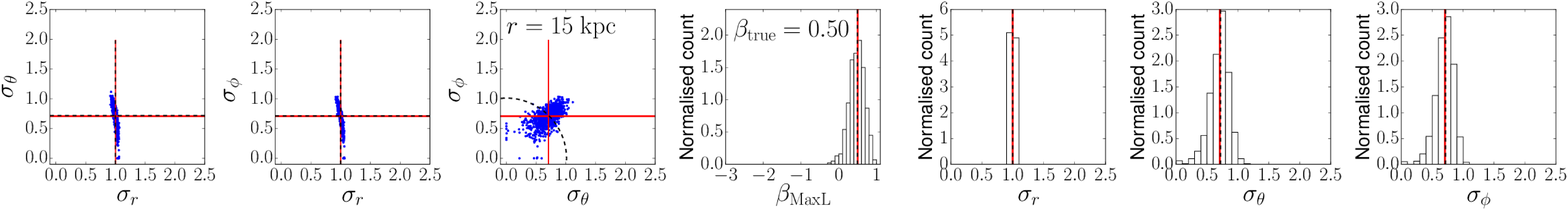} \\
	\includegraphics[angle=0,width=0.9\columnwidth]{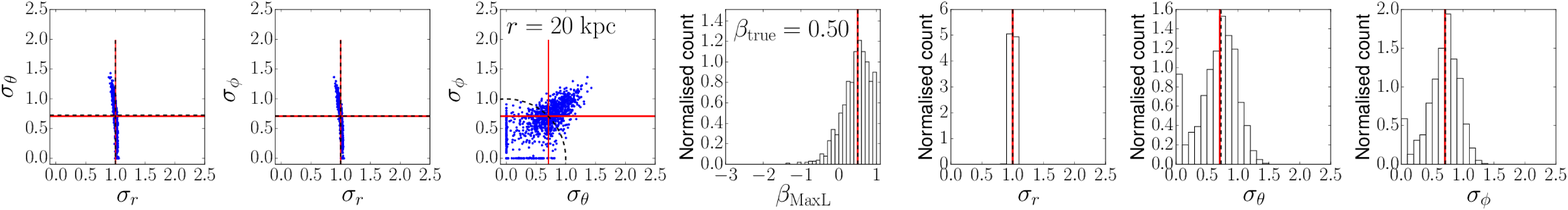} \\
	\includegraphics[angle=0,width=0.9\columnwidth]{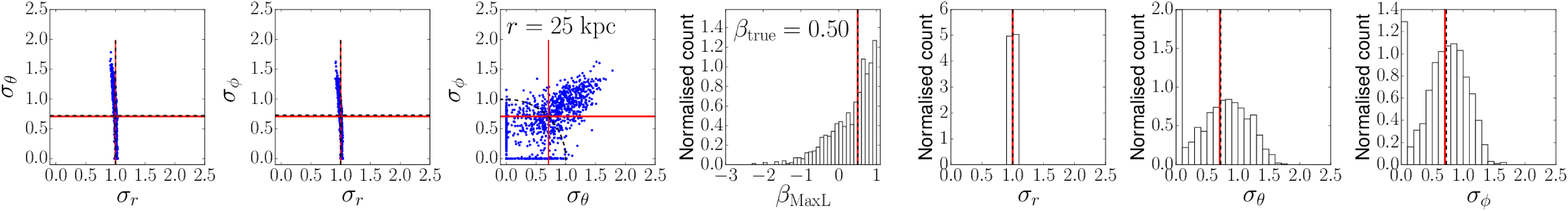}
\end{center}
\caption{
Figure showing the same as in Figure \ref{fig:showCaseBetap0.5_MaxL}, but with the use of $N=10^4$ stars in each mock catalogue. 
}
\label{fig:N1e4showCaseBetap0.5_MaxL}
\end{figure}

\begin{figure}
\begin{center}
	\includegraphics[angle=0,width=0.9\columnwidth]{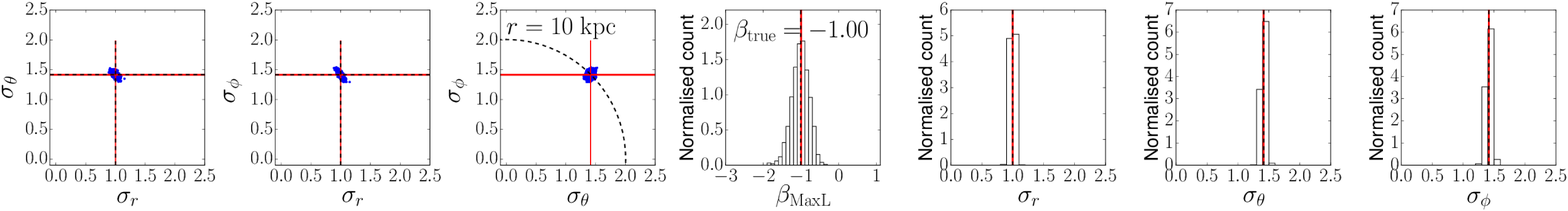} \\
	\includegraphics[angle=0,width=0.9\columnwidth]{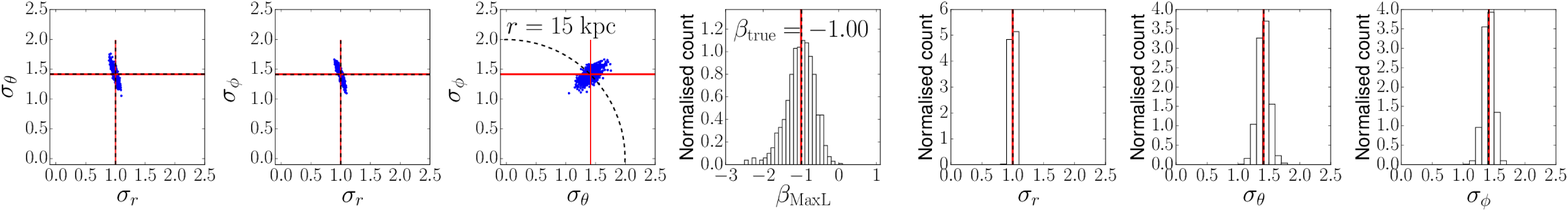} \\
	\includegraphics[angle=0,width=0.9\columnwidth]{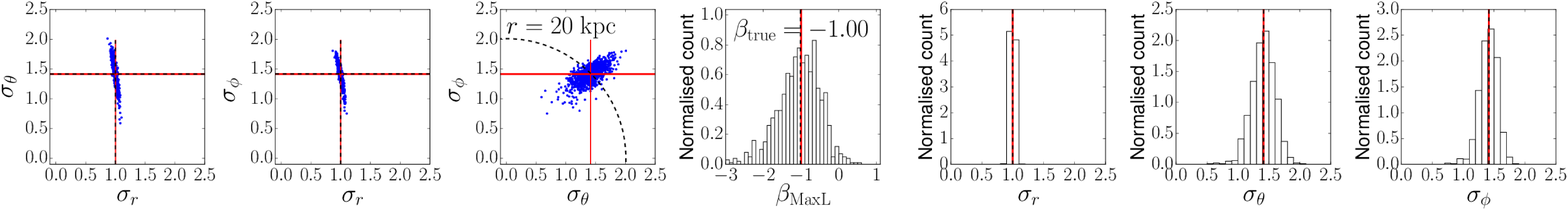} \\
	\includegraphics[angle=0,width=0.9\columnwidth]{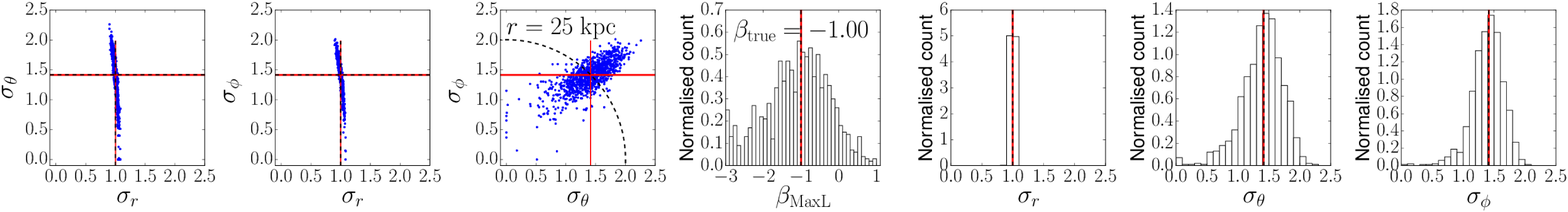}
\end{center}
\caption{
Figure showing the same as in Figure \ref{fig:showCaseBetam1.0_MaxL}, but with the use of $N=10^4$ stars in each mock catalogue. 
}
\label{fig:N1e4showCaseBetam1.0_MaxL}
\end{figure}

First, we did exactly the same analyses as in Section \ref{section:showCase} by using these mock catalogues. 
Figures \ref{fig:N1e4showCaseBetap0.5_MaxL} and \ref{fig:N1e4showCaseBetam1.0_MaxL}  
show the distributions of $\beta_{\rm MaxL}$ 
for $\beta_{\rm true}=0.5$ and $\beta_{\rm true}=-1$, respectively. 
A comparison of Figures \ref{fig:showCaseBetap0.5_MaxL} and \ref{fig:N1e4showCaseBetap0.5_MaxL}  
suggests that the performance of the maximum-likelihood method is better when $N=10^4$  than when $N=10^3$ at all the radii explored here. 
From Figure  \ref{fig:N1e4showCaseBetap0.5_MaxL}, 
we note that the histogram of $\beta_{\rm MaxL}$ is peaked at around $\beta_{\rm true}=0.5$ at $r\leq20 \kpc$, 
and that the median value of $\beta_{\rm MaxL}$ coincides with $\beta_{\rm true}$ even at $r=25 \kpc$. 
Also, Figure \ref{fig:N1e4showCaseBetam1.0_MaxL} suggests that 
the peak of the histogram of $\beta_{\rm MaxL}$ as well as the median value of $\beta_{\rm MaxL}$ coincide with 
$\beta_{\rm true}=-1$ at all the radii explored here. 
The peaked histogram of $\beta_{\rm MaxL}$ at $r=25 \kpc$ seen in Figure \ref{fig:N1e4showCaseBetam1.0_MaxL} 
is in contrast to the highly flattened histogram at $r=25 \kpc$ seen in Figure \ref{fig:showCaseBetam1.0_MaxL}.

\begin{figure}
\begin{center}
	\includegraphics[angle=0,width=0.24\columnwidth]{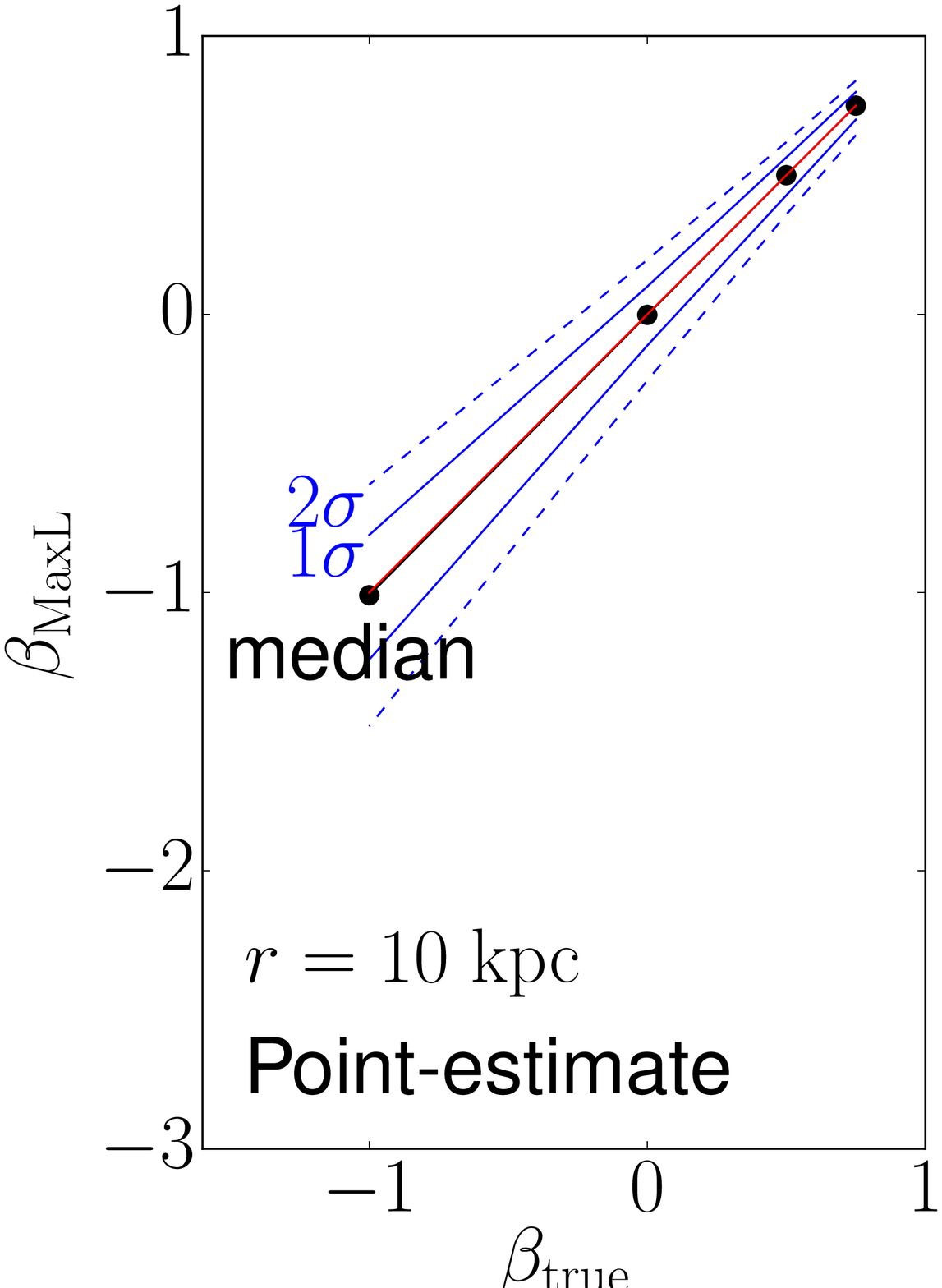}
	\includegraphics[angle=0,width=0.24\columnwidth]{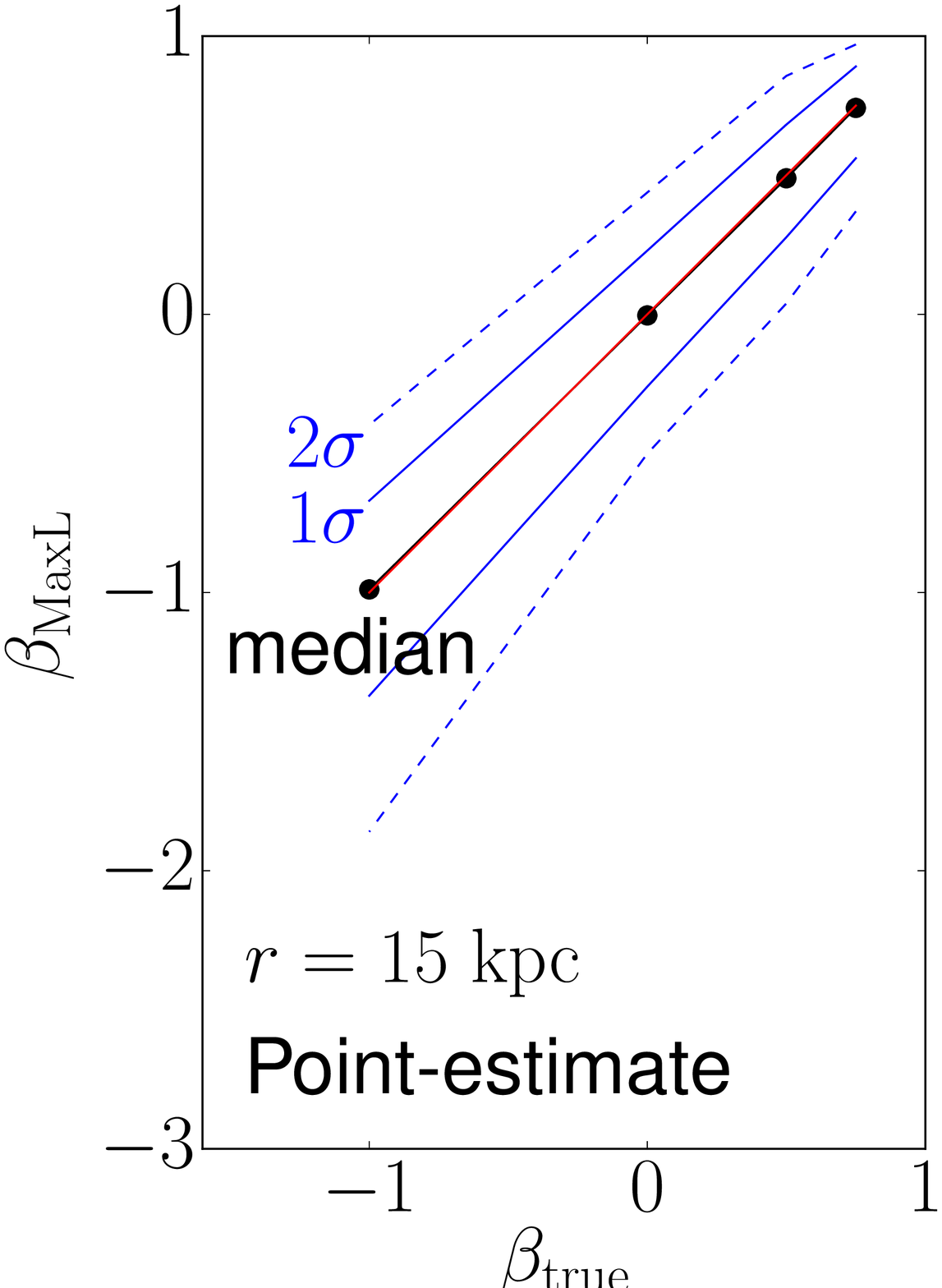}
	\includegraphics[angle=0,width=0.24\columnwidth]{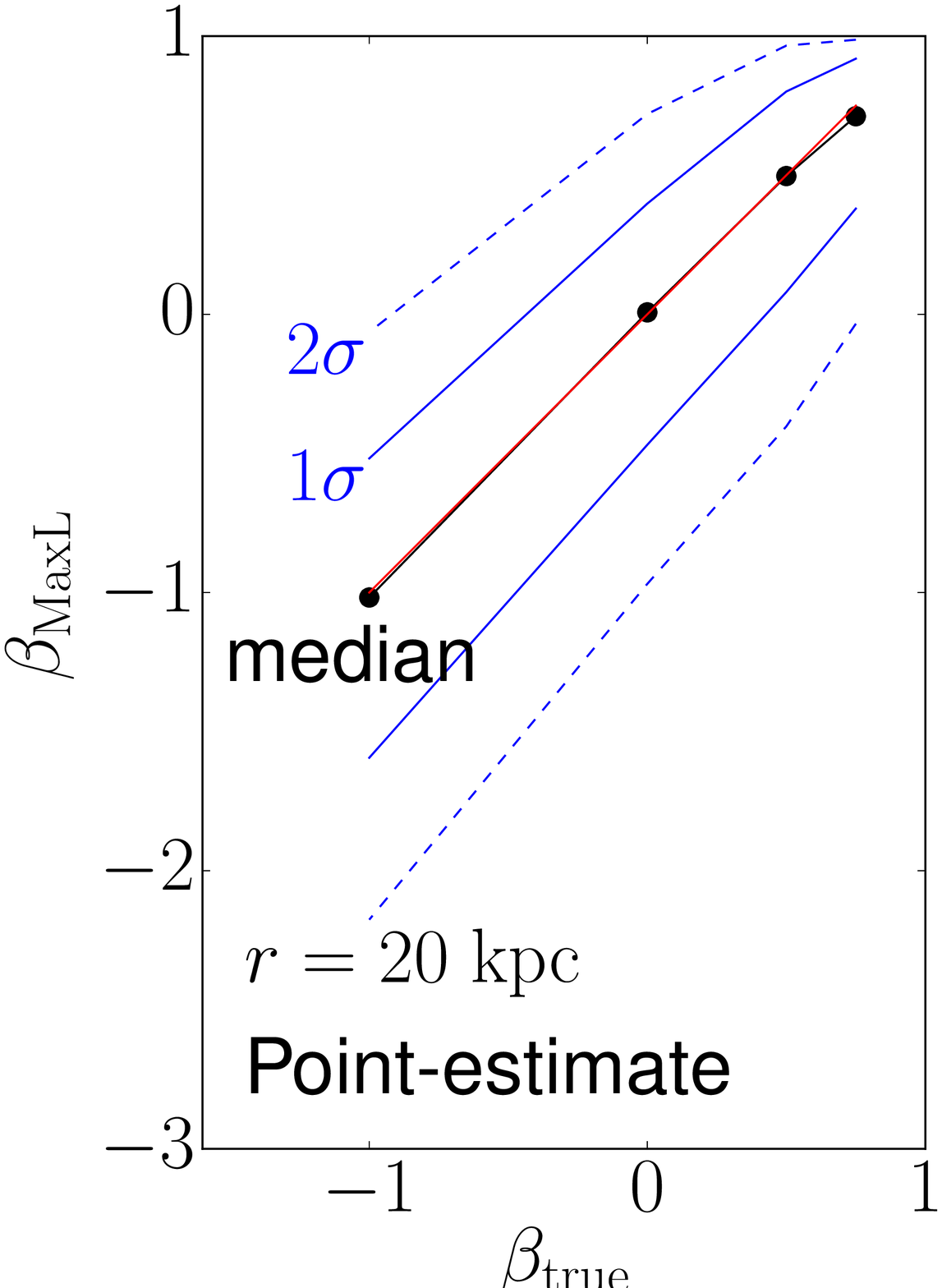}
	\includegraphics[angle=0,width=0.24\columnwidth]{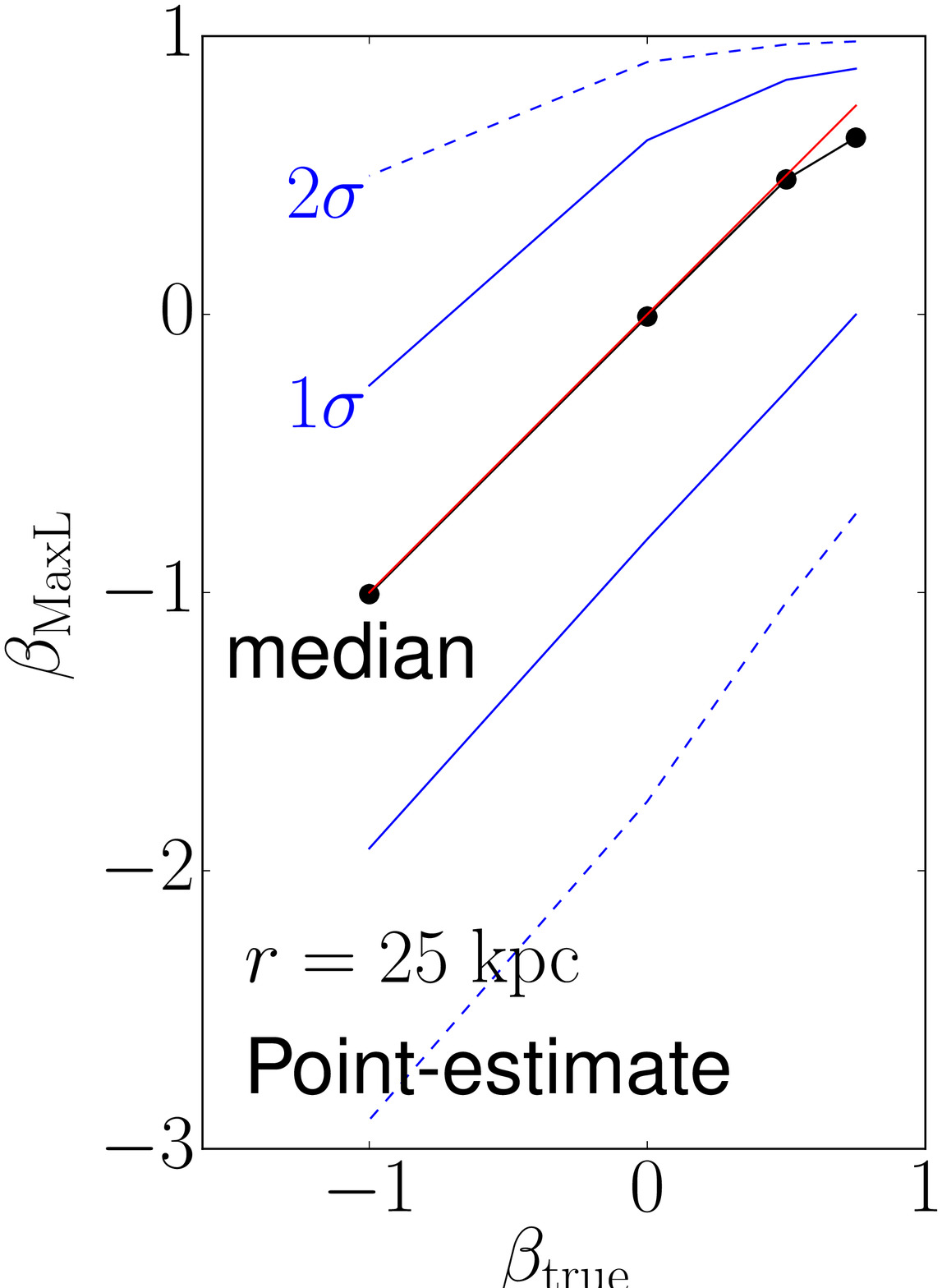} 
\end{center}
\caption{
Figure showing the same as in Figure \ref{fig:fixedrGC_bootstrap_MaxL}, but with the use of $N=10^4$ stars in each mock catalogue. 
}
\label{fig:N1e4fixedrGC_bootstrap_MaxL}
\end{figure}

Secondly, we did the same analyses as in Section \ref{section:MaxL_fixedrGC} 
by using the mock catalogues with $N=10^4$ stars.
Figure \ref{fig:N1e4fixedrGC_bootstrap_MaxL} shows the 
distribution of $\beta_{\rm MaxL}$ as a function of $\beta_{\rm true}$ 
for different Galactocentric radius $r$ of sample stars. 
We see that the median value of $\beta_{\rm MaxL}$ almost perfectly coincides with $\beta_{\rm true}$ at $r \leq 25 \kpc$. 
Also, we found that the one- and two-$\sigma$ ranges of the posterior distribution of $\beta_{\rm MaxL}$ 
for the case of $N=10^4$ seen in Figure \ref{fig:N1e4fixedrGC_bootstrap_MaxL} 
are significantly smaller than the corresponding ranges 
for the case of $N=10^3$ seen in Figure \ref{fig:fixedrGC_bootstrap_MaxL}.

These results indicate that 
the maximum-likelihood method can in principle reliably estimate $\beta$ 
at $r\leq 25 \kpc$ if we have $N=10^4$ stars at a given radius.

\section{Two distribution function models} \label{appendix:twoDF}

Here describe some details on the two distribution function models used in Section \ref{section:other_estimates}, 
$f_{\rm const}(E,L)$ and $f_{\rm OM}(E,L)$,
which are functions of energy $E$ and total angular momentum $L$. 
In the following, we assume that the potential of the Milky Way is spherical and is expressed as 
$\Phi (r)  = v_0^2 \ln (r / r_0)$ with $(r_0, v_0) = (220 \kms, 8 \kpc)$.

\subsection{Constant $\beta$ model}

The model with const $\beta(r)$ is given by 
\eq{
f_{\rm const}(E,L) = A \exp  \left[ -(\alpha+2) \frac{E}{v_0^2} \right] \left( L_{\rm max}^2 (E) - K L^2\right).
}
Here, $L_{\rm max}(E) = r_0 v_0 \exp(2E/v_0^2 - 1)$ is the angular momentum of a star with energy $E$ moving on a circular orbit. 
We note that $f_{\rm const}(E,L) \geq 0$ is always satisfied if $K \leq 1$. 
The density profile of this distribution function is given by $\rho(r) \propto r^{-\alpha}$ ($\alpha>0$). 
The velocity anisotropy is governed by $(K, \alpha)$ and expressed as 
\eq{
\beta(r) = \beta_{\rm const} = \frac{K}{\frac{1}{2} (\alpha+2)^{7/2} \alpha^{-5/2} \exp(-1) - K} . 
}
The probability density that a star at $\vector{x}$ characterized by $Q_r = \vector{e}_{\rm los} \cdot \vector{e}_r $ 
has a line-of-sight velocity $v_{\rm los}$ is expressed as 
\eq{
P(\vlos | \vector{x} , K, \alpha) 
&= \frac{1}{\sqrt{2 \pi} v_0} \exp \left[ - \frac{\alpha}{2} \left( \frac{\vlos}{v_0} \right)^2 \right] 
\frac
{1}
{(\alpha+2)^{5/2} - 2 K \alpha^{3/2} \exp(1) } \\
&\times \left\{
\alpha^{1/2} (\alpha+2)^{5/2} 
- K \alpha^{3/2} (\alpha+2)^{1/2} \left[1 + Q_r^2 + (\alpha+2) (1-Q_r^2) \left( \frac{\vlos}{v_0} \right)^2 \right]  
\exp \left[ 1 - \left( \frac{\vlos}{v_0} \right)^2 \right] 
\right\} .
}
In Section \ref{section:other_estimates}, 
the mock catalogues with $\beta_{\rm const}=0.25$ and $-0.42$ are generated by assuming $(K, \alpha) = (0.84, 2)$ and $(-3, 2)$, respectively. 
Given the mock data, the local fitting method finds the pair $(K, \alpha)$ that maximizes the likelihood.

\subsection{Osipkov-Merritt model} \label{appendix:OM}

Osipkov-Merritt model is a broad class of distribution functions 
that only depends on $Q=E + L^2/(2 r_a^2)$ with $r_a$ a constant. 
Here we adopt a family of functions of the form 
\eq{
f_{\rm OM}(E,L) = A \exp  \left[ -\alpha \frac{E + L^2/(2 r_a^2)}{v_0^2} \right] 
}
with $\alpha>0$. 
The density profile of this distribution function is given by 
\eq{
\rho(r) = (2\pi)^{3/2} v_0^3 \alpha^{-3/2} 
A {\left(\frac{r}{r_0}\right)}^{- \alpha } 
{\left (1+\frac{r^2}{r_a^2} \right)}^{-1} .
}
The velocity anisotropy is given by $\beta(r) = r^2 / (r_a^2 + r^2)$. 
The probability density that a star at $\vector{x}$ characterized by $Q_r = \vector{e}_{\rm los} \cdot \vector{e}_r$ 
has a line-of-sight velocity $v_{\rm los}$ is expressed as 
\eq{
P( \vlos | \vector{x}, r_a, \alpha ) 
=  \frac
{1}{\sqrt{2 \pi} \sigma_{\rm los}}
\exp \left[ -\frac{\vlos^2}{2\sigma^2_{\rm los}} \right],
\label{eq:PDF_vlos}
}
where the line-of-sight velocity dispersion is given by 
\eq{
\sigma^2_{\rm los} = \frac{ (r_a^2 + r^2 Q_r^2) } { (r_a^2 + r^2) } \cdot \frac{v_0^2}{\alpha} .
\label{eq:sigma_vlos}
}
In Section \ref{section:other_estimates}, 
the mock catalogues are generated by assuming $(r_a, \alpha) = (10 \kpc, 2)$. 
Given the mock data, the local fitting method finds the pair $(r_a, \alpha)$ that maximizes the likelihood.

\end{document}